\documentclass[fleqn,usenatbib]{mnras}
\usepackage{orcidlink}
\usepackage{newtxtext,newtxmath}
\usepackage[T1]{fontenc}
\usepackage[normalem]{ulem}

\DeclareRobustCommand{\VAN}[3]{#2}
\let\VANthebibliography\thebibliography
\def\thebibliography{\DeclareRobustCommand{\VAN}[3]{##3}\VANthebibliography}

\usepackage{graphicx}	% Including figure files
\usepackage{amsmath}	
\usepackage{threeparttable}

\def\kt{\mathrm{kT}} 
\def\ab{\mathrm{Z}} 
\def\nh{\mathrm{N}_{\mathrm{H}}} 
\def\kev{\mathrm{keV}}

\def\nhunit{10^{20}\mathrm{cm}^{-2}}
\def\sx{\mathrm{\Sigma}_\mathrm{x}}

\def\ficm{\mathrm{F_{\mathrm{ICM}}}} 
\def\fbck{\mathrm{F_{\mathrm{bck}}}}

\def\evmap{\mathrm{N}}
\def\vfmap{\mathrm{w}}
\def\bckmap{\mathrm{B}}

\def\evcub{\mathrm{n_{evt}}}
\def\eecub{\mathrm{E}}
\def\bckcub{\mathrm{N_{bck}}}

\newcommand{\simgt}{\,\hbox{\lower0.6ex\hbox{$\sim$}\llap{\raise0.6ex\hbox{$>$}}}\,}
\newcommand{\simlt}{\,\hbox{\lower0.6ex\hbox{$\sim$}\llap{\raise0.6ex\hbox{$<$}}}\,}

\title[]{Deep insights into Abell 2163: unveiling the treasure trove of ICM plasma physics}

\author[R. Santra et al.]{
R. Santra,$^{1,2}$\thanks{E-mail: ramananda1999@gmail.com}\orcidlink{0009-0002-0373-570X}
R. Kale,$^{1}$\orcidlink{0000-0003-1449-3718}
S. Giacintucci,$^{3}$\orcidlink{0000-0002-1634-9886}
H. Bourdin,$^{4,5}$
R. Jain,$^{1}$\orcidlink{0009-0002-9692-3899}
A. Botteon,$^{2}$\orcidlink{0000-0002-9325-1567}
G. Brunetti$^{2}$\orcidlink{0000-0003-4195-8613}
\\
$^{1}$National Centre for Radio Astrophysics, Tata Institute of Fundamental Research, Pune 411007, India\\
$^{2}$INAF - IRA, Via Gobetti 101, I-40129 Bologna, Italy\\
$^{3}$Naval Research Laboratory, 4555 Overlook Avenue SW, Code 7213, Washington, DC 20375, USA\\
$^{4}$Dipartimento di Fisica, Università di Roma Tor Vergata, Via della Ricerca Scientifica 1, I-00133 Roma, Italy\\
$^{5}$INFN, Sezione di Roma 2, Università di Roma Tor Vergata, Via della Ricerca Scientifica, 1, Roma, Italy
}

\pubyear{\the\year{}}

\begin{document}
\label{firstpage}
\pagerange{\pageref{firstpage}--\pageref{lastpage}}
%\linenumbers
\maketitle

\begin{abstract}
Nonthermal emission observed in galaxy clusters provides a direct probe into the plasma physics of the intra-cluster medium (ICM) under extreme conditions. We report the first detailed analysis of the giant radio halo in the merging galaxy cluster Abell 2163, using upgraded Giant Metrewave Radio Telescope (uGMRT) and Very Large Array (VLA) observations. Combining radio data (300--1400 MHz) with archival X-ray data offers a unique opportunity to study the complex ICM physics of the cluster. The sensitive uGMRT observations map the halo emission for the first time out to an extent of $\sim$3.3 Mpc, up to $r_{500}$, and also effectively recover other diffuse sources. The radio surface brightness profile is well fitted with an exponential function up to r$_{500}$, with an evolution of e-folding radius over frequencies (larger at low frequencies). The spatially resolved spectral index map reveals fluctuations and outward radial steepening of the average spectral index. Radio and X-ray surface brightness are well correlated, with a correlation slope of $\sim$0.70 for the halo, and $\sim$0.40 for the ridge. The correlation slope varies from cluster centre to outskirts, suggesting the magnetic field and thermal gas density scaling relation (B $\propto n_{e}^{0.5}$) should be reassessed, provided that the re-acceleration efficiency is constant. We propose that diffuse lobes at the periphery could serve as a reservoir for seed electrons, behind the radio halo emission, with an estimated acceleration efficiency reaching $\sim$0.1\% in the external regions. Additionally, a major E-W merger is suggested, leading to turbulence in the ICM and generating the halo.

\end{abstract}

% Select between one and six entries from the list of approved keywords.
% Don't make up new ones.


%%%%%%%%%%%%%%%%%%%%%%%%%%%%%%%%%%%%%%%%%%%%%%%%%%

%%%%%%%%%%%%%%%%% BODY OF PAPER %%%%%%%%%%%%%%%%%%

\section{Introduction} \label{sec:intro}

Galaxy clusters grow by accretion of matter from the surrounding galaxy groups and other clusters through filaments, inducing turbulence within the intracluster medium (ICM), the largest baryonic component of the galaxy cluster \citep[e.g.,][]{1982ARA&A..20..547F}. As a result, substantial energy (10$^{63}$ $-$ 10$^{64}$ erg) is released during these mergers, leading to the acceleration of cosmic rays (CR) and the amplification of the magnetic field within the ICM. These processes are observed through diffuse radio emissions in galaxy clusters, generally classified as radio halos and relics \citep[see reviews;][]{2012A&ARv..20...54F, 2019SSRv..215...16V, 2023JApA...44...38P}. These radio sources have steep spectra\footnote{S$_{\nu}$ $\propto$ $\nu^{\alpha}$, where S$_{\nu}$ is the flux density at the frequency $\nu$, and $\alpha$ is the spectral index} ($\alpha$ \textless -1), indicating the particle acceleration to be inefficient. Although synchrotron radio emissions have been detected throughout the cluster volumes, the physical  mechanisms are still under investigation \citep{2014IJMPD..2330007B, doi:10.1126/sciadv.abq7623}.

Radio relics are diffuse sources at galaxy cluster outskirts, thought to trace outgoing merger shock waves, with high linear polarization (up to 60 $-$ 70\%) and aligned magnetic fields \citep[e.g.,][]{2008A&A...486..347G, 2010Sci...330..347V, Kale12,2021ApJ...911....3D,2021A&A...646A.135R,pal25}. It is accepted that part of the shock's kinetic energy powers relic radio emission via Diffusive Shock Acceleration (DSA) of cosmic ray electrons \citep[DSA;][]{1998A&A...332..395E}. There is ongoing debate about whether electrons accelerate from a thermal pool (keV) to relativistic energies (GeV) \citep[standard scenario;][]{1998A&A...332..395E,2007MNRAS.375...77H} or from an existing mildly relativistic electron population \citep[re-acceleration;][]{2005ApJ...627..733M,2011ApJ...734...18K,2016ApJ...823...13K}. The standard scenario is challenged by the non-detection of gamma rays from clusters \citep[e.g.,][]{vazza15,vazza16} and by energetic arguments, if particles are accelerated directly from the thermal pool \citep{2020A&A...634A..64B}.

Giant radio halos extending on Mpc scales are preferentially found in the centre of massive and disturbed systems \citep[for reviews;][]{2019SSRv..215...16V}. The currently favoured scenario for the generation of the radio halo involves the re-acceleration of the pre-existing relativistic electrons, due to turbulence originating in cluster mergers \citep[re-acceleration models][]{2001MNRAS.320..365B,petrosian2001nonthermal,2006AN....327..557C, 2007MNRAS.378..245B, 2016MNRAS.458.2584B}. The contribution from secondary electrons due to hadronic collisions \citep[e.g.,][]{1999NuPhS..70..495B} has been constrained to be subdominant from Fermi-LAT observations and by arguments based on the energetics of CRs \citep[e.g.,][]{2012MNRAS.426...40B, 2021A&A...648A..60A}, yet secondary electrons may contribute significantly if they are re-accelerated by turbulence during mergers \citep[e.g.,][]{2005MNRAS.363.1173B,2011MNRAS.412..817B,2024arXiv240813846N}. Although the current observational data support turbulent re-acceleration model, the energy cascading from Mpc scales down to smaller ones is still poorly understood \citep[e.g.,][]{2014IJMPD..2330007B}. Furthermore, the origin of the seed electrons remainsnuncertain, with cluster radio galaxies being a natural reservoir of non-thermal components in the ICM \citep[e.g.,][]{2024Galax..12...19V}. Recent radio observations reveal an increasing complexity of the phenomenology of diffuse sources. In particular, diffuse radio emission has been recently detected up to large distances from the cluster centre \citep[e.g.,][]{2020ApJ...897..115S, 2021A&A...646A.135R, 2022SciA....8.7623B, 2022Natur.609..911C,2023A&A...678A.133B,salunkhe25,rajpurohit25} and between pairs of clusters in pre-merger phases \citep[e.g.,][]{2019Sci...364..981G,2020MNRAS.499L..11B, 2022A&A...668A.107D, 2023A&A...679A.107B, 2024A&A...685L..10P,danhu24}, thus confirming the presence of non-thermal components and efficient physical processes that amplify the magnetic field at such large scales. \citet{2022Natur.609..911C} reported on the discovery of a new type of diffuse source in four merging clusters (ZwCl 0634.1+4750, A665, A697, and A2218) that surrounds classical giant radio halos, and has been called megahalo, with an extension of the diffuse emission up to scales $\sim$ r$_{500}$ (which marks the radius that encloses a mean overdensity of 500 times the critical density of the universe at the cluster redshift) and exhibits a shallower surface brightness radial profile than that of their embedded radio halo, are characterized by a very steep spectrum ($\alpha$ \textless -1.6, between 50 and 144 MHz). Not many giant radio halos have been followed with the deep exposure to map the faint extended emission in the cluster outskirts, however, the sensitive new generation of radio telescopes such as the LOw Frequency ARray (LOFAR), upgraded Giant Metrewave Radio Telescope (uGMRT), Meer Karoo Array Telescope (MeerKAT), and Karl G. Jansky Very Large Array (VLA) will play a crucial role in constraining properties of the emission in these peripheral regions.

Studying targets with uncommon features can shed light on the physical properties that underlie standard classification schemes for galaxy clusters. The massive galaxy cluster Abell 2163 (hereafter A2163) is a favourable target for investigating the role of intricate environmental and dynamic conditions at different physical scales. This work presents the deep uGMRT and archival VLA radio observations of the highly disturbed cluster A2163. We have also used archival XMM-\textit{Newton} data sets to better understand the connection between the cluster dynamics and large-scale diffuse emission. The XMM-\textit{Newton} data include archival datasets (P.I. H. Bourdin) with a total exposure (MOS and PN) of 246.7 ks on the cluster (after cleaning for background flares). The XMM-\textit{Newton} images and temperature maps have not been published before and are presented here for the first time.

\begin{table}
  \centering
  \caption{Summary of uGMRT observations.}
  \begin{tabular}{@{}lcc@{}}
    \hline
      &  Band 3  & Band 4 \\
    \hline
Frequency range & 300-500 MHz & 550-750 MHz  \\ 
 
 No. of channels & 2048  & 2048 \\
 
 Bandwidth: total/effective & 200/156 MHz & 200/177 MHz \\

 No of Antennas & 29 & 30 \\
 
 On source time & 10 Hr.  & 10 Hr.  \\

 Largest angular scale & 1920$''$  & 1020$''$  \\

 Shortest baseline (in units of $\lambda$) & 150  & 200  \\
   \hline
  \end{tabular}
  \label{obs_tab}
  \begin{tablenotes}
\item Notes: The effective bandwidth is measured on the target-only ms file.
\end{tablenotes}
\end{table}

\subsection{Abell 2163} \label{Abt_target}

A2163 is one of the largest and most highly disturbed clusters, with a redshift of 0.203 \citep{1994ApJ...436L..71M}. It's mass (M$_{500}$) is 1.6 $\times$ 10$^{15}$ M$_{\odot}$ \citep{planck14}. Optical studies reveal a complex galaxy distribution with two maxima and an elongated, flat shape stretching along E-W \citep{1997ApJ...482..648S, 2004ApJ...613...95C}. \citet{2008A&A...481..593M} conducted a detailed optical analysis, revealing two main components: a central structure (A) and a northern component (B). The galaxy density distribution of the primary cluster (A2163-A) displays strong elongation in the E $-$ W direction and a luminosity-dependent bimodal distribution. \citet{2012A&A...540A..61S} reported a bimodal dark matter distribution in the central mass clump (A2163-A), with a 3:1 mass ratio between its components. Additionally, Abell 2163-B has been identified through weak lensing as an independent entity with a mass of 2.7 $\times$ 10$^{14}$ M$_{\odot}$.

At X-ray wavelengths, this cluster shines brightly \citep{1994ApJ...436L..71M}. Initial temperature measurements pointed to a global temperature of 15 keV \citep{1992ApJ...390..345A}. Previous surveys revealed that the gas within the cluster was not homogeneously heated, with a noticeable temperature difference at its core \citep{1994ApJ...436L..71M,1996ApJ...456..437M}. \citet{2009ApJ...704.1349O} reported a cold front very close to the cluster centre. \citet{2011A&A...527A..21B} studied the remarkable hot gas properties using \textit{Chandra}, identifying a separate cold gas clump, reminiscent of features seen in the ``Bullet cluster'' \citep{2002ApJ...567L..27M}. Later, \citet{2018A&A...619A..68T} reported the discovery of three shock fronts at the NE and SW positions, using Suzaku observations. The detection of many substructures, both in the periphery and close to the central region, signifies the complex dynamical motion of the thermal gas in this cluster. Later, \citet{2021ApJ...906...87R} reported \textit{NuSTAR} observations at 3 $-$ 30 keV, and estimated a limit on the average magnetic field strength to be \textgreater 0.22$\mu$G and \textgreater 0.35$\mu$G, combining with the radio observations from the VLA.

\subsubsection{Previous radio studies}

A2163 remains considerably unexplored in the radio regime. \citet{2001A&A...373..106F} reported a giant radio halo in the central region ($\sim$ 2.9 Mpc) and a radio relic along the northeast (NE) side, located 2.2 Mpc away, with the VLA 1.4 GHz. \citet{2004A&A...423..111F} performed a spectral study between 332 and 1400 MHz, yielding spectral indices of -1.19 $\pm$ 0.02 for the radio halo and -1.02 $\pm$ 0.02 for the relic. However, their spatially resolved spectral index analysis suffered from poor resolution of 60$''$ ($\sim$ 200 kpc), hindering insights into the origin of the radio halo. 

\citet{2020ApJ...897..115S} carried out the legacy GMRT observations of this cluster at 150, 323, and 610 MHz, and reported the discovery of a linear, bright ``ridge'' at the centre of the cluster. The largest extent of the ``ridge'' was 800 $\times$ 200 kpc$^{2}$. The integrated spectral index of the ``ridge'' was reported as -1.28 $\pm$ 0.05, situated between the two merging sub-clusters, possibly suggesting synchrotron plasma re-acceleration via a merger shock. However, the limited quality of the legacy GMRT data sets prevents us from gaining further insight into the ``ridge''. Their 332 and 610 MHz observations suffered from a severe flux density loss at the shorter baselines, leading to a non-detection of the full extent of the halo. Their estimated spectral index for the radio halo was -2.09$\pm$ 0.20, steeper than \citet{2004A&A...423..111F}, attributed to the significant flux density loss at shorter baselines. Due to the very high Signal-to-Noise ratio (SNR) of the radio halo emission extending to large radii, this cluster is an ideal laboratory to test the particle acceleration models.

The paper is organised as follows. The uGMRT observations and data analysis procedures are explained in Sec.\ref{data_analysis}. In Sec.\ref{results}, we present the uGMRT continuum images at different resolutions and discuss newly discovered features from our study, and Sec.~\ref{xray-analysis} presents the results from the X-ray analysis. The results obtained from non-thermal vs thermal correlations are described in Sec.\ref{th-nonth}. The results are discussed to incorporate the possible theoretical explanation in Sec.\ref{discussion}. We summarised our results and findings in Sec.\ref{sum}. Throughout this paper, we have adopted a flat $\Lambda$CDM cosmology with H$_{0}$ = 70 km s$^{-1}$, $\Omega$ $_{m}$ = 0.3, $\Omega_{\Lambda} =0.7$. At the redshift of A2163, 1$''$ corresponds to a linear scale of 3.3 kpc.

\begin{table*}
  \centering
  \caption{Properties of the radio images.}
  \begin{tabular}{@{}ccccccc@{}}
    \hline\hline
    & Name & Beam Size ($''$, $^{\circ}$) & Robust & \textit{uv} range & \textit{uv}taper ($''$) & map rms ($\mu$Jybeam$^{-1}$)  \\
    \hline
    uGMRT band 3 & IMG1 & 8$''$ $\times$ 7$''$, 60.5  & 0.5 & None & None& 25.0   \\ 

 & IMG2 & 16$''$ $\times$ 16$''$, 0.0 & 0 & \textgreater 0.2k$\lambda$& 15$''$ & 34.0  \\

   & IMG3 & 35$''$ $\times$ 35$''$, 0.0  & 0 & \textgreater 0.2k$\lambda$ & 30$''$ & 75.0\\

   & IMG4 & 45$''$ $\times$ 45$''$, 0.0  & 0 & \textgreater0.2 k$\lambda$& 45$''$ & 87.3 \\

 \hline

uGMRT band 4 & IMG5 & 5$''$ $\times$ 4$''$, 60.9  & 0.5 & None & None& 8.5   \\ 

 & IMG6 & 16$''$ $\times$ 16$''$, 0.0 & 0 & \textgreater 0.2k$\lambda$& 15$''$ & 16.7  \\

  & IMG7 & 35$''$ $\times$ 35$''$, 0.0  & 0 & \textgreater0.2 k$\lambda$& 30$''$ & 46.9 \\

  & IMG8 & 45$''$ $\times$ 45$''$, 0.0  & 0 & \textgreater0.2 k$\lambda$& 45$''$ & 61.2\\
  
\hline

VLA (C+D) & IMG9 & 16$''$ $\times$ 16$''$, 0.0  & 0 & None & None& 35.0  \\ 

 & IMG10 & 35$''$ $\times$ 35$''$, 0.0  & 0 & \textgreater 0.2k$\lambda$ & None& 58.0  \\
\hline
\end{tabular}

\begin{tablenotes}
    \item Notes: IMG1, IMG5, and IMG9 are made using the visibilities without subtracting the point sources; the rest are made using point source-subtracted visibilities. The rms noise value quoted here is estimated from a region close to the phase centre.
\end{tablenotes}

\label{img_summary}
\end{table*}

\section{Observations and data analysis} \label{data_analysis}

We present the data analysis procedure for the uGMRT observations. We used the same 1.4 GHz VLA observation, as \citet{2001A&A...373..106F} (project AF328). The observations were carried out in the C-array configuration for about 4 hours, and twice in the D-array configuration for a total of 3.4 hours, with two intermediate frequencies (IF), centred at 1365 MHz and 1435 MHz, with a bandwidth of 25 MHz per IF for the C-array observations and 50 MHz per IF for the D-array observations. We refer to \citet{2021ApJ...906...87R} for the VLA data analysis of A2163.

\subsection{uGMRT}
Abell 2163 was observed with the uGMRT (proposal code: 44\_084) at 400 and 650 MHz using the GMRT Wideband Backend (GWB) \citep{2017CSci..113..707G}. We used the uGMRT online RFI (Radio Frequency Interference) filtering \citep{2022JAI....1150008B,2023JApA...44...37B}, to mitigate broadband RFI (e.g., powerline sparking) during the observations in both frequency bands. The observational parameters are summarised in Table~\ref{obs_tab}.

We processed uGMRT data with \texttt{CASA} using the \texttt{CAPTURE\footnote{\url{https://github.com/ruta-k/CAPTURE-CASA6}}} pipeline \citep{2021ExA....51...95K}, specifically developed for the reduction of the GMRT continuum data. After this initial flagging, the flux density of the primary calibrator was set according to the flux scale of \citet{2017ApJS..230....7P}. After following the standard calibration routines and applying some flagging, the calibrated target source data were split and further flagged using automated flagging modes. Special care has been taken to remove the narrow band RFI using the \texttt{aoflagger} \citep{2012A&A...539A..95O}, and some manual flagging at the short baseline at both frequencies. Target visibilities were imaged using the \texttt{CASA} task \texttt{tclean} with wide-field and wide-band imaging algorithms, with \texttt{nterms=2}. We carried out 6 rounds of phase only and 4 rounds of amplitude calibration to obtain a good gain solution for the target. For further details on the data analysis procedures for uGMRT, please follow \citet{2021ExA....51...95K,2022MNRAS.514.5969K}.

\subsubsection{Radio imaging and flux density errors}

The central region of A2163 is contaminated by weak/faint discrete sources like unresolved point sources and radio galaxies, causing complications in the measurement of the flux densities of the radio halo. We have created models for discrete sources by applying a \textit{uv}-cut of \textgreater 4.0 k$\lambda$ (angular scale of 50$''$). At this scale, we are filtering out the diffuse emission from the halo or ridge. The clean components corresponding to the compact sources were subtracted from the observed data. After subtraction, the data file was imaged again using \texttt{tclean} with a \textit{uv} baseline of \textless 10 k$\lambda$, to highlight the extended emission. We used the \texttt{multi-scale} to map the large-scale diffuse emission properly, and set \texttt{nterms = 3} in \texttt{tclean}. To account for the uncertainty introduced by point-source subtraction (following \citealt{2021A&A...650A..44B}), we compared the halo flux density derived after \textit{uv}-plane model subtraction with that obtained by algebraically subtracting the flux densities of compact sources from the total (halo + sources). From this comparison, we estimate the subtraction error to be $\sim$8\% for uGMRT and $\sim$5\% for VLA, which are included in the quoted flux density uncertainties. The image properties are summarised in Table.~\ref{img_summary}.

The overall flux density scale for all observations was checked by comparing the spectra of compact sources. We followed a similar strategy, explained in \citet{2020ApJ...897..115S} to check the errors of the flux density scale by comparing the values with the NVSS (1400 MHz) and TGSS - ADR1 (147 MHz). We show the comparison in Appendix.~\ref{fluxscale-com}. We combine all the images according to the flux scale of \citet{2017ApJS..230....7P}. We assume a flux density uncertainty is $10\%$ for both the uGMRT frequencies \citep{2017ApJ...846..111C}, and 2.5\% for VLA frequencies. The flux density uncertainty ($\Delta S$) is defined by:
\begin{equation}
    \Delta S = \sqrt{(f.S)^{2} + N_{\rm beam}.(\sigma_{\rm rms})^{2} + (\sigma _{\rm sub}.S)^{2} },
    \label{eq-flux-err}
\end{equation}
where $S$ is the flux density, $f$ is the absolute flux density calibration error, $N_{\rm beam}$ is the number of beams, $\sigma_{\rm rms}$ is the rms noise, and $\sigma_{\rm sub}$ is the point source subtraction error. The images were corrected for the primary beam using the task \texttt{ugmrtpb}\footnote{\url{https://github.com/ruta-k/uGMRTprimarybeam-CASA6}} for the GMRT, and \texttt{PBCOR} for VLA.

\begin{table}
\centering
\caption{Effective exposure time of each XMM-Newton -EPIC observation\label{fig:xmm_exposures}}
\begin{tabular}{lccc}
\hline\hline
Instruments & EMOS1 & EMOS2 & EPN \\ 
\hline
Med. live time (ks) & 122.1 & 122.2 & 101.7 \\ 
Med. on time (ks) & 92.3 & 95.4 & 59.0 \\ 
Time selection (\%) & 75.6 & 78.1 & 58.0 \\ 
\hline

\end{tabular}
\end{table}

\subsection{XMM-Newton}

X-ray observations allowed us to map the surface brightness and thermal structure of the hot gas component in A2163. For this purpose, we used the deepest XMM-Newton pointing of A2163 (obs-id: 0694500101), performed with the European Photon Imaging Camera (EPIC), using both the MOS (Metal Oxide Semiconductor CCD) and PN (pn-junction CCD). A list of effective exposure times associated with each EPIC focal instrument is reported in Table \ref{fig:xmm_exposures}.

EPIC imaging and spectroscopy were performed after rebinning of photon event positions, $k,l$, to an angular resolution of 3.5$''$, while photon event energies, $e$, were rebinned in energy intervals whose width increases as a function of their central value in the range 10-100 eV. X-ray spectra assume an ICM emission, $\ficm (\kt,\ab,\nh,e)$, which were modelled using the Astrophysical Plasma Emission Code (APEC; \citealt{smith01}), with the element abundances of \citet{grevesse98}. These spectra were redshifted and corrected for absorption by the Galactic interstellar Medium, assuming absorption cross sections of \citet{balucinska92} and a Galactic $\nh$ that we set to $15.65 \times \nhunit$ from measurements averaged near A2163 in the Leiden/Argentine/Bonn Survey of Galactic HI \citep{kalberia05}. Following the approach detailed in \citet{bourdin08}, an effective exposure, $\eecub(k,l,e)$, and a background noise model, $\bckcub(k,l)~\fbck(k,l,e)$, have been associated with each coordinate of the position-energy event cube, $\evcub(k,l,e)$. The background noise model includes a quiescent particle background spectrum and three astrophysical components that are weighted in position and energy by the effective exposure. Spatial variations of the fluorescence lines and the continuum of the particle background spectrum are modelled as detailed in \citet{2013ApJ...764...82B}. Astrophysical components of the background noise include the unresolved Cosmic X-ray Background that we model using a power-law with spectral index $\Gamma=1.42$, and the galactic foreground, which we modelled using two thermal bremsstrahlung spectra with temperatures $\mathrm{T}_1=0.099~\kev$ and $\mathrm{T}_2=0.248~\kev$. We jointly fitted the normalisations of these three components within an annular region that encircles A2163, with an innermost radius exceeding $1.4 \times r_{500}$.

X-ray surface brightness, $\sx(k,l)$, has been mapped in the "soft" energy band, $\Delta E = [0.3-2.5]$ keV, characterised by its high signal-to-noise. The energy band chosen for the analysis stems from a complex set of reasons that balance the detector sensitivity and effective area, as well as the particle background. The surface brightness map results from the exposure corrected residual separating the event map, $\evmap(k,l) = \sum_{e \in \Delta E} \evcub(k,l,e)$, from the background noise, $\bckmap(k,l) = \sum_{e \in \Delta E}  \bckcub(k,l)~\fbck(k,l,e)$. In this computation, diffuse emission has been interpolated in place of bright point sources, which were identified and masked using \textsc{SExtractor} \citep{1996A&AS..117..393B}.  Moreover, the exposure correction, $\vfmap(k,l)$, relies on an averaged spectrum of the observed cluster atmosphere, $\ficm (\bar{\kt}, \bar{\ab}, e)$, which we fitted over the radii range $[0.15, 0.75] \times r_{500}$:

\begin{eqnarray} \label{equ:sx}
  	\vfmap(k,l) &=& \frac{\max_{k,l}{\sum_{e \in \Delta E} \ficm (\bar{\kt}, \bar{\ab}, e) \eecub(k,l,e)}} {\sum_{e \in \Delta E} \ficm (\bar{\kt}, \bar{\ab}, e) \eecub(k,l,e)} \nonumber \\
	\sx(k,l) &=& \vfmap(k,l) \times \left[ \evmap(k,l) - \bckmap(k,l) \right]. 
\end{eqnarray}

X-ray temperature was mapped using a spectral-imaging algorithm that combines spatially weighted likelihood estimates of the projected intra-cluster medium temperature with a curvelet analysis. This algorithm can be seen as an adaptation of the spectral-imaging algorithm presented in \citet{bourdin15} for the X-ray data. Briefly, temperature log-likelihoods are first computed from spectral analysis in each pixel of the maps and are then spatially weighted with kernels that correspond to the negative and positive parts of B3-spline wavelet functions. We use this method to derive wavelet coefficients of the temperature features and their expected fluctuation from spatially weighted Fisher information. We use such coefficients to derive wavelet and curvelet transforms that typically estimate features of apparent size in the range of [3.5, 60] arcseconds. We finally reconstructed a temperature map from denoised curvelet transforms, inferred through 4-$\sigma$ soft thresholding of the curvelet coefficients. For each pixel of this map, temperature uncertainties were inferred from a distribution of 100 temperature values that result from Monte-Carlo runs of the algorithm on mock EPIC observations of an X-ray source, whose surface brightness and temperature distribution mimicked A2163.

\begin{figure*} 
    %\centering
    \includegraphics[width=8.8cm, height = 9.4cm]{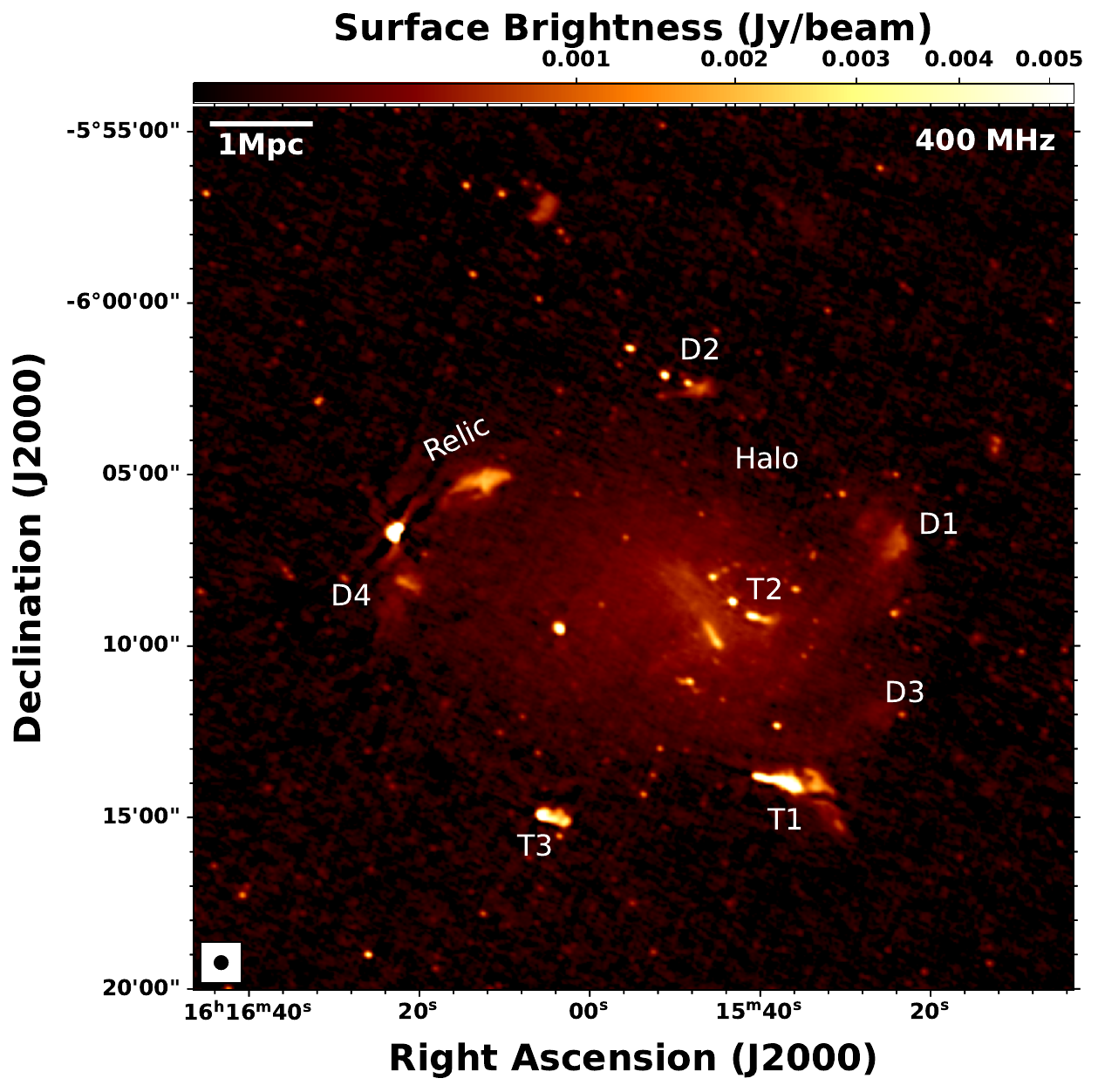}
    \includegraphics[width=8.3cm, height=9.4cm]{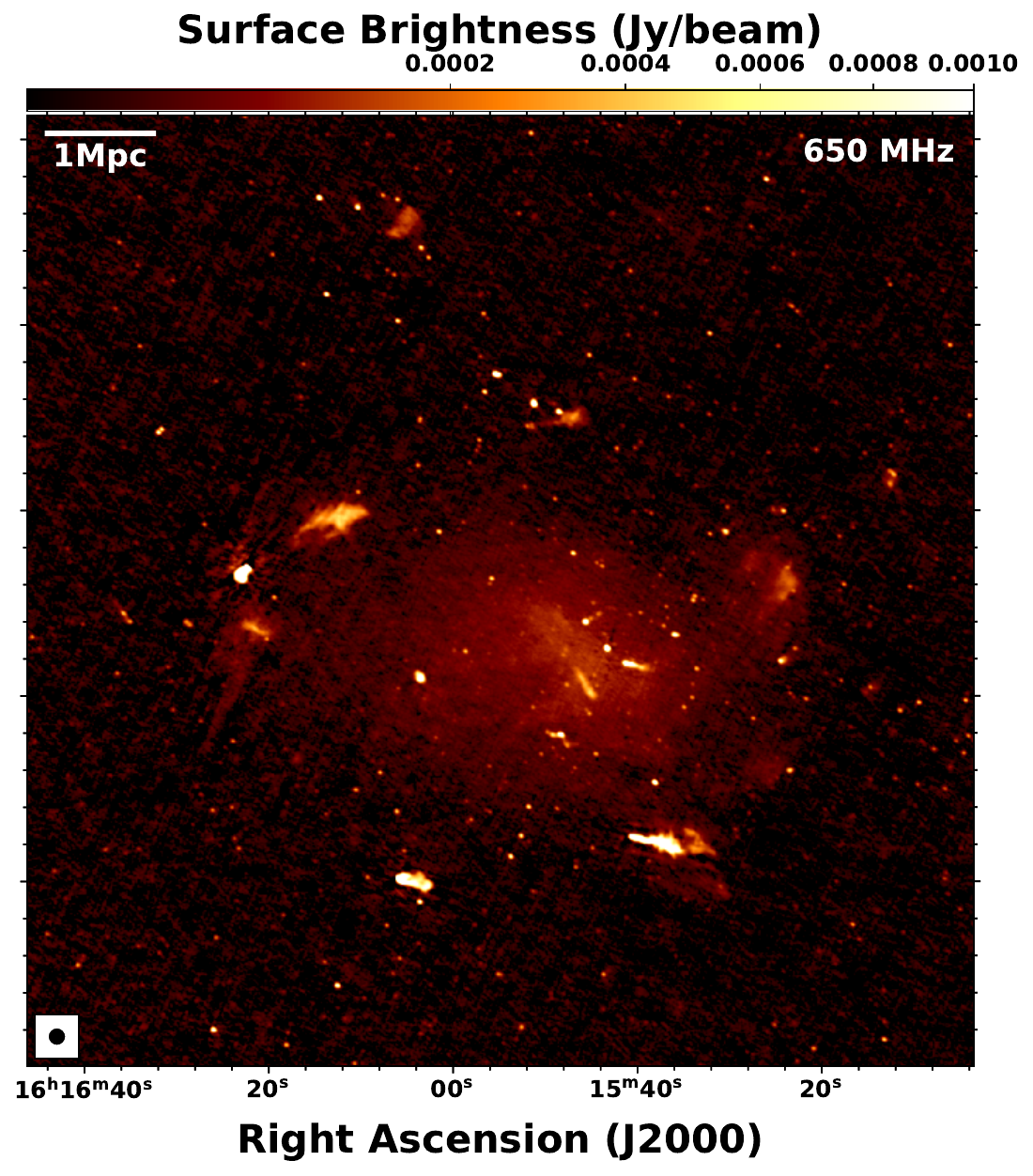}
    
    \caption{\textit{Left:}uGMRT band 3 full resolution image IMG1 obtained with robust$=0.5$ is shown here. The rms is $\sigma_{\rm rms}$ $=$ 25 $\mu$Jybeam$^{-1}$ and the beam is 8$''$ $\times$ 7$''$ . We have labelled the relevant extended sources: the relic at the NE outskirts, and other diffuse sources and tailed galaxies. The low-surface brightness gigantic radio halo emission is seen in the central region. \textit{Right:}uGMRT band 4 full resolution image IMG5 obtained with robust$=0.5$ is shown here. The rms is $\sigma_{\rm rms}$ $=$ 8.5 $\mu$Jybeam$^{-1}$ and the beam is 5$''$ $\times$ 4$''$.}
    \label{A2163_intro}
\end{figure*}

\begin{figure*}
    \centering
    \includegraphics[width=0.70\textwidth]{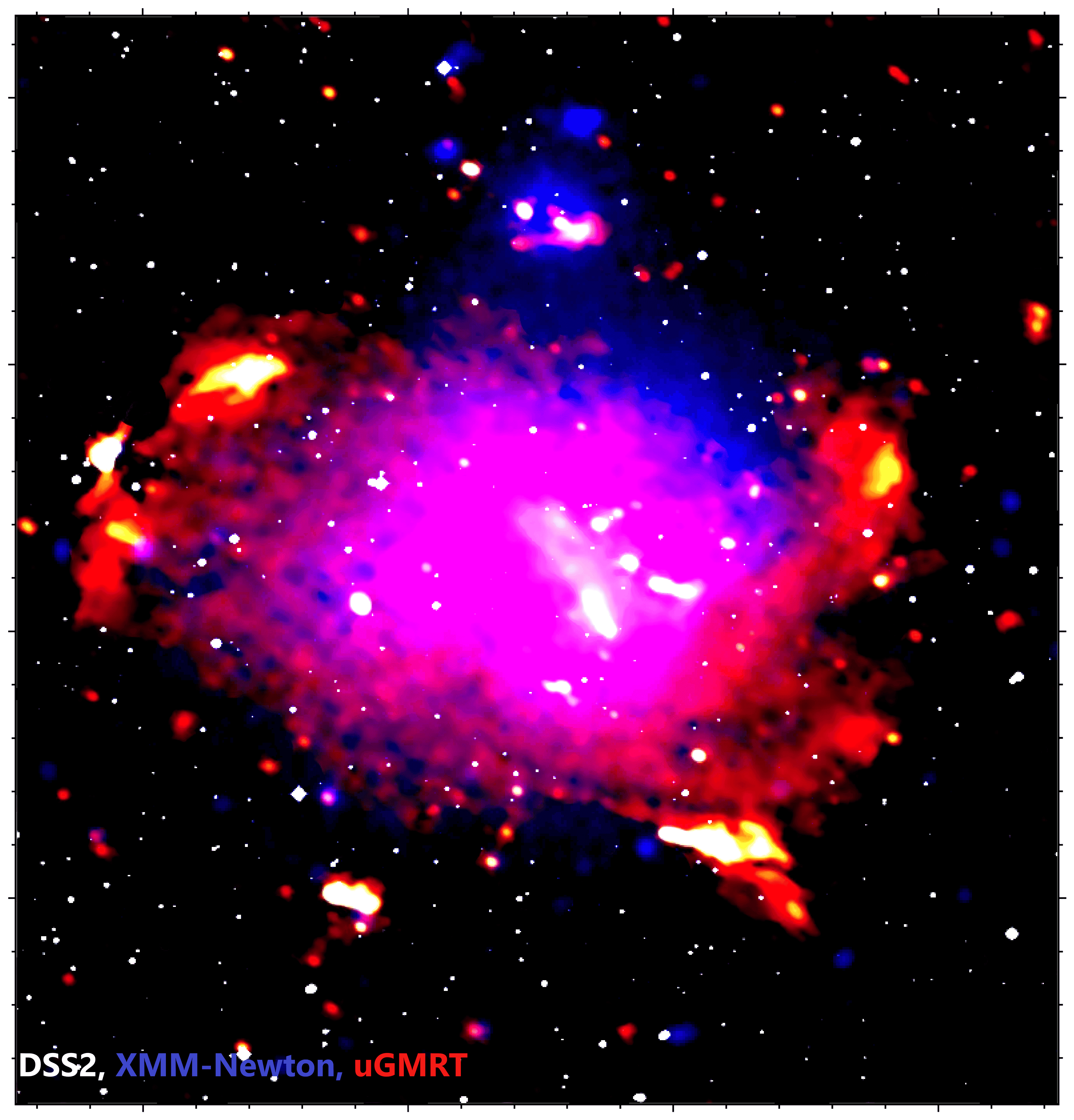}
    \caption{Multi-wavelength view of the central region of the A2163 is shown. The white intensity shows the optical image of the A2163 field in the DSS2 red. The intensity in blue shows the \textit{XMM-Newton} X-ray image at 0.3$-$2.5 keV, which is smoothed to 5$''$ using a Gaussian kernel. The radio emission from the uGMRT band 3 image is shown in red, with resolution 16$''$ $\times$ 16$''$.}
    \label{A2163_RGB}
\end{figure*}

\begin{figure*}
    \centering
    \begin{tabular}{cc}
        \includegraphics[width=0.45\textwidth]{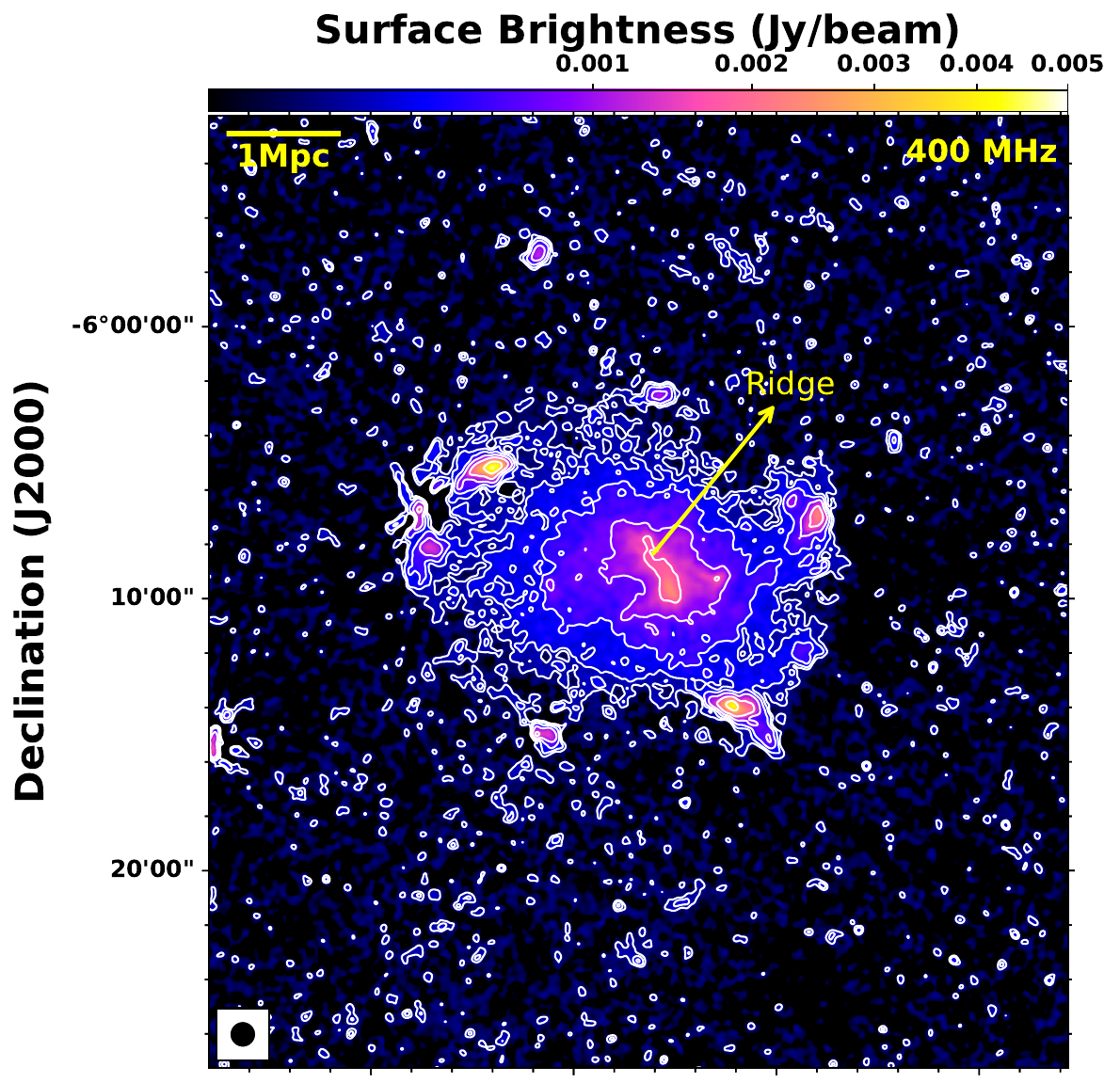} &
        \includegraphics[width=0.37\textwidth]{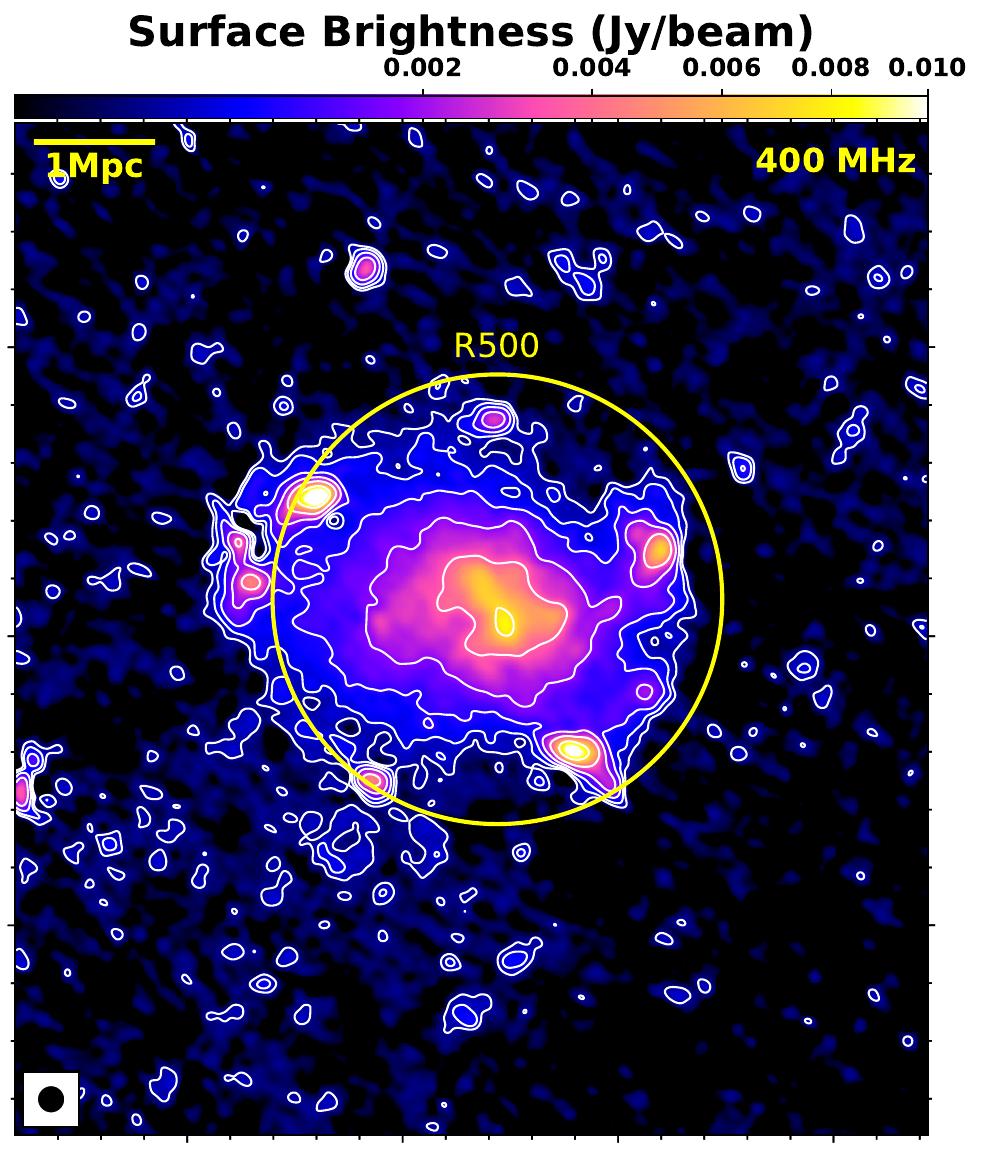} \\
        \includegraphics[width=0.45\textwidth]{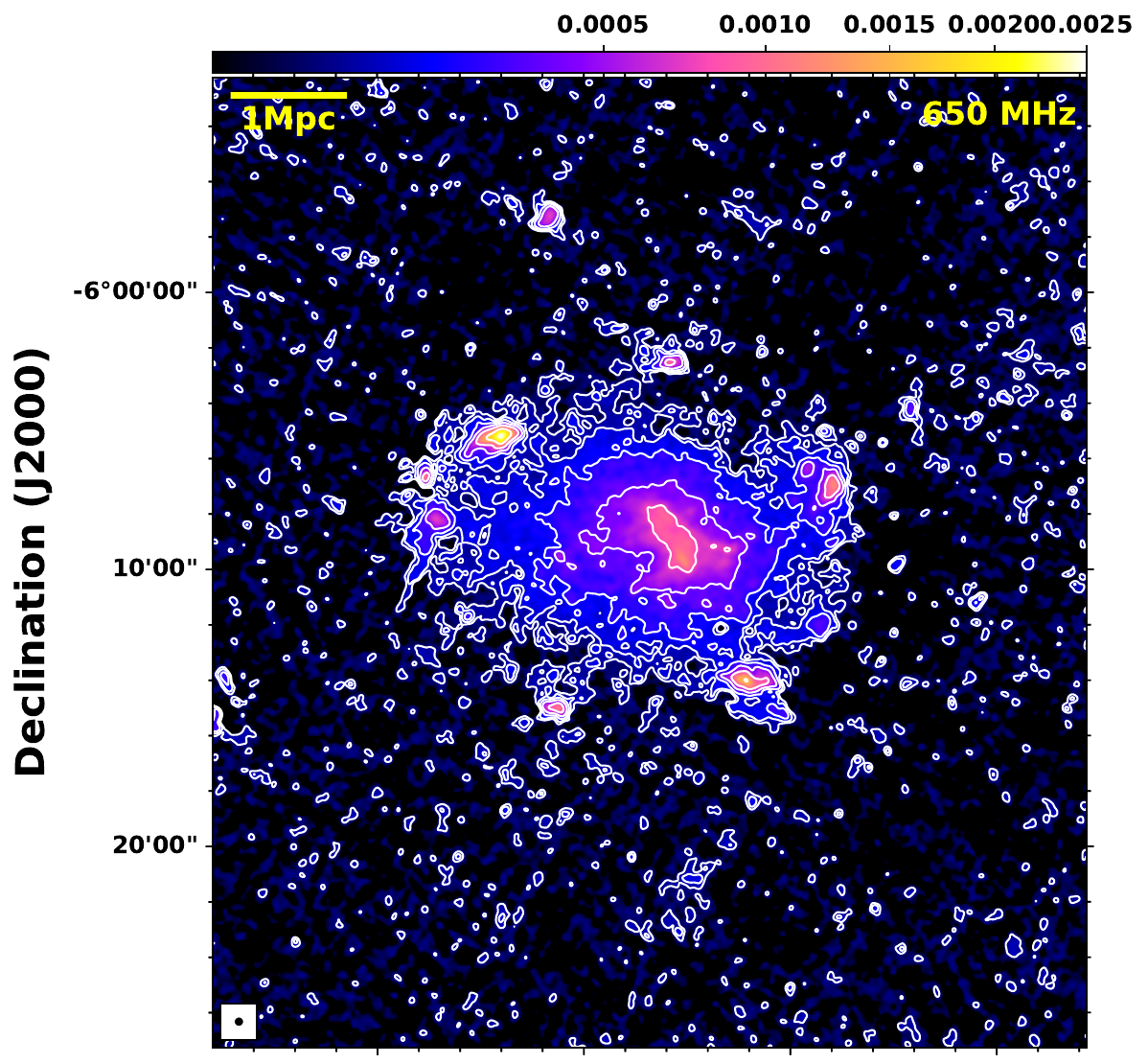} &
        \includegraphics[width=0.37\textwidth]{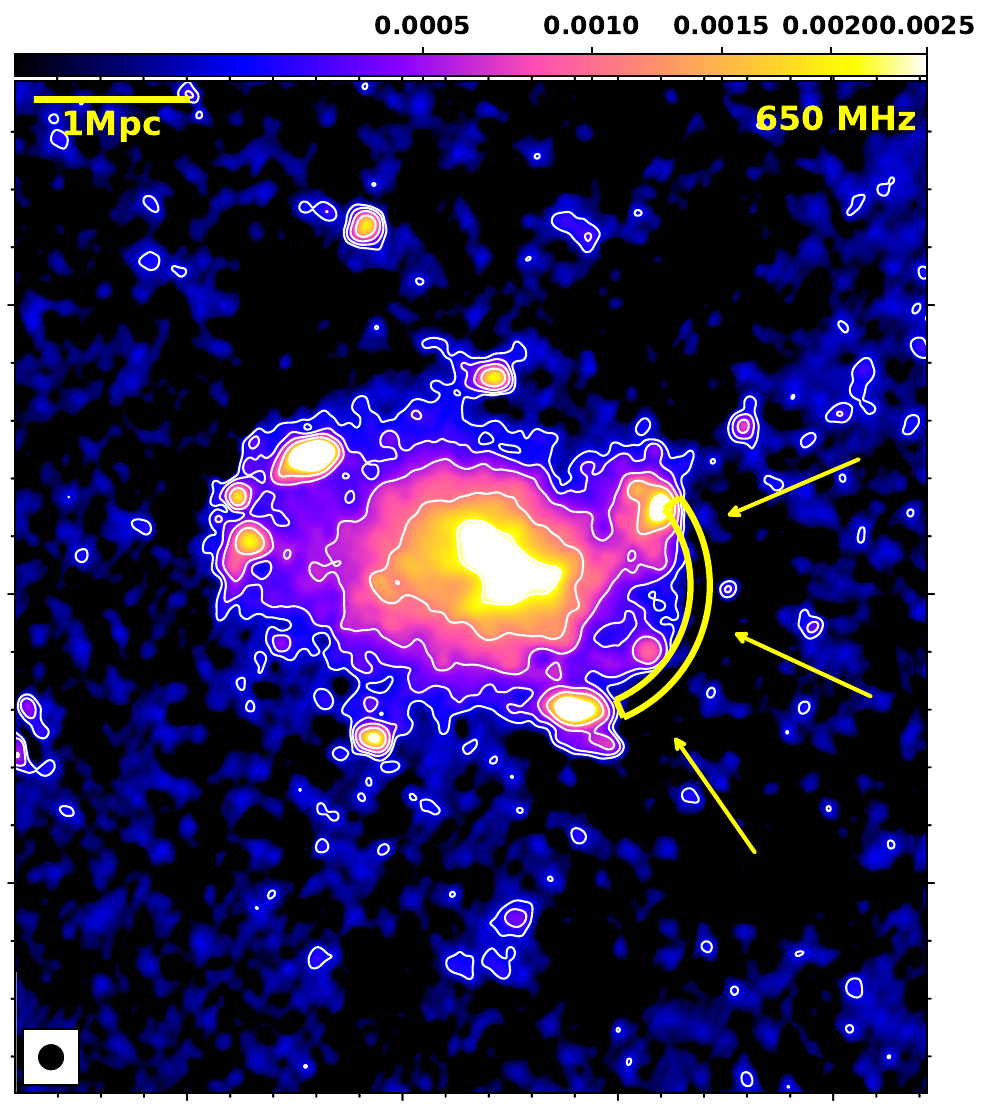}\\
        \includegraphics[width=0.45\textwidth]{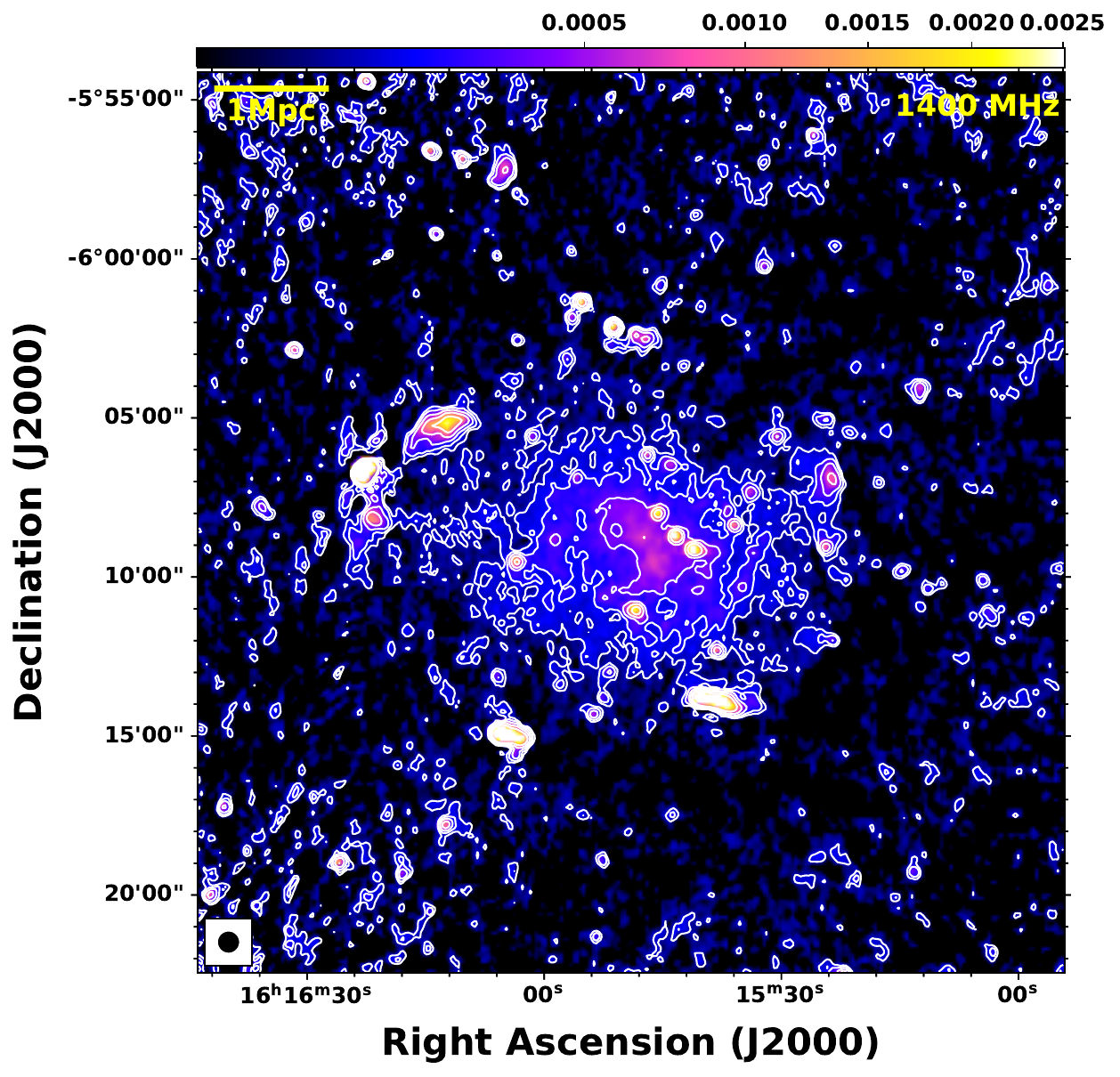} &
        \includegraphics[width=0.36\textwidth]{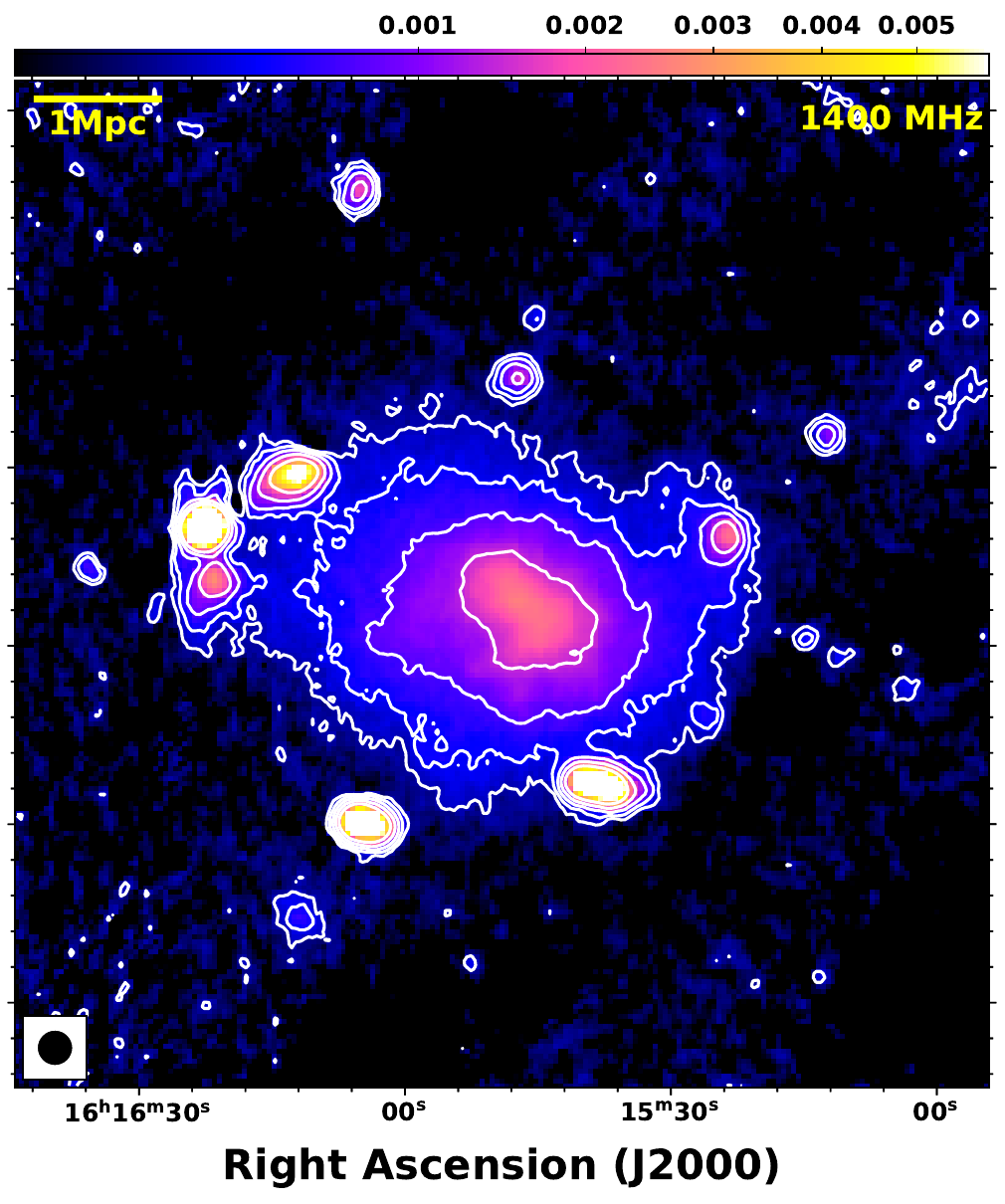} \\
    \end{tabular}
    \caption{High (left) and low (right) resolution uGMRT band3 (top), uGMRT band4 (middle), and VLA 1.4 GHz (bottom) images of Abell 2163 are shown. The low-resolution images are created at a common 35$''$ resolution (IMG4, IMG8, IMG10). The beam size is indicated in the bottom left corner of each image. The yellow circle indicates the r$_{500}$ of the cluster, and the ridge emission is also labelled. The presence of a sharp edge at the SW outskirts is also shown (see the middle panel, right image). The contour levels starts with 3$\sigma_{rms}$ $\times$ [1,2,4...], where $\sigma_{rms}$ is reported in Table.~\ref{img_summary}.}
    \label{A2163_gmrt_vla}
\end{figure*}

\section{Results} \label{results}

\subsection{Components of the diffuse emission}

We have obtained the most sensitive image of the A2163 cluster using the uGMRT wide bandwidth capability. At high resolution, we achieved $\sigma_{\rm rms}$ = 25 $\mu$Jy~beam$^{-1}$ and 8.5 $\mu$Jy~beam$^{-1}$ at 400 and 650 MHz respectively (see Table.~\ref{img_summary}). Our images are six times more sensitive than previously published images, at nearly similar frequencies \citep{2020ApJ...897..115S}. The radio emission from the A2163 field consists of several different components, either associated with the ICM or originating from the interaction of the radio galaxies with the environment. In Figure.~\ref{A2163_intro}, the diffuse emission from the A2163 cluster field is shown, without subtracting point sources. The most relevant features are labelled. In this section, we present the emission from the diffuse sources as imaged by uGMRT and archival VLA. An analysis of their properties, including radio images at other frequencies and a comparison with the X-ray emission from the ICM, is presented in the following sections.

\subsubsection{Radio halo}

The 400 MHz uGMRT image of the galaxy cluster A2163 is presented in Figure.~\ref{A2163_intro}, where we have adopted the source labelling convention of \citet{2021ApJ...906...87R}. In \citet{2020ApJ...897..115S}, the full extent of the radio halo at 332 and 610 MHz was not recovered due to limited \textit{uv}-coverage at short baselines. In contrast, our deep observations have succeeded in detecting emission out to larger scale, reaching r$_{500}$ (1680 kpc). The combination of high sensitivity and improved angular resolution allowed us to investigate the radio halo emission in unprecedented detail. Figure.~\ref{A2163_RGB} provides a multi-wavelength composite of the cluster, where the central halo emission (shown in red) exhibits a morphology closely resembling that of the XMM-\textit{Newton} X-ray thermal emission (shown in blue). In addition, we identify 33 discrete sources in band 3 and 39 in band 4 (above the 4$\sigma_{\rm rms}$ threshold) embedded within the diffuse component, contributing approximately 20\% (136.3 mJy at band 3) and 22\% (75.64 mJy at band 4) of the total halo flux density after subtraction (quoted in Table.~\ref{int-spec-tab}), highlighting that, if not properly removed, they would lead to a significant overestimate of the diffuse halo emission. The halo shows elongation along the east–west axis, corresponding to the merger axis. Unlike some other well-known giant halos \citep[e.g.,][]{2020ApJ...897...93B,2021A&A...646A.135R,2022ApJ...933..218B}, we do not detect any filamentary substructures within the halo region. Instead, the emission appears relatively uniform and seems to be connected to other diffuse sources, tailed radio galaxies, and the relic in the low-resolution image. At lower frequencies, the halo is slightly more extended, whereas at higher frequencies it appears contracted, consistent with spectral steepening toward its periphery. The largest linear size (LLS) of the halo is measured to be 3.3 Mpc (17$'$) at 400 MHz and 3.0 Mpc (15$'$) at 650 MHz, making it nearly twice as extended as previously reported at comparable frequencies. The halo encompasses an area of about 3.3 $\times$ 2.8 Mpc$^{2}$, 3.0 $\times$ 2.2 Mpc$^{2}$, and 2.7 $\times$ 1.8 Mpc$^{2}$ at 400 MHz, 650 MHz and 1.4 GHz, respectively, in low-resolution (35$''$) maps. The emission is asymmetric relative to the cluster centre, with a smaller extent ($\sim$1.4 Mpc) toward the west compared to the east ($\sim$1.9 Mpc) in all frequency bands. In the point-source-subtracted images ($\sim$35$''$ resolution), the diffuse emission is observed to extend toward the northern sub-group. A distinct radio edge is visible in the low-resolution map (marked with a yellow arrow in Figure.~\ref{A2163_gmrt_vla}) on the western side, where the emission exhibits a sharp drop rather than a gradual decline. This abrupt boundary, beyond which radio emission is not detected, may be associated with a shock front, similar to features observed in A520 \citep{2005ApJ...627..733M,2018ApJ...856..162W}, A754 \citep{2024arXiv240618983B}, and the bullet cluster \citep{2014MNRAS.440.2901S,sikhosana23,2023A&A...674A..53B}.

\subsubsection{Radio ridge}

\begin{figure}
	\includegraphics[width=0.98\columnwidth] {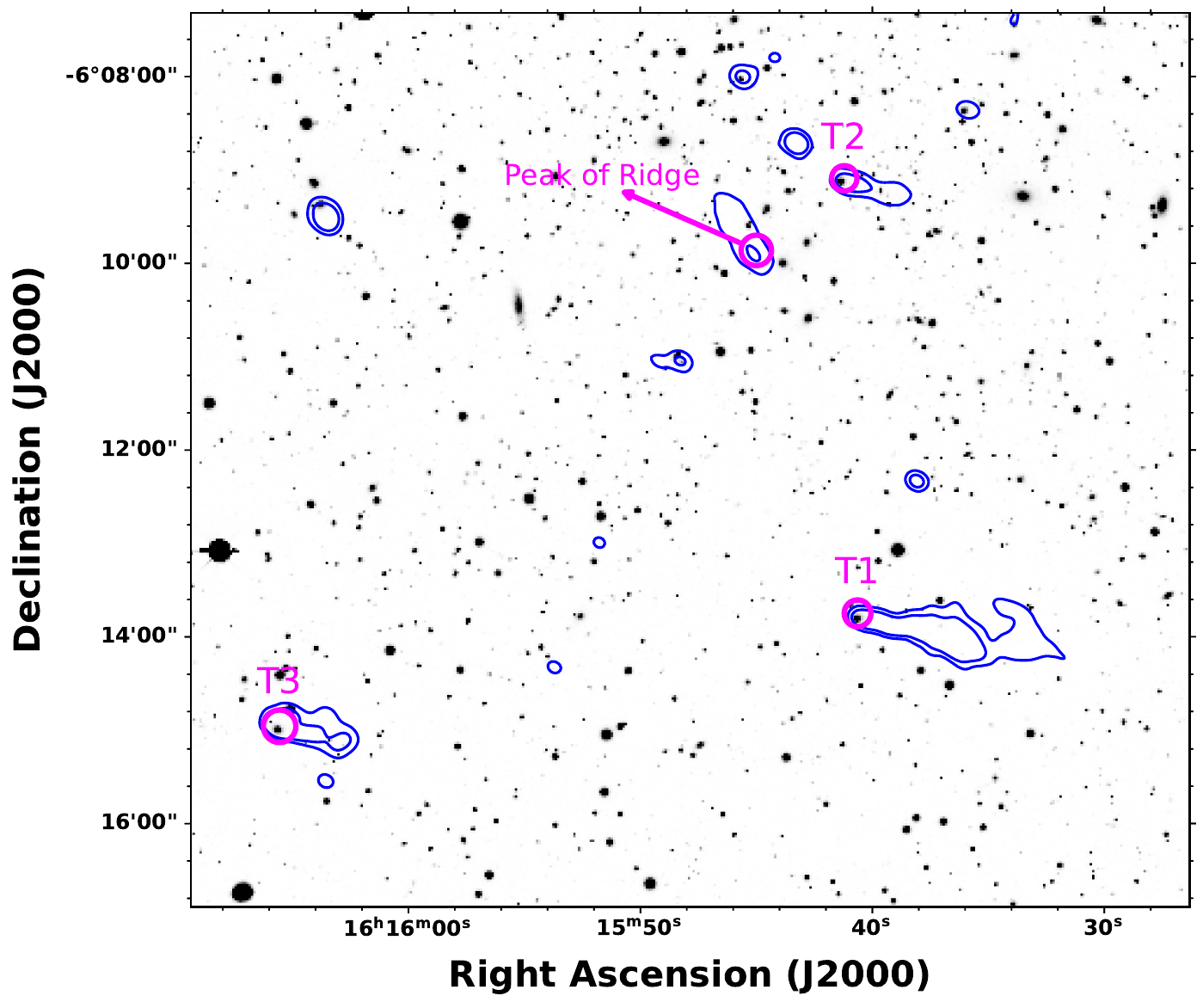}
        \caption{DESI i-band image (greyscale) cutout of the A2163 field shown, overlaid with the radio emission at 650 MHz, with a beam size of 5$''$ $\times$ 4$''$ (IMG5). The contour levels are drawn at [90, 270]$\times$ 3$\sigma_{\rm rms}$, where $\sigma_{\rm rms} = 8.5$ $\mu$Jybeam$^{-1}$, to show the overlay more effectively. The magenta circle indicates the position of the peak of the tailed radio sources and the coincidence of a galaxy in optical.}
    \label{a2163_op_image}
\end{figure}

\citet{2020ApJ...897..115S} reported the discovery of a linear, flat-spectrum `ridge', with no optical counterpart in the DESI (Dark Energy Spectroscopic Instrument) image, for the peak of the ridge emission (shown in Figure.~\ref{a2163_op_image}). We recovered the ridge emission at the centre of the halo in A2163 with higher significance at both frequencies. The ridge is elongated along the north-south direction and is not exactly perpendicular to the east-west elongation, but is slightly tilted. It is situated at the epicentre of the merging activity, between the two subclusters, as detected in the weak-lensing analysis. We do not see any filamentary substructures or components in the ridge emission. A tailed galaxy is seen to be present at that location in the high-resolution image; however, in the compact source-subtracted image, the ridge emission is mostly diffuse. The ridge has a size of 5$'$ $\times$ 2$'$ (990 $\times$ 396 kpc$^{2}$) and is circumscribed by white contours (Figure.~\ref{A2163_gmrt_vla}). The width of the ridge is higher in our uGMRT images, compared to what is reported in \citet{2020ApJ...897..115S}. The ridge is characterised by a higher surface brightness than the surrounding diffuse emission, which makes it stand out from the radio halo and may provide an extra budget of seed electrons in those regions. The surface brightness (the flux densities are quoted in Table.~\ref{int-spec-tab}) of the ridge is $\sim$4 times higher than that of the surrounding radio halo emission. The projected separation between the peak of the ridge emission and the cluster centre (lensing mass peak from 
the dark matter distribution) is 200 kpc, indicating that the brightest part of the diffuse emission does not coincide with the cluster centre. The low-resolution VLA image (35$''$) shows the high surface brightness at the centre; however, the ridge boundary is indistinguishable at this low resolution.

\subsubsection{Relics and the other sources}

\citet{2001A&A...373..106F} classified the NE peripheral extended source as a radio relic (we label it as such in our Figure.~\ref{A2163_intro}), due to the absence of any obvious optical counterpart; however, the possibility that its nature is a diffuse lobe or a wide-angle-tailed galaxy remained. Later, \citet{2018A&A...619A..68T} detected a merger shock at that location, strengthening the fact that it is a radio relic. To the NE of the cluster centre, this source has an arc-shaped morphology. The radio relic has an angular size of $\sim$ 3.75$'$ at 400 MHz and 2.80$'$ at 650 MHz, corresponding to $\sim$ 750 kpc and 554.4 kpc at the cluster redshift, with flux densities quoted in Table.~\ref{int-spec-tab}. The relic is seen to overlap with the halo emission and is situated at a distance of 1.68 Mpc from the cluster centre. A slight extension is seen in the downstream region (towards the cluster centre), where the magnetic field may have been amplified by the shock passage. The surface brightness of the relic is very uniform, with no filamentary substructures. In the low-resolution images (35$''$), the radio emission, which may be part of the halo emission, is seen to extend beyond the outer edge of the relic. That emission is 4-5 times fainter than the radio halo emission and extends up to 200 kpc from the outer boundary of the relic. The outer edge of the relic does not coincide ($\sim$ 150 kpc offset) with the position of the X-ray detected shock (1.5 Mpc) front reported by \citet{2018A&A...619A..68T}. This offset may be due to the poor resolution of the Suzaku data, which has been used to identify the position of the shock.

In addition to the relic, the cluster is known to host several complex radio galaxies and diffuse sources at the periphery. In the band 3 and band 4 images, we recovered these sources. We have tabulated their flux densities and integrated spectral indices in Table.~\ref{TG_table}, and all of them have slightly steeper (\textless $-$0.75) spectral indices. Note that the spectral indices/flux densities are likely contaminated by the diffuse emission (halo) in which the sources are embedded. The D1 lobe is seen to be connected to a stream of emission from the extended radio halo, and the D4 lobe is extended along the southern direction, away from the radio halo emission. One compact source is associated with D2 (possibly the host), and the extended emission is elongated toward the east, as seen clearly in Figure.~\ref{A2163_RGB}. The tailed galaxies T1, T2, and T3 all have their jets extended towards the direction of the merger axis, indicating the jet-ICM interactions. The high-resolution image of T3 at 650 MHz shows the presence of a two-sided jet, bending at a very narrow angle. The source T1 shows an arc-shaped connection with a point source below. T1 is connected with a diffuse lobe with a very complex and bent morphology. The width of T1 increases slightly after a certain distance from its head, which could indicate some complex energisation of the particle, loss of collimation, or disruptions. All tailed galaxies have optical counterparts in the DESI i-band image (Figure.~\ref{a2163_op_image}), with a spectroscopic redshift of 0.225 (T1), 0.207 (T2), and 0.222 (T3).

\begin{table*}
  \centering
  \caption{Properties of the diffuse lobes and tailed sources }
  \begin{tabular}{@{}ccccccccc@{}}
    \hline\hline

& Source & RA & DEC & S$_{400 \rm MHz}$ & S$_{650 \rm MHz}$ & S$_{1400 \rm MHz}$ & Spectral index  \\
&  & hh mm ss & $^{\circ}$ $'$ $''$ & (mJy) & (mJy) & (mJy)     \\
    \hline
& T1 & 03 12 53.2  & +08 23 12 & 146.86 $\pm$ 15.60 & 74.23 $\pm$ 1.60 & 33.22 $\pm$ 3.20 & -1.03 $\pm$ 0.15\\ 
 
  & T2 & 03 12 53.9  & +08 23 05 & 27.52$\pm$ 2.60  & 20.80 $\pm$ 2.61 & 10.48 $\pm$ 1.80 & -0.75 $\pm$ 0.09  \\
 
 & T3 & 03 12 56.9 & +08 22 11   & 56.80 $\pm$ 0.30 & 40.38 $\pm$ 0.60 &  29.72 $\pm$ 4.00 & -0.52 $\pm$ 0.07 \\

 & D1 & 03 12 57.5 & +08 22 09 & 28.09$\pm$ 2.90  & 17.93 $\pm$ 2.50 & 6.39 $\pm$ 0.6 & -1.09 $\pm$ 0.02  \\
 
 & D2 & 03 12 57.3 & +08 21 37 & 13.77 $\pm$ 3.50  & 6.67 $\pm$ 0.56 & 3.31 $\pm$ 0.70  & -0.95 $\pm$ 0.21  \\

& D4 &  03 12 58.3& +08 21 14 & 15.32 $\pm$ 1.80  & 6.50 $\pm$ 0.78 & 3.56 $\pm$ 0.42  & -1.01 $\pm$ 0.30  \\

 \hline
  \end{tabular}
  \label{TG_table}
\begin{tablenotes}
    \item Notes: Flux densities were estimated from the 3$\sigma$ contour region by integrating over the region.
\end{tablenotes}
\end{table*}

\subsection{Integrated spectrum} \label{int_spec}

We have calculated the flux densities of the halo and relic and estimated the integrated spectral indices. Accurate flux density measurements of extended sources with interferometers are challenging for several reasons: (1) missing short \textit{uv} distances due to which the flux of the extended structures gets resolved out; (2) different \textit{uv}-coverage among telescopes; (3) different weighting schemes, different resolution, and improper deconvolution, and (4) improper subtraction of flux density contributions from unrelated sources. If not done carefully, these issues may significantly affect the morphology of the emission as well as the integrated spectral index. \citet{2017ExA....44..165D} had shown that faint and extended emissions may be partially lost if unmodelled during calibration steps. For imaging, we adopted a minimum baseline of 200$\lambda$, which can recover sources in the sky of size 32$'$, corresponding to 6.3 Mpc at the cluster redshift. Also, for such a large extended emission, the recovery percentage is $\sim$ 80\% at a redshift of 0.2 and -6$^{\circ}$ declination \citep{2017ExA....44..165D}. The flux density values for the halo and relic are tabulated in Table.~\ref{int-spec-tab}.

\begin{figure}
	\includegraphics[width=\columnwidth]{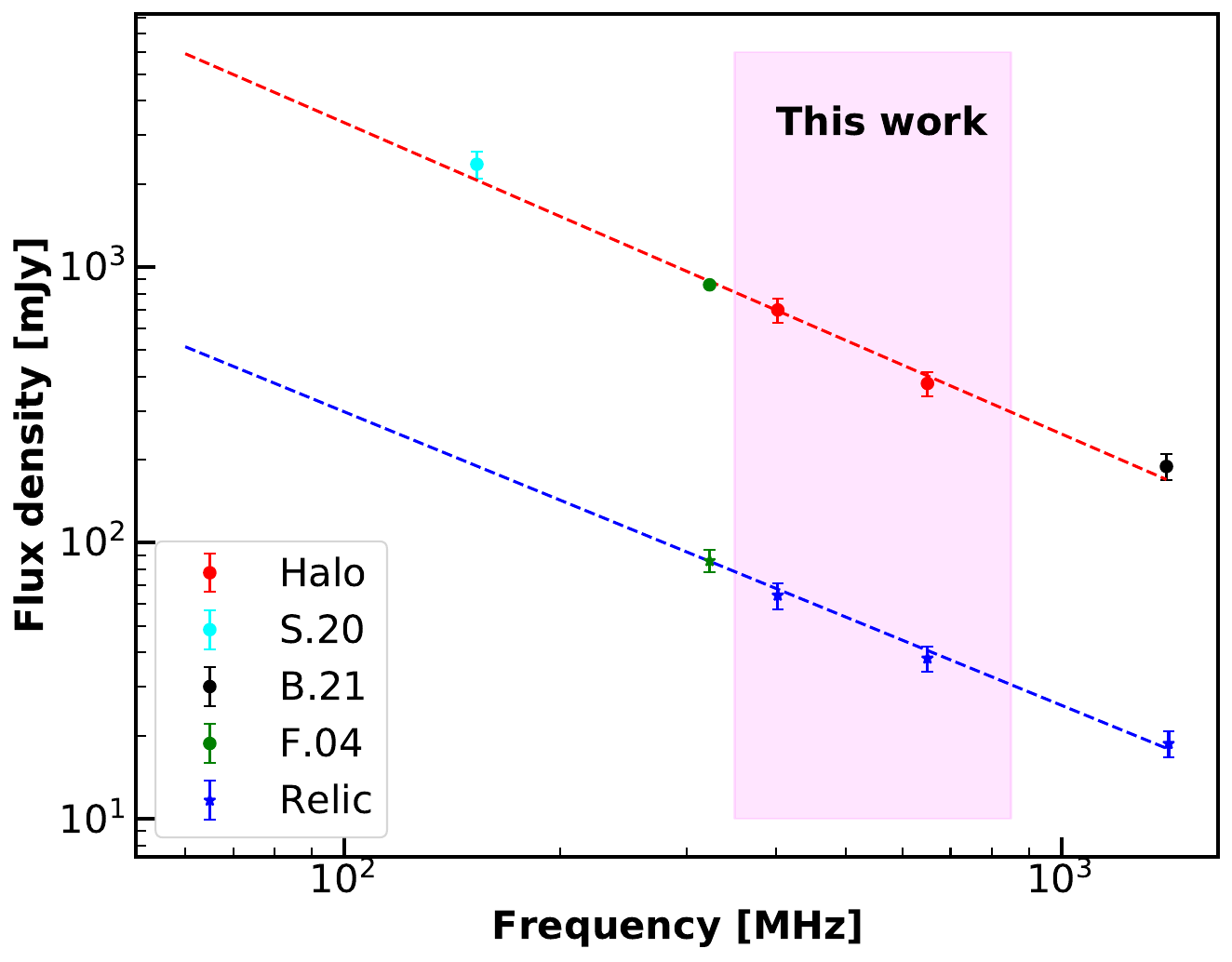}
        \caption{Integrated spectrum of the radio halo and relics from literature data and our work is shown. The magenta sector indicates the observations discussed in this work. The dashed line shows a single power-law fit to the spectrum. The error bars are estimated using the equation.~\ref{eq-flux-err} for our analysis.}        
    \label{int_spec_halo_relic}    
\end{figure}

Figure.~\ref{int_spec_halo_relic} presents the integrated radio spectra for both the halo and the relic. For the halo, the data are well described by a single power-law model over the frequency range 153--1400 MHz, yielding an integrated spectral index of $-$1.19 $\pm$ 0.09, which falls within the typical range observed for giant radio halos. The spectral index value cited here is calculated between 400--1400 MHz and agrees closely with the measurements reported by \citet{2004A&A...423..111F, 2020ApJ...897..115S} (green and cyan points in Figure.~\ref{int_spec_halo_relic}). In estimating the halo’s flux density, we have included the ridge emission; excluding this contribution changes the halo-only spectral index between 400 and 650 MHz to $-$1.34, which is marginally steeper than the $-$1.19 obtained when the ridge is included. From our measurements, the ridge itself has a spectral index of $-$1.36 $\pm$ 0.08 (400--650 MHz), in agreement with the mean value reported by \citet{2020ApJ...897..115S}. We further examined the eastern and western portions of the halo, divided at the cluster centre as indicated in the inset of Figure.~\ref{A2163_halo_SB}, to determine whether each region shares the same integrated spectral index as the entire halo or exhibits differences. The eastern part has an average spectral index of $-$1.12 $\pm$ 0.02, while the western part has $-$1.25 $\pm$ 0.05. These findings suggest that the overall power-law spectrum of the halo may result from the combined contribution of multiple components, similar to the case of MACSJ0717+3745 reported by \citet{2021A&A...646A.135R}.

The integrated spectrum of the relic is also well fitted by a single power-law of spectral index -1.02 $\pm$ 0.03. For a stationary shock, the integrated spectral index, $\alpha_{\rm int}$, is steeper by 0.5 than the injection spectral index, $\alpha_{\rm inj}$, 
\begin{equation}
    \alpha_{\rm int} = \alpha_{\rm inj} - 0.5
\end{equation}

Assuming the Diffusive Shock Acceleration to be the origin of the particle acceleration in the relic region, the Mach number (M$_{\rm S}$) is related to the injection spectral index $\alpha_{\rm inj}$ as 

\begin{equation}\label{eq.4}
    \rm M_{\rm S} =\sqrt{\frac{2\alpha_{\rm inj} -3}{2\alpha_{\rm inj} +1}} = 5.4\pm0.02 
\end{equation}

The very high Mach number derived for the relic in A2163 may result from the fact that shock surfaces with higher Mach numbers contribute more to the observed emission than those with lower Mach numbers. Moreover, the radio-inferred Mach number is particularly sensitive to the high-value tail of the Mach number distribution with high values \citep[e.g.,][references therein]{wittor21, 2022MNRAS.509.1160I}. Using the flux density at 1.4 GHz, we have estimated a radio power P$_{1.4 \rm GHz}$ = (2.13 $\pm$ 0.06) $\times$ 10$^{24}$ W~Hz$^{-1}$. With this value, the relic in A2163 fits well in the radio power vs. LLS relation observed for other known relics \citep[e.g.,][references therein.]{chatterjee24}.

\begin{table}
  \centering
  \caption{Flux density estimates for the radio halo and relic.}
\begin{tabular}{@{}ccccc@{}}
    \hline
     Source & Freq. & Flux density & Ref. & Spectral index \\
      \hline\hline
    Halo + Ridge &153 & 2356 $\pm$ 260 &  S.20  \\
      & 325 & 861.0 $\pm$ 10.0 & F.01 & \\
    
    &402 & 698.0 $\pm$ 69.0& This work & $-$1.19 $\pm$ 0.09 \\
    
    &650 & 348.0 $\pm$ 34.0&  This work &  \\
    
    &1410& 189.0 $\pm$ 10.0 &  B.21 & \\

    \hline

    Ridge &402 & 117.4 $\pm$ 13.0 & This work & $-$1.36 $\pm$ 0.08 \\
    
    &650 & 70.5 $\pm$ 9.0 & This work & \\

    \hline 
    
    Relic & 325 & 86.0 $\pm$ 2.0& F.04 &  \\
    & 400 & 64.4 $\pm$ 7.0& This work & \\
    & 650 & 38.1 $\pm$ 4.0& This work & $-$1.02 $\pm$ 0.03 \\

    & 1410 & 18.0 $\pm$ 0.3& This work & \\
    \hline
\end{tabular}
     
\begin{tablenotes}
    \item Notes: The references are S.20: \citet{2020ApJ...897..115S}, F.01: \citet{2001A&A...373..106F}, B.21: \citet{2021ApJ...906...87R}, F.04: \citet{2004A&A...423..111F}.
\end{tablenotes}
     
  \label{int-spec-tab}
\end{table}

\subsection{Radio surface brightness profile} \label{SB_prof_halo}

The high sensitivity and deep exposures of recent low-frequency telescopes have been able to map the diffuse emission on scales roughly corresponding to r$_{500}$ of the four clusters. \citet{2022Natur.609..911C} reported that the radial SB profiles of those radio halos flatten in the outskirts, while steeper in the inner regions. The large size, and high SNR of the halo make A2163 an ideal test-bed to search for distinct components of the radio halo. To explore this, we used only the 400 MHz image (with tailed galaxies, relics, and ridges masked) at a 45$''$ resolution due to the detection of the radio halo to the largest extent among the three frequencies. We manually masked these sources in the image plane, not from the \textit{uv} domain. We extracted the surface brightness profile using circular annuli of 22.5$''$ width (half the FWHM of the restoring beam) up to the 2.9 $\sigma_{rms}$ contour level along two different directions (e.g., east and west) to account for potential biases related to geometrical effects. The uncertainties in the profiles are estimated considering both systematic errors, primarily from flux scale calibration (assumed to be 5–10\%), and statistical uncertainties arising from the rms noise level of a radio image. The centre of the annuli was chosen as the peak of the radio surface brightness.

The surface brightness profile of radio halos is commonly described by the exponential law of the form (assuming a spherical morphology of the radio halo): 

\begin{equation} \label{halo_model_eq}
    \rm I(r) = \rm I_{0} e^{-\frac{r}{r_{e}}}
\end{equation}

where I$_{0}$ is the central surface brightness and r$_{e}$ is the e-folding radius, at which the surface brightness is I$_{0}$/e. Although the simplistic exponential model has traditionally offered a reasonable approximation of radio halo radial profiles \citep[e.g.,][]{2006AN....327..565O, 2009A&A...499..679M}, recent observations with new-generation telescopes revealing substructures and multiple components have underscored its limitations \citep[e.g.,][]{2021A&A...646A.135R, 2023A&A...674A..53B}, especially since merging systems often display highly extended and irregular features.

\begin{figure}
	\includegraphics[width=\columnwidth]{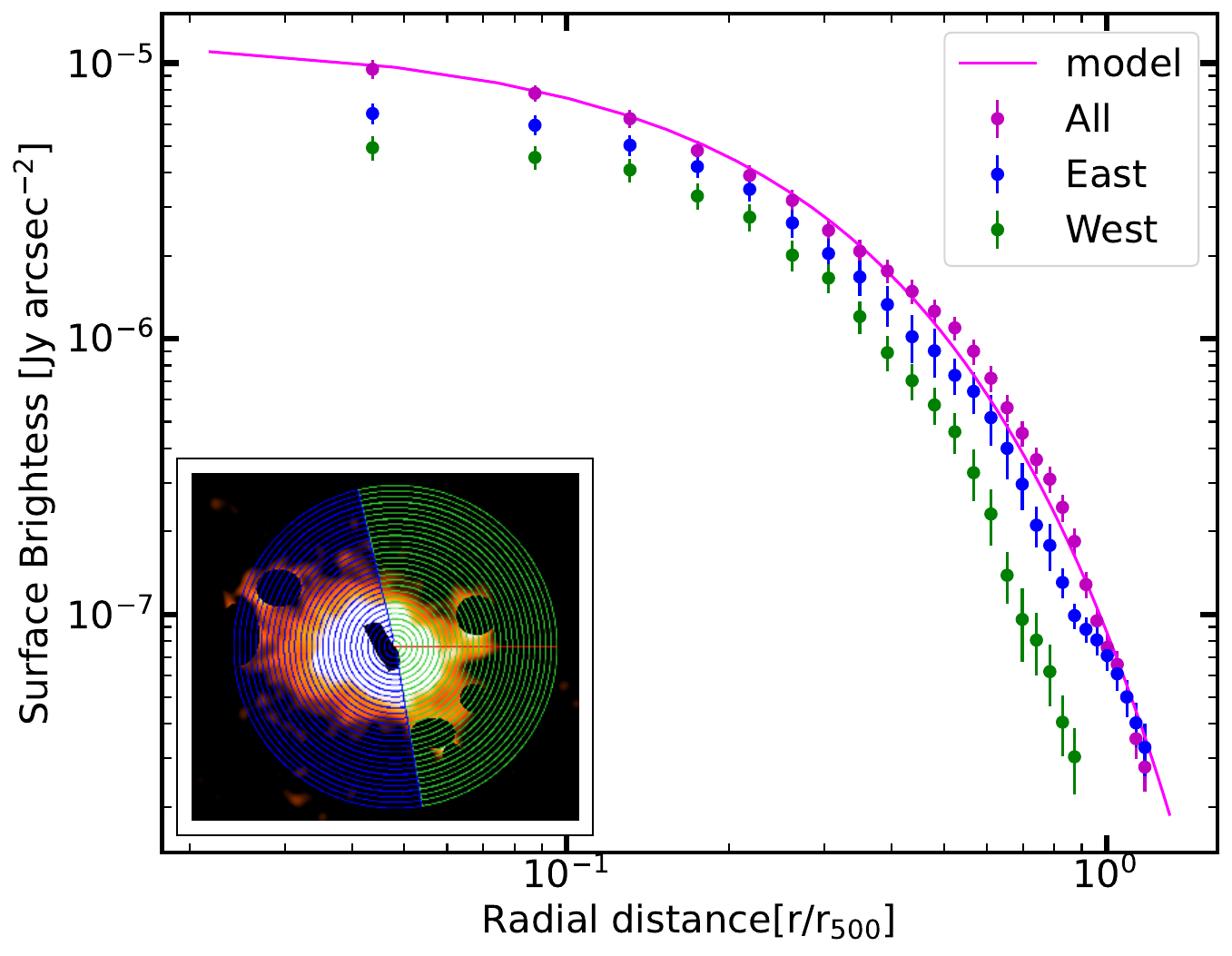}
        \caption{The radial profile of the radio surface brightness for the halo emission is shown at 400 MHz. The two different colours indicate the eastern and western sub-parts of the halo, and the annotated image shows the annular region, which has been used to estimate the average surface brightness. Since the halo emission is less extended in the western part, we have not put any points for those regions. The magenta solid line shows the model (equation.~\ref{halo_model_eq}) fitted to the radial profile for the total halo emission.}        
    \label{A2163_halo_SB}    
\end{figure}

In Figure~\ref{A2163_halo_SB}, we present the radial surface brightness profiles of the radio halo, normalised by the cluster's r$_{500}$. Each data point represents the average radio brightness measured within a concentric annulus. The corresponding central surface brightness and the e-folding radius obtained from the exponential fit are listed in Table~\ref{radio_SB_fit_param}. Notably, the radio halo emission is well described by a single exponential profile that extends up to r$_{500}$, despite the simplified assumption of symmetrical morphology. This is particularly remarkable given the cluster’s dynamically disturbed state and the presence of several substructures. We do not find any hint of secondary components at larger radii in the surface brightness profile, even when examining the eastern and western regions of the cluster. \citet{2021A&A...647A..50C} also report the central surface brightness and e-folding radius values at 1400 MHz. Their findings show that the best-fit e-folding radius varies with observing frequency, being larger at lower frequencies—consistent with results from other extended radio halos \citep[e.g.,][]{rajpurohit25}, suggesting a more centrally peaked emission at low frequencies. It is also noteworthy that the surface brightness profile shown here extends to a larger radius ($\sim$1700 kpc) compared to that in \citet{2021A&A...647A..50C} ($\sim$1300 kpc). The potential impact of systematics on the non-detection of a second component will be discussed in a later section.

\begin{table}
  \centering
  \caption{Fitting results obtained from an exponential fit}
  \begin{tabular}{@{}ccccc@{}}
    \hline
      Freq. (MHz) & $\chi_{r}^{2}$ & I$_{0}$ & r$_{\rm e}$ & Ref.  \\
    \hline\hline

    400 & 1.19 & 12.26 $\pm$ 4.10 & 566 $\pm$ 20 & This work \\

    1400 & 1.16 &2.18$\pm$ 0.11 & 402$\pm$ 10  & C.21 \\

    \hline
  \end{tabular}
  \label{radio_SB_fit_param}
\begin{tablenotes}
    \item Notes: Column (2): reduced $\chi^{2}$ value. Column (3): central brightness in units of $\mu$Jyarcsec$^{-2}$. Columns (4): e-folding radii in units of kpc. The reference C.21: \citet{2021A&A...647A..51C}
\end{tablenotes}
\end{table}

\subsection{Detection of radio surface brightness edge}

Dynamical motions in the ICM can create distinctive features in X-ray images that map the thermal bremsstrahlung emission from galaxy clusters, such as sharp surface brightness discontinuities caused by shocks and cold fronts \citep{2007PhR...443....1M}. Recently, \citet{2023A&A...674A..53B} observed similar discontinuities in the radio surface brightness profile of MGCLS\footnote{\url{http://mgcls.sarao.ac.za/}} clusters, coinciding with the locations where such discontinuities are detected in X-rays. This similarity suggests a significant interaction between thermal and non-thermal components in galaxy clusters. MeerKAT is crucial for detecting these sharp discontinuities due to its high resolution and extended source sensitivity. Our deep and sensitive uGMRT images offer an opportunity to search for possible radio surface brightness discontinuities in the diffuse radio halo emission.

\begin{figure}
    \includegraphics[width=\columnwidth]{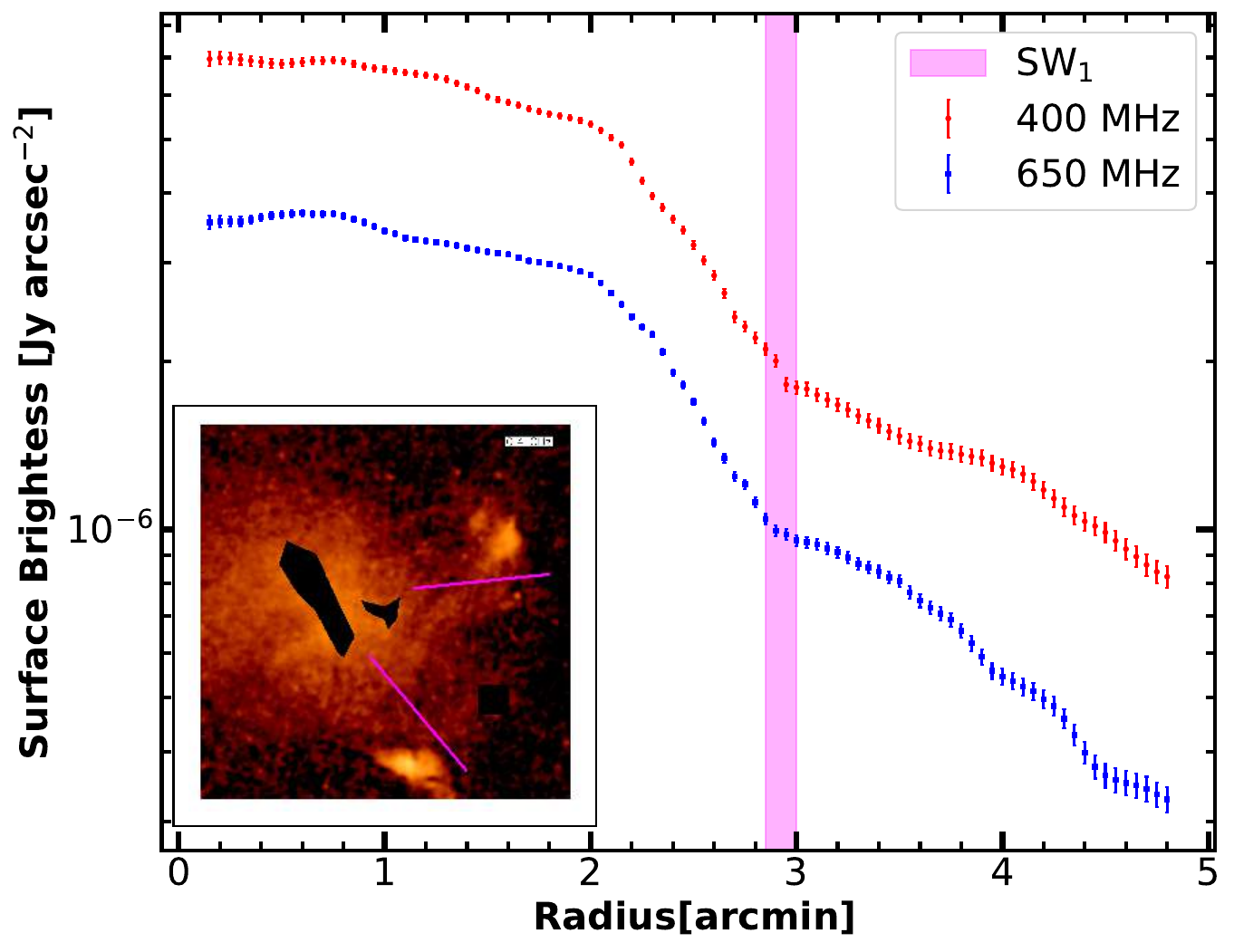}
    \caption{Radio surface brightness profiles extracted from the discrete source-subtracted uGMRT image. The magenta coloured segment in the inset panel denotes the sector used to extract the surface brightness profiles plotted following the different colours for different frequencies. The magenta region indicates the shock wave at that location, reported by \citet{2018A&A...619A..68T}.} 
    \label{SB_edge_halo}
\end{figure}

For better visualisation of the discontinuities, in Figure.~\ref{SB_edge_halo} we present the radio surface brightness profiles extracted from the discrete source-subtracted image (masking the residuals of the tailed galaxies, diffuse lobes, and the ridge) across the southwestern edge. The beam size of the radio image is 10$''$ $\times$ 10$''$. We followed the procedures described in \citet{2023A&A...674A..53B} (1$''$ radial binning, which is smaller than the resolution of the radio image, meaning that the data points are correlated) for extracting the profiles. The profile shows some discontinuity at a distance of 3$'$, where \citet{2018A&A...619A..68T} reported the presence of a merger shock, with a Mach number of 1.8. The profiles show rapid surface brightness declines over small scales. In this case, the radio surface brightness drops by a factor of $\sim$ 3$-$4 within $\sim$ 100 kpc, very similar to the jump reported in the Abell 2744 E sector by \citet{2023A&A...674A..53B}. The jump is clearly seen in 400 and 650 MHz. To our knowledge, this is the first time we have detected a radio surface brightness discontinuity in this system. The non-thermal emission from the halo depends on the cosmic ray electron density and magnetic field strength, and determining what causes the jump is not trivial.

\citet{2018A&A...619A..68T} reported another shock front (at a distance of 6.5$'$, with a Mach number of 3.2$_{-0.7}^{+0.6}$), where the radio halo emission appears to be bounded by this front. We observed a sharp radio halo edge at very low resolution ($\sim$35$''$) at that location across all frequencies (indicated by the yellow arrow in Figure~\ref{A2163_gmrt_vla}). Similar cases of shock fronts enclosing halo emission have been reported in other clusters \citep[e.g.,][]{2005ApJ...627..733M,brown11,2019SSRv..215...16V, 2024arXiv240618983B}. We emphasise that our observations were deep and that we successfully recovered a significant fraction of the large-scale emission in the eastern direction. However, negative patches (with sizes of $\sim$300 kpc, at the $-$4$\sigma$ level) are present nearby, which makes it difficult to detect any potential pre-shock emission. The sharp discontinuity is only visible in the lower-resolution (35$''$) images. At higher resolutions, the edge is not detectable. The surface brightness rapidly declines within 200 kpc (61$''$) of the shock front's position. \citet{2018ApJ...856..162W} presented detailed modelling of a similar radio edge at the bow shock in A520, simulating the pre-shock radio emission. Their analysis provided evidence to rule out the adiabatic compression model as the origin of the observed edge; however, their analysis was limited by image artefacts in the upstream region. To draw further conclusions about the nature of these edges, more samples need to be studied using the sensitive spectro-polarimetric capabilities of next-generation telescopes.

\begin{figure*}
    \includegraphics[width=9cm, height = 7.8cm]{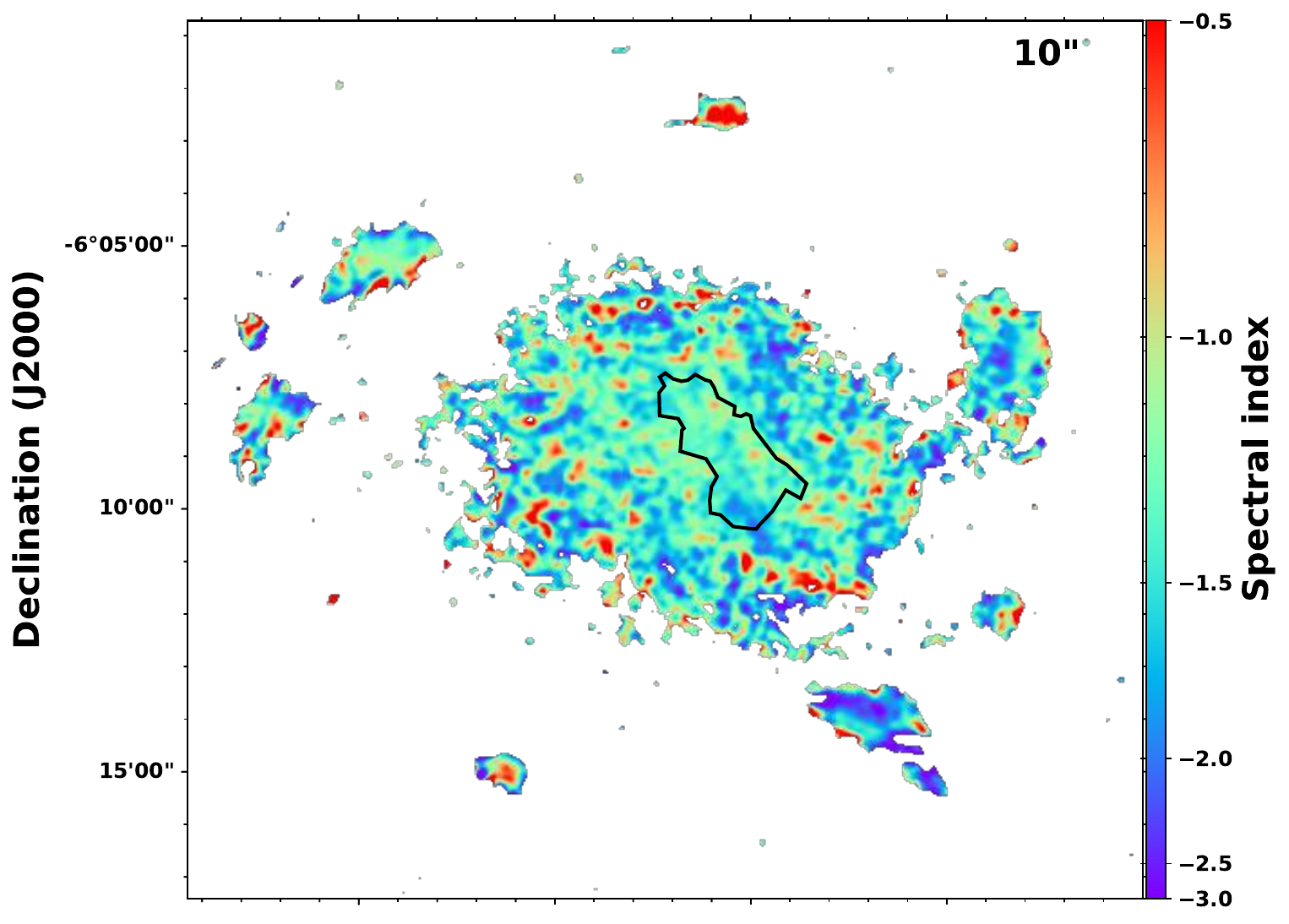}
    \includegraphics[width=\columnwidth, height=7.8cm]{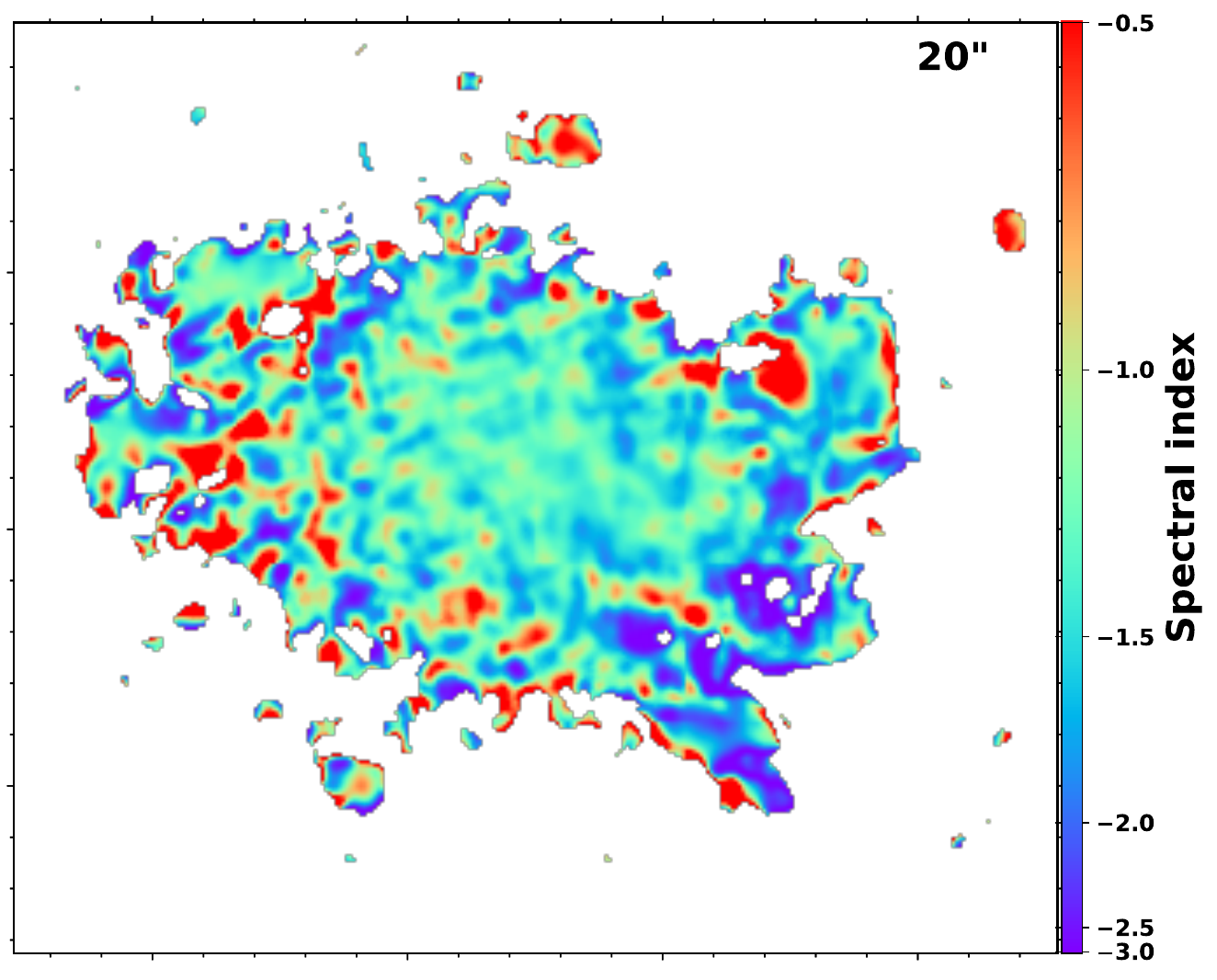}
    \includegraphics[width=9cm, height = 8cm]{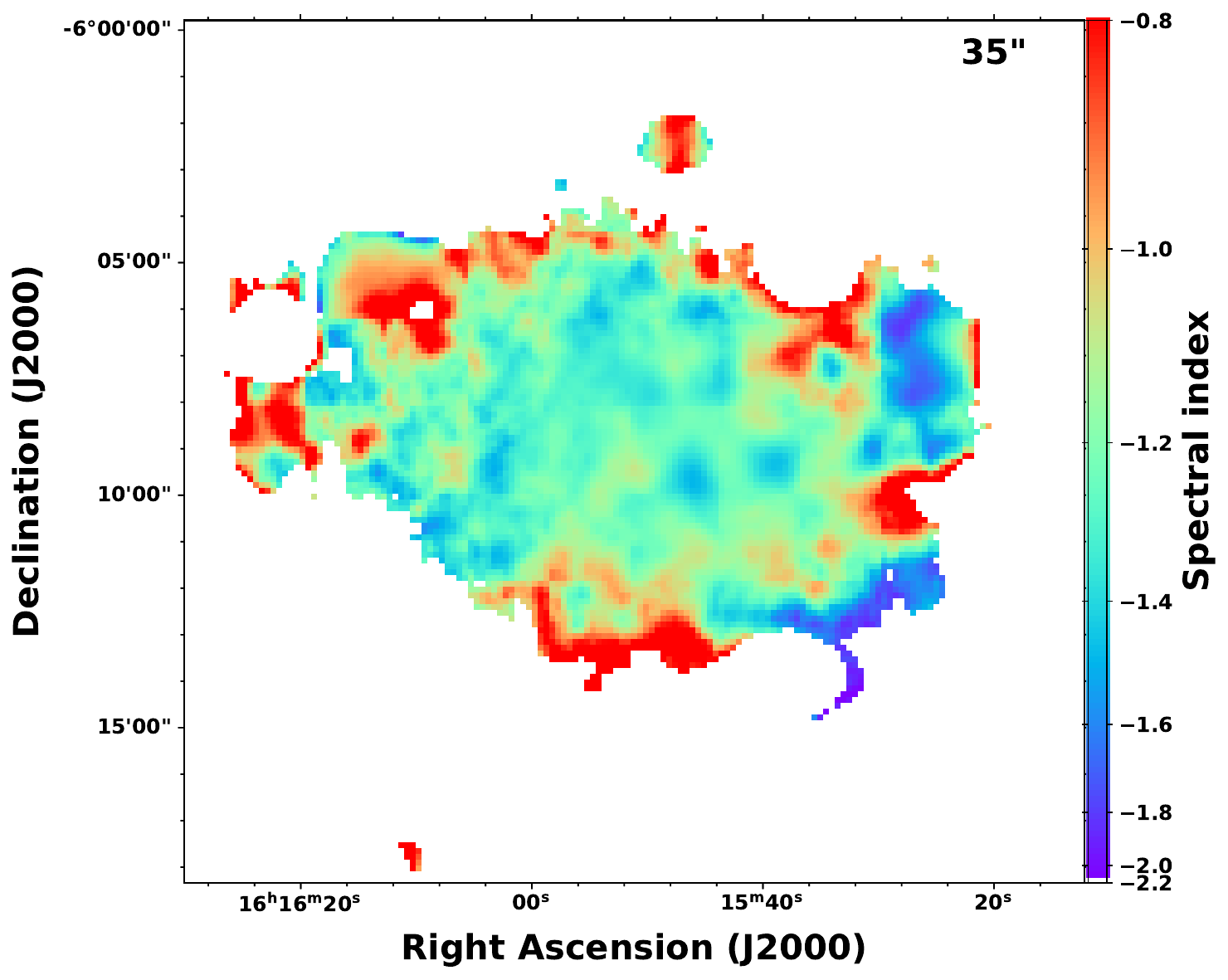}
    \includegraphics[width=\columnwidth, height=8cm]{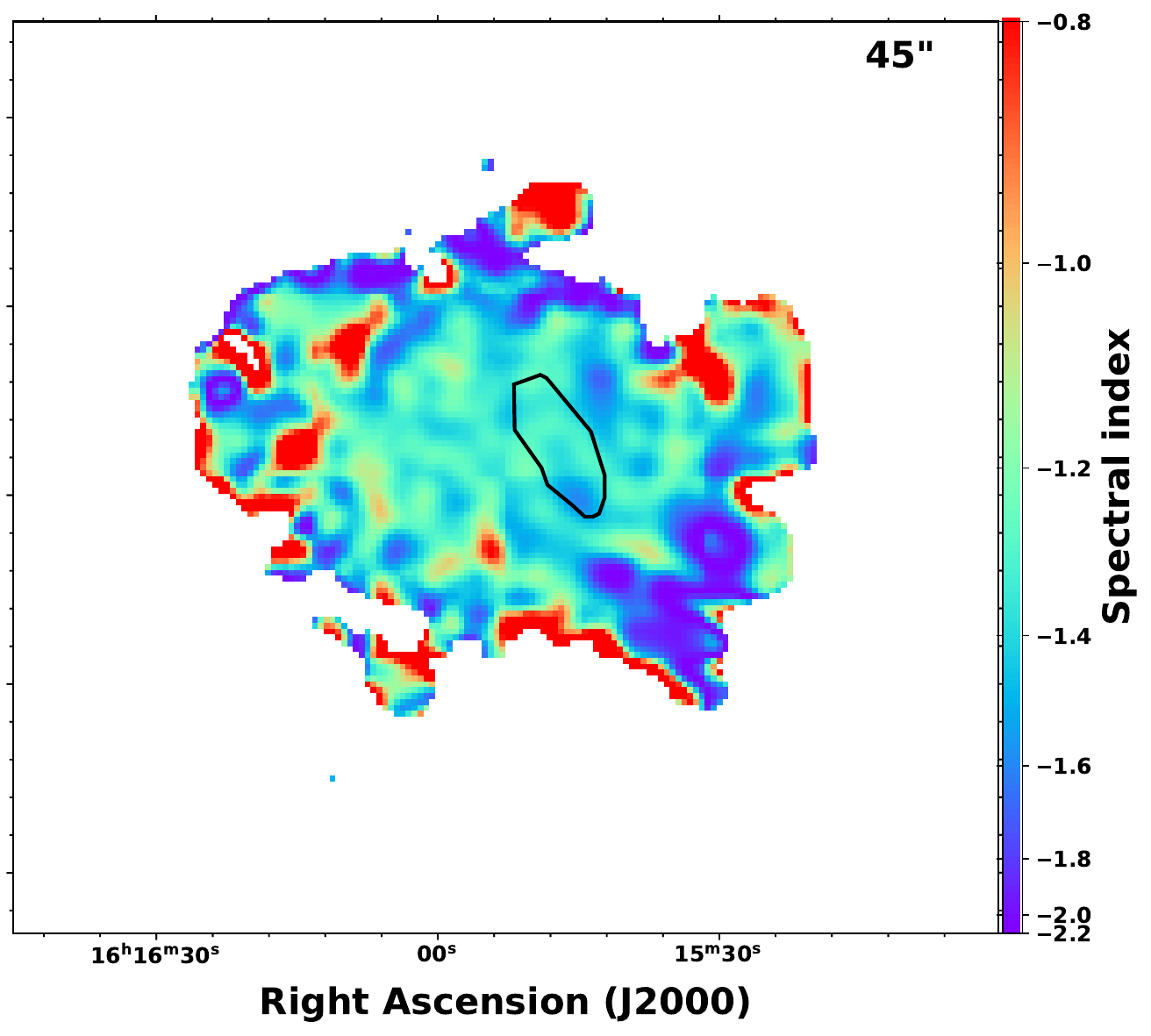}
     
    \caption{\textit{Upper left:} The spectral index map between the 400 and 650 MHz is shown at a 10$''$ resolution. The black polygon marked the position of the ridge. \textit{Upper right:} The same is shown at a 20$''$ resolution. \textit{Lower left:} Resolved spectral map for the uGMRT (400 MHz) and VLA (1400 MHz) is shown at 35$''$ resolution, using the point source subtracted visibilities. \textit{Lower right:} The same is shown at 45 $''$ between the two uGMRT frequencies, and the black polygon marks the ridge location.}
    \label{spec_map}    
\end{figure*}

\subsection{Resolved spectral map} \label{resv-spec-study}

\begin{figure*}
    \includegraphics[width=\columnwidth, height= 7.0cm]{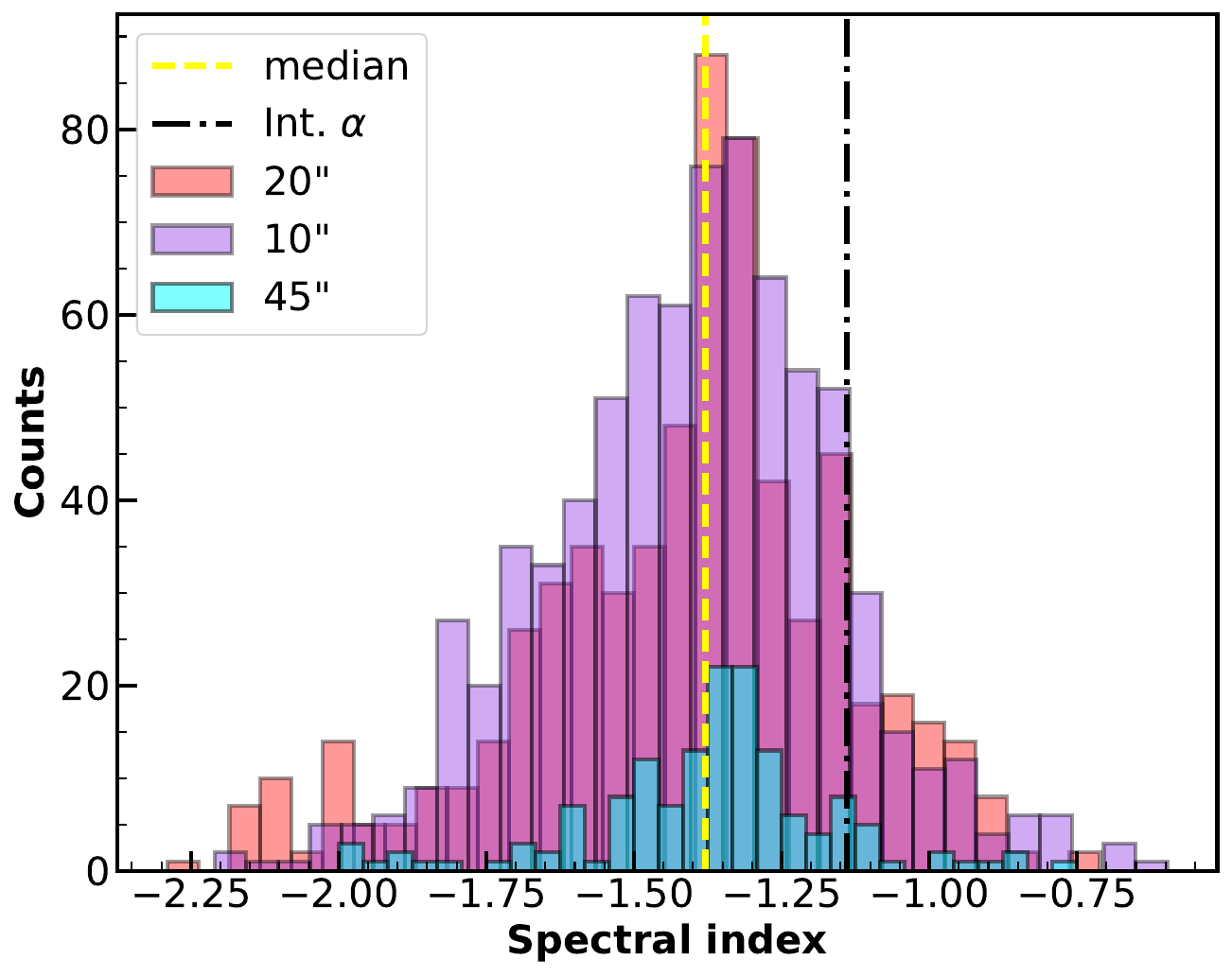}
    \includegraphics[width=\columnwidth, height=7.0cm]{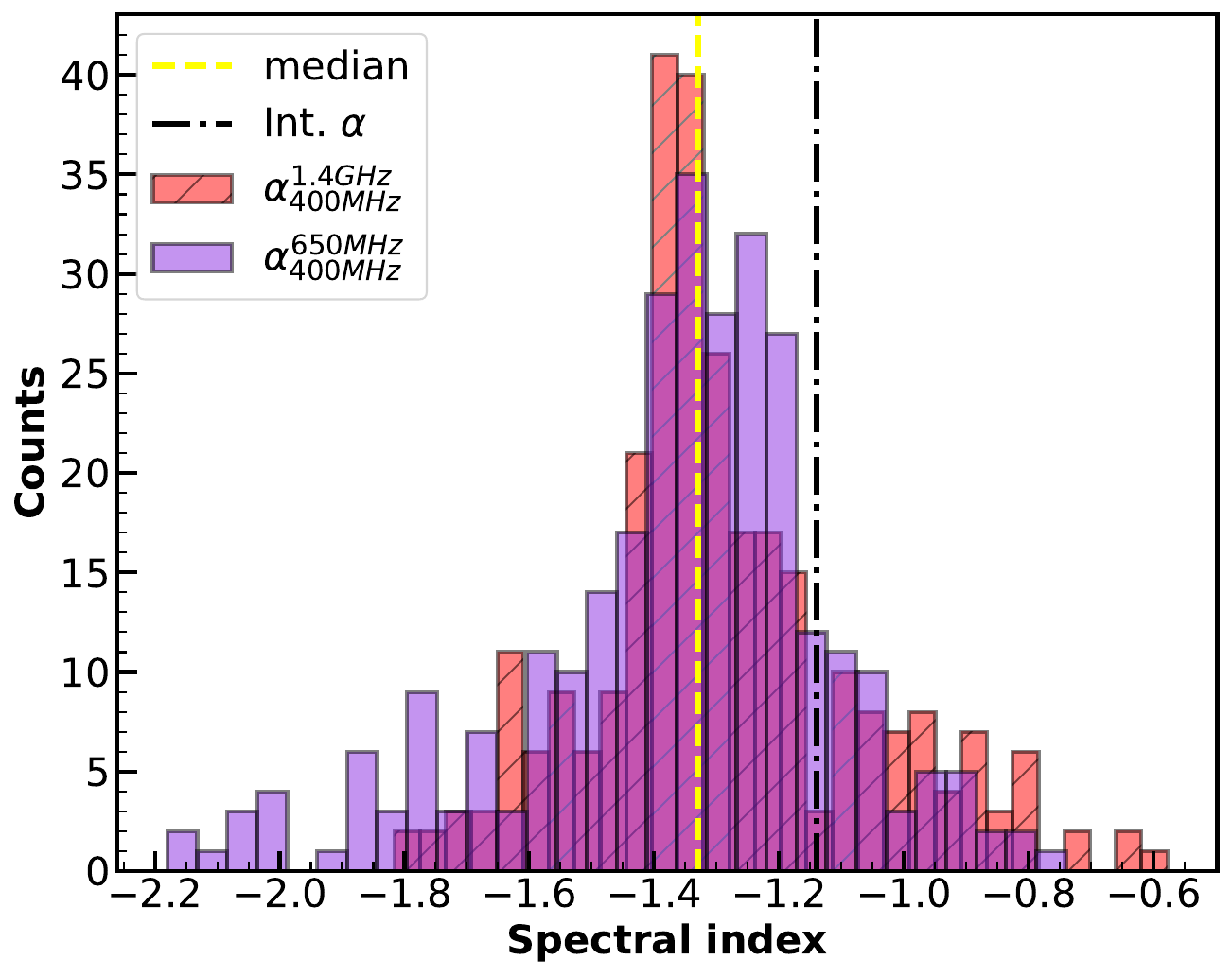}
    \caption{\textit{Left:} Histogram of the spectral index distribution for the halo in A2163 between 400 MHz and 650 MHz is shown for three different (10$''$, 20$''$, 45$''$) resolutions. The values for the spectral indices were obtained by sampling the halo into beam-sized square boxes. The yellow dotted line is the median value of the spectral index, $\alpha_{\rm med}$ = -1.42. The black dashed-dotted line shows the integrated spectral index value for the halo. \textit{Right:} The same is shown for the uGMRT and VLA frequencies at a resolution of 35$''$, and the median and integrated spectral index are denoted by the dotted lines.} 
    \label{spec-dist}
\end{figure*}

Spatially-resolved spectral index information is crucial in understanding the different physical processes related to mergers and the origin of the Mpc-scale diffuse emission at the cluster centre and in the outskirts \citep[e.g.,][]{2016ApJ...818..204V, 2024ApJ...962...40S}. Our uGMRT and VLA images have allowed us to study the resolved spectral map for the radio halo region with a spatial scale spanning over one order of magnitude. The spectral index maps are created using the point source-subtracted visibilities from uGMRT and VLA datasets. The images were made with a similar inner \textit{uv} cut-off (\textgreater 0.2 k$\lambda$) and a uniform weighting to match the spatial scale of the radio emission at both 400 and 650 MHz. The archival VLA observations were not deep enough and lacked longer baselines, restricting the \texttt{uv}-range to 0.2 $-$ 16 k$\lambda$ for the spectral index map between uGMRT and VLA. We also created a T-T plot (shown in the Appendix.~\ref{A2163_tt}), a useful tool for comparing the flux density at two frequencies from different instruments because of their ability to account for any variations in the short spacing. We also searched for any astrometry positional shift between the VLA and uGMRT, but not found any significant offset. For each image, the pixels with values below 3$\sigma_{\rm rms}$ are blanked; while this ensures reliable measurements, it may introduce biases by preferentially excluding regions with intrinsically flat or steep spectra due to sensitivity limits at different frequencies. For further information on the spectral index calculation via pixel-to-pixel comparison, we refer to \citet{doi:10.1126/sciadv.1701634}. 
 
We produced spectral index maps at three different resolutions (10$''$, 20$''$, and 45$''$) using the uGMRT frequency bands, as well as a low-resolution (35$''$) spectral map combining the uGMRT and VLA data, presented in Figure.~\ref{spec_map}. The associated error maps are shown in Figure.~\ref{spec_map_err} (Appendix.~\ref{err-map}). The halo exhibits localised areas where the spectral index deviates noticeably from the integrated value. In the 10$''$ resolution image, substantially small-scale fluctuations are visible across the halo, which tend to average out when the resolution is decreased. At 20$''$ resolution, a steep-spectrum feature appears to connect the D1 structure to the radio halo, suggesting lobe ageing followed by re-energisation within the halo region. Toward the periphery, the spectral index becomes steeper, and the combined uGMRT–VLA low-resolution maps reveal an overall patchy distribution in the central regions. The 45$''$ resolution image predominantly shows a uniform spectral index of $\sim-$1.3 throughout most of the halo, with a few steeper patches located in the southwestern outskirts. While \citet{2004A&A...423..111F} reported indications that the western halo has a flatter spectrum than the eastern side, our high-sensitivity spectral mapping does not confirm such a trend. The solid black contour in Figure.~\ref{spec_map} delineates the ridge from the halo, with the ridge showing spectral index values between $-$1.0 and $-$1.4, in close agreement with \citet{2020ApJ...897..115S}. The ridge had previously been interpreted as lying near the merger’s epicentre, potentially explaining a slightly flatter spectrum. As we go from 10$''$ to 45$''$ resolution, small-scale fluctuations are smoothed out, yet the ridge shows an overall uniform spectral index distribution, consistent with its integrated value.

The high-resolution spectral index map shows some intrinsic fluctuations; therefore, we followed \citet{2017ApJ...845...81P} to check the spectral index distribution over the extent of the radio halo (note that we have masked the ridge region for extracting the spectral index values) for 10$''$, 20$''$, and 45$''$ using the uGMRT maps (Figure.~\ref{spec-dist}, left). We find that the distribution is asymmetric for all three different resolutions and the median values peak at \textless$\alpha_{10''}$\textgreater = $-$1.42, \textless$\alpha_{20''}$\textgreater = $-$1.39, \textless$\alpha_{45''}$\textgreater = $-$1.35, with a standard deviation of $\sigma_{10''}$ = 0.21, $\sigma_{20''}$ = 0.19, and $\sigma_{45''}$ = 0.14 respectively. To understand the level of variation in the values, we fit a simple zeroth order polynomial to the histogram data and obtain reduced chi-squared values of $\chi_{10''}^{2} = 1.80$, $\chi_{20''}^{2} = 1.50 $, and $\chi_{45''}^{2} = 1.21$, stipulating that some intrinsic fluctuations exist and can be confirmed by comparing the median error value of the distributions with the corresponding standard deviations. We subtracted the median spectral index error ($\sim$ 0.12) in quadrature from each standard deviation and estimate the intrinsic scatter to be $\sigma_{\rm sub, 10''}$ $\sim$ 0.17, $\sigma_{\rm sub, 20''}$ $\sim$ 0.15, and $\sigma_{\rm sub, 45''}$ $\sim$ 0.07. The decrease in $\sigma$ with increasing beam size reflects the smoothing of spatial fluctuations in the spectral index due to averaging over a larger number of emitting regions within the beam. We have also shown the spectral index distribution for maps (35$''$) between 400 and 650 MHz, and 400 and 1400 MHz (Figure.~\ref{spec_map}, right), to check for any variations. We find that the median spectral index peaks around \textless$\alpha_{\rm median}$\textgreater = -1.35, with a standard deviation of 0.21 for the 400 and 1400 MHz spectral map. Therefore, we have studied the presence of fluctuations in the radio halo region, originating from turbulence in the ICM, over a physical resolution of 36 to 126 kpc. The distribution of spectral index for the halos, as reported by \citet{2020ApJ...897..115S}, peaks around -2.26, contrasting with ours, likely due to significant flux loss at 332 MHz in their analysis.

\begin{figure}
    \includegraphics[width=\columnwidth]{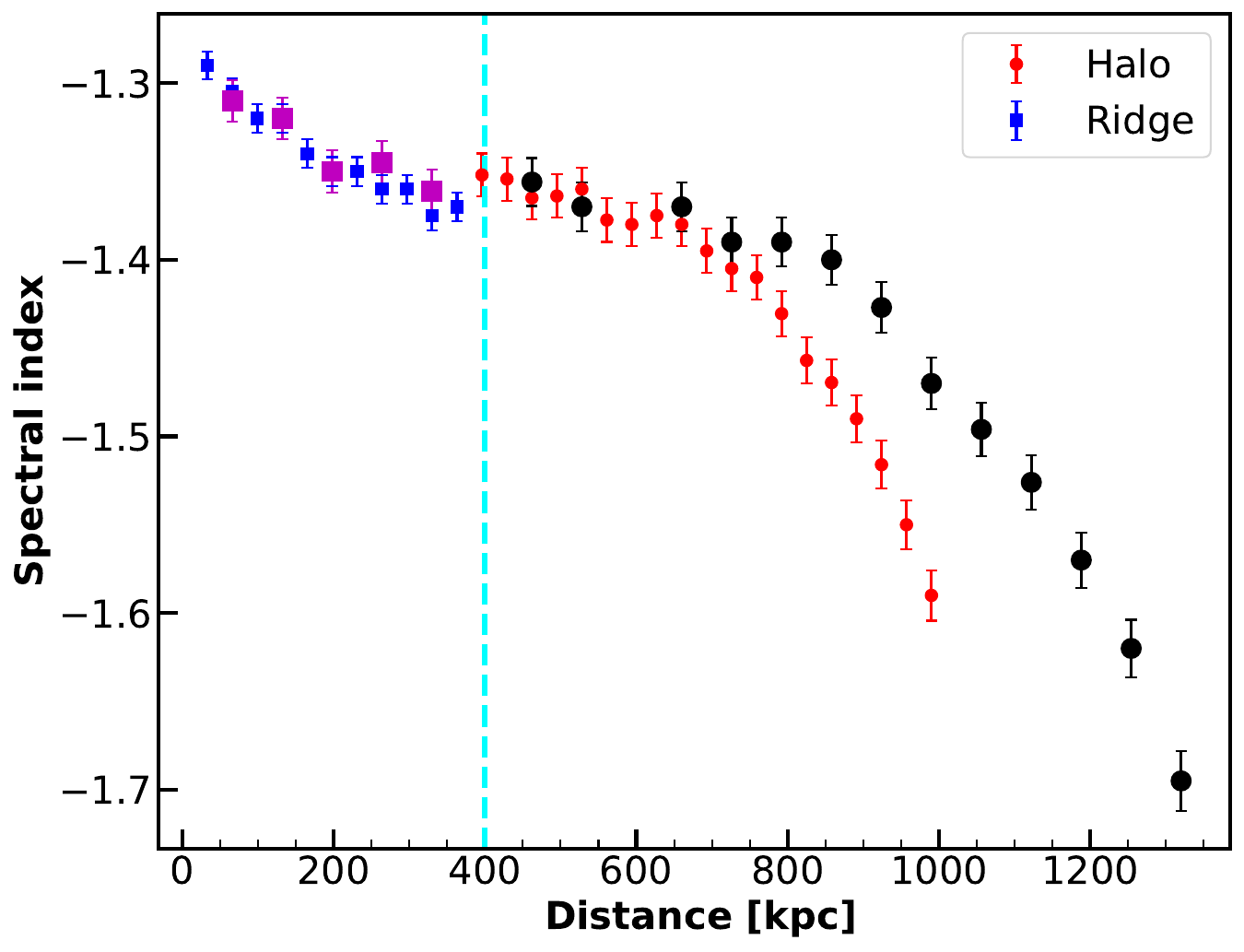}
    
    \caption{The radial profile for the spectral index is shown for the ridge (square) and halo (circle) for two different resolutions: 10$''$, 20$''$ (big). The profile for the ridge is shown along its width, using the length-averaged boxes. Excluding the ridge and tail galaxies, the spectral information for the halo is extracted from the circular annular region. Magenta symbols mark the 20$''$ ridge measurements, while black symbols mark the 20$''$ halo measurements. The cyan dashed line shows the boundary between the halo and the ridge.} 
    \label{rad-prof-halo}
\end{figure}

\citet{2004A&A...423..111F} speculated a spectral steepening (-1.0 to -1.6) for the eastern sub-cluster along the southeast sector. Later, \citet{2020ApJ...897..115S} studied the spectral variation over the radius and reported a radial steepening from -1.49 in the central region to -2.55 at the outskirts. We also checked for the radial variation of the spectral index using our high-resolution spectral map obtained with uGMRT. We masked the residual of the tailed galaxies, the ridge, and the other diffuse lobes at the periphery. The diffuse halo emission was sampled with circular concentric annuli, where the centre of the annulus is chosen as the cluster centre. The radial profile shows a signature of steepening ($-$1.4 to $-$1.6) from the central position to the outskirts (Figure.~\ref{rad-prof-halo}). The results are shown for two different resolutions in (10$''$, 20$''$) images, and the radial evolution is overall similar for all cases. The 10$''$ profile shows the trend most strongly among the two. The steepening of the spectral index at the outer regions is also supported by the change of the e-folding radius over frequency, being higher (lower frequency) to lower (higher frequency). We speculate that the apparent steepening in the outskirts may correspond to regions with less efficient turbulent re-acceleration or lower magnetic fields \citep[e.g.,][]{2001MNRAS.320..365B}, although we note that these regions also coincide with larger uncertainties.

Using the 10$''$ spectral map, we extracted the length-averaged spectral index profile along the width of the ridge to search for possible trends. The profile changes from $-$1.28 to $-$1.38 over the width; however, the trend is not very striking (Figure.~\ref{rad-prof-halo}). The spectral index change may be attributed to the averaging of some steep spectral patches at the southern end of the ridge, where the tailed radio galaxy was present.

\begin{figure*}
    \includegraphics[width=9.5cm, height = 9cm]
    {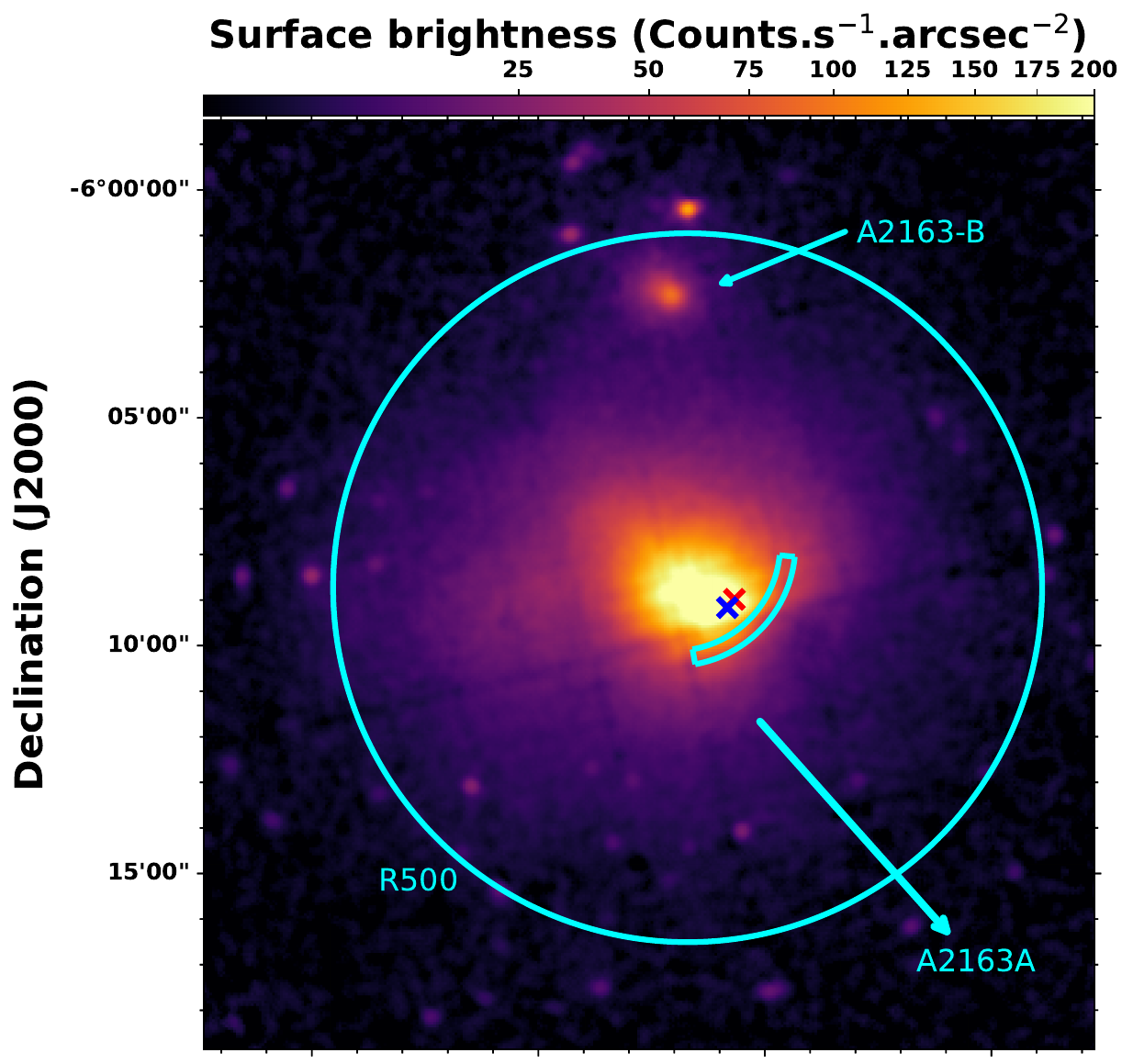}
    \includegraphics[width=8cm, height=9cm]{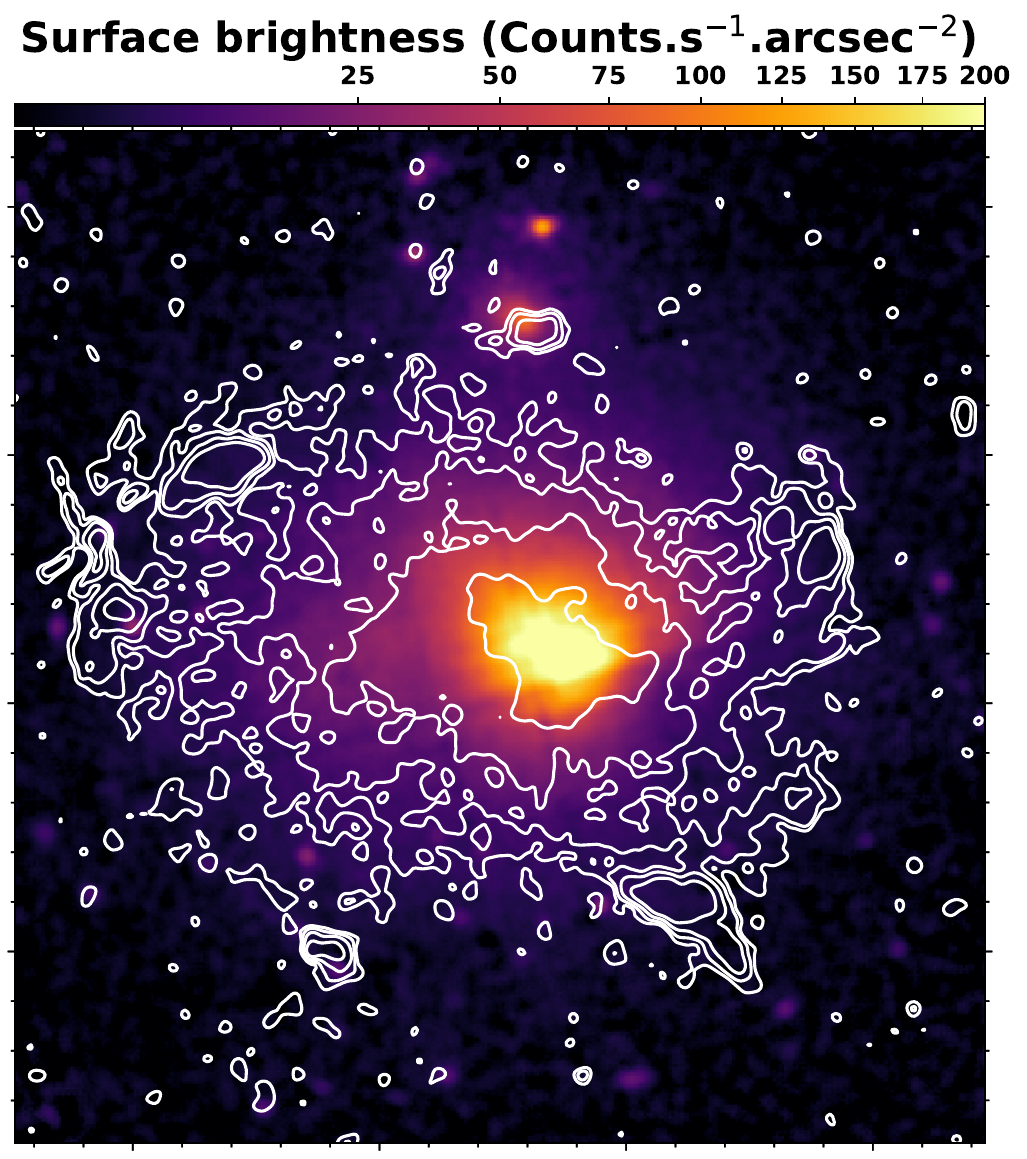}
    \includegraphics[width=9.0cm, height = 9.3cm]{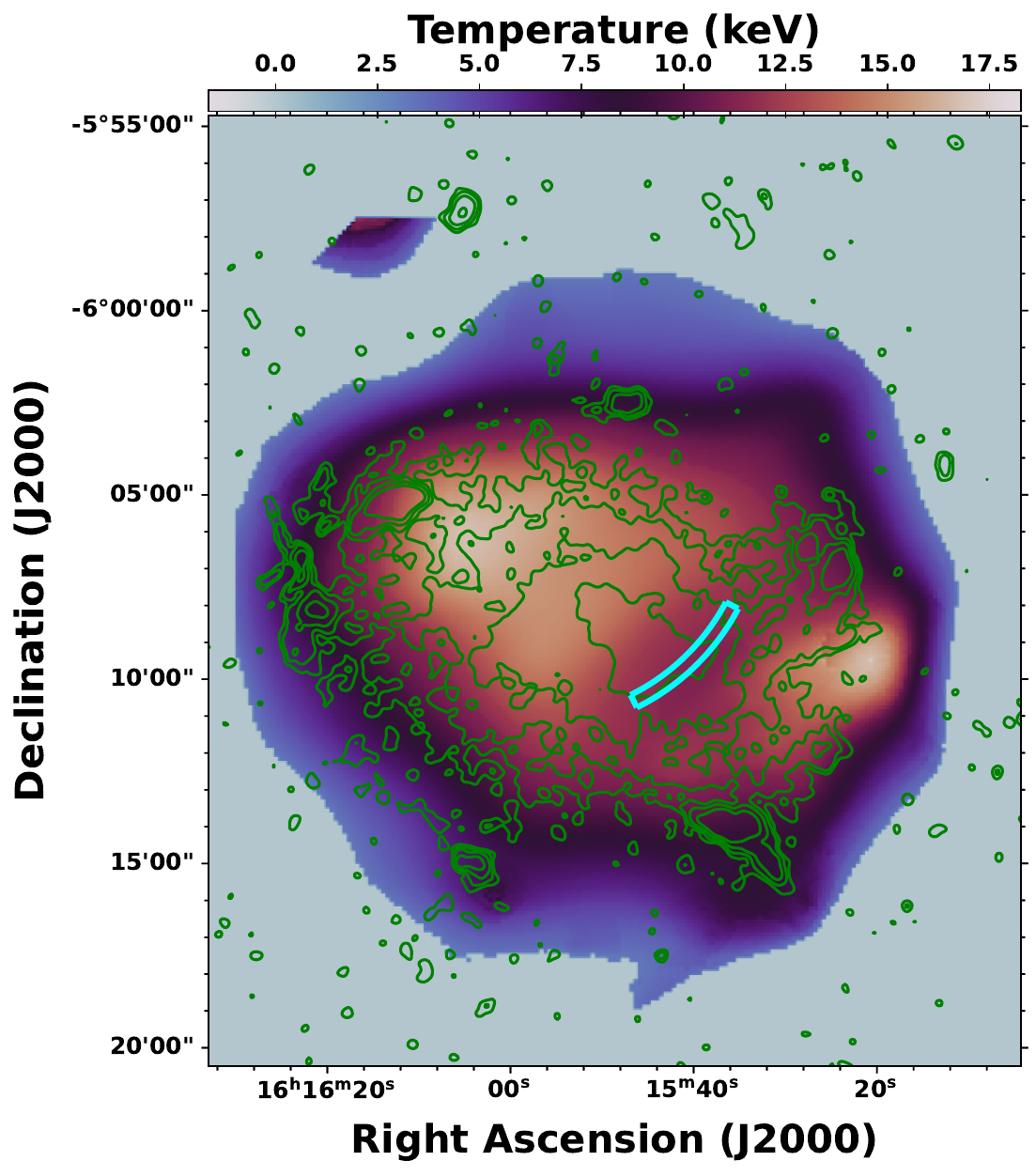}
    \includegraphics[width=8cm, height=9.3cm]{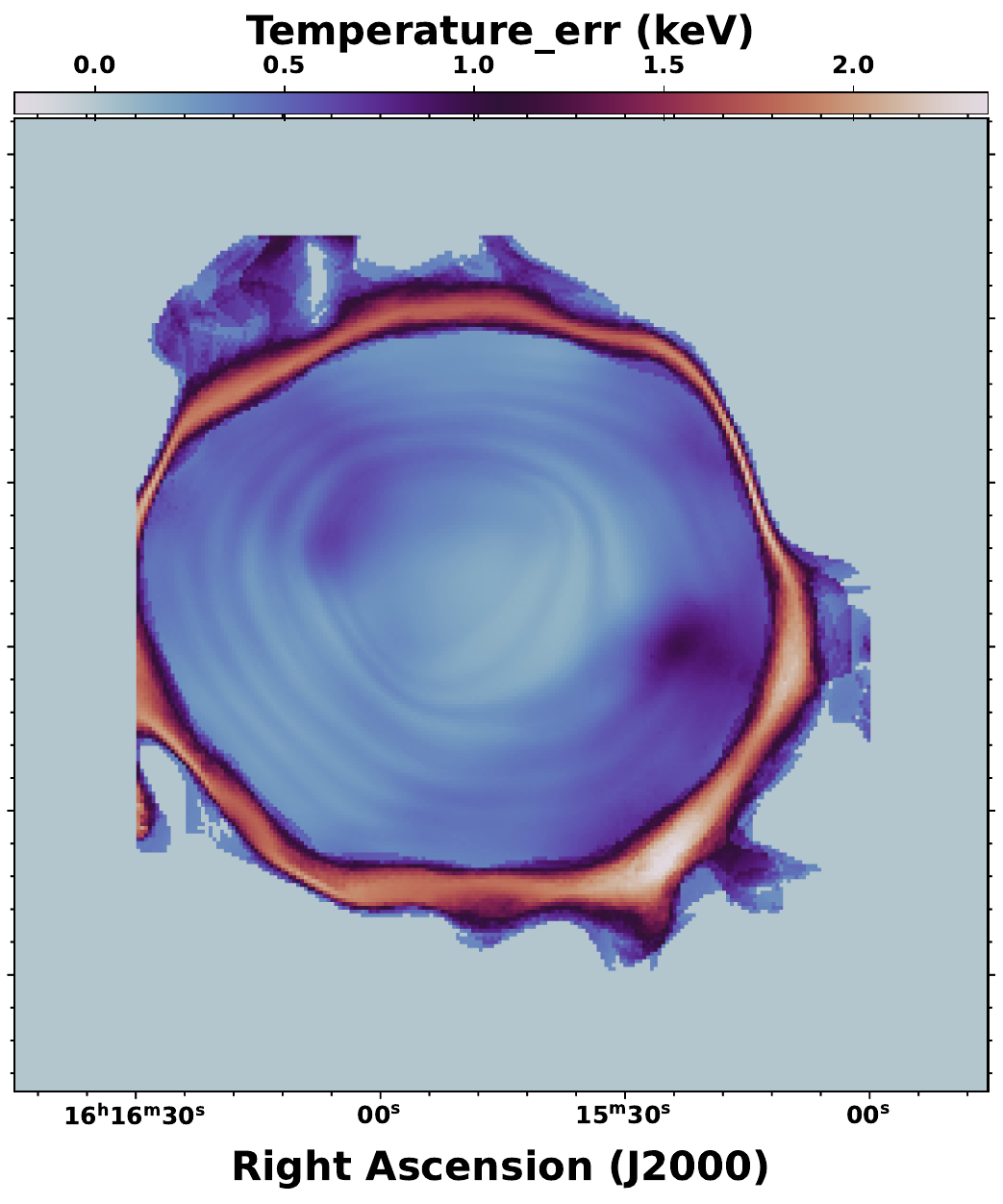}
     
    \caption{\textit{Upper left:} The exposure-corrected, background-subtracted XMM-\textit{Newton} is shown, smoothed with a Gaussian Full-width half maximum of 5$''$. The cyan circle indicates the r$_{500}$ of the cluster, and the northern subgroup is also mentioned. The arc-shaped section indicates the cold front close to the centre. The red and blue `x' shaped region indicates the peak of the X-ray and radio surface brightness. \textit{Upper right:} The same is shown, with an overlay of the radio emission (white contours) at 400 MHz. The contours level starts from 3$\sigma_{\rm rms}$ $\times$ [1,2,4..], where $\sigma_{\rm rms}$ = 34 $\mu$Jy/bm, at a resolution of 16$''$. \textit{Lower left:} temperature map extracted from the XMM-\textit{Newton} surface brightness map is shown, overlaid with the radio emission at 400 MHz (similar contour level to the previous). The magenta sector indicates the cold front. Most of the radio emission occurs at a comparable hotter part of the cluster. \textit{Lower right:} Here, the error map for the temperature is shown. At the edge region, the errors are high compared to the central regions.}
    \label{xray_SB_map}
\end{figure*}

\section{X-ray analysis} \label{xray-analysis}

\subsection{Surface brightness map}

X-ray observations of A2163 suggest a complex and irregular morphology of the ICM, as revealed by the \textit{XMM-Newton} surface brightness map. The diffuse emission from the thermal gas has been detected up to $\sim$ r$_{500}$ (shown in Figure.~\ref{xray_SB_map}). The \textit{Chandra} observations (90 ks.) were less deep compared to \textit{XMM-Newton} (246 ks.); therefore, the emission has not been detected significantly in the outskirts, also partially due to the smaller field-of-view of the \textit{Chandra} \citep{2011A&A...527A..21B}. The emission declines radially outward. The main component (A2163-A) and a northern subcluster (A2163-B) are also detected quite well and appear to be connected by a faint trail of gas. The X-ray emission from the main cluster is asymmetrical, with a roughly elongated morphology extending along the east-west directions. The bright central region of the cluster also appears to be elongated along the southwest to northeast direction. The cluster exhibits a single surface brightness concentration at the centre, although optical observations reveal two different peaks at the location of two BCGs \citep{2008A&A...481..593M}. A convex-shaped feature (cyan sector) is seen very close to the centre; it is associated with a cold front, as reported by \citet{2009ApJ...704.1349O}.

In Figure.~\ref{xray_SB_map}, we compare the X-ray morphology of the cluster with that of radio emission at 400 MHz. The radio halo emission largely follows the X-ray surface brightness distribution of the \textit{XMM-Newton} maps. In both radio and X-ray, the main cluster is elongated along the east-west direction. Radio emission from the halo extends over the entire region of the detected X-ray emission. The radio emission associated with the A2163-B is also seen, and in very low-resolution images, it is connected to the radio halo emission. The peak of the radio and X-ray surface brightness does not coincide, showing an offset of $\sim$ 20 kpc. The ridge is morphologically correlated with the high X-ray surface brightness region at the centre, and overlaps with the cold front. \citet{2011A&A...527A..21B} also reported a cold gas clump in the west direction, a scenario with features resembling those detected in the so-called ``Bullet cluster'' \citep{2002ApJ...567L..27M}, and the ridge is just situated east of that feature, indicating a very complex interplay of the thermal gas and non-thermal plasma at the cluster centre. The emission from the halo extends further in the west direction, where the X-ray emission is fainter. The halo emission also seems to be bounded (discussed in previous sections) by the shock front, as reported by \citet{2018A&A...619A..68T}.

\subsection{Temperature maps}

A2163 is a very disturbed cluster; therefore, a standard radial profile analysis would only provide a partial view of the temperature distribution. The resulting projected temperature map with overlaid radio contours and  corresponding uncertainties, is shown in Figure.~\ref{xray_SB_map}. The disturbed morphology of the cluster is highlighted by the temperature variation, particularly in the core region. Despite the `global’ temperature of the cluster being 15 keV, it is not isothermal at this temperature. The cluster shows an asymmetric temperature distribution, with two distinct subclumps, separated by a cool temperature strip ($\sim$ 8 keV) between them. Both subclumps have similar average temperature of $\sim$ 16 keV. The intersection of the two subclumps, the region surrounding the cool temperature strip, shows a uniform temperature value of $\sim$ 12 keV, indicating the shocked-heated gas due to the merger. The overall spatial temperature distribution is stretched along the east-west direction, in line with the scenario in which the merger had happened in this direction. The errors associated with the temperature are minimal ($\sim$ 0.2 keV) in the central region, whereas at the outer region, it reaches $\sim$ 1.9 keV. Radio halo emission correlates well with the hotter part of the cluster (within 0.5$\times$r$_{500}$) and also with the comparatively colder ($\sim$ 9 keV) gas distribution in the outskirts. The eastern sub-cluster is larger compared to the western clump; similarly, the radio halo is more extended in the eastern part, compared to the western part. The ridge is spatially correlated with the high-temperature ($\sim$ 13 keV) regions of the eastern sub-clump; however, some parts are also co-spatial with the cool temperature strip in projection. The spatially resolved projected temperature map reveals many new interesting features, along with the maps published by \citet{2001ApJ...563...95M,2004ApJ...605..695G,2011A&A...527A..21B}, although we obtain higher resolution and statistical quality in the maps using different algorithms.

\begin{figure}
	\includegraphics[width=\columnwidth]{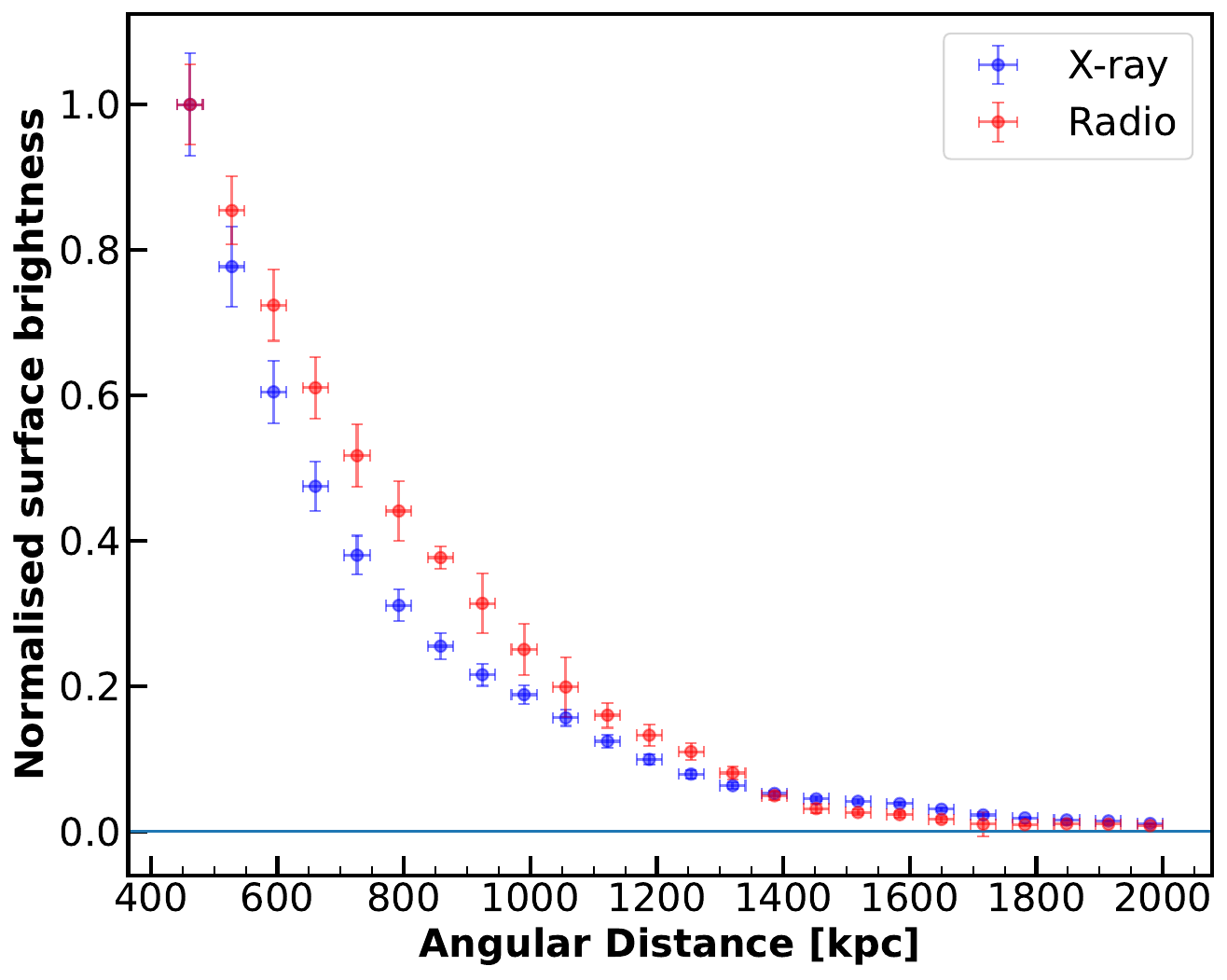}
        \caption{Radio (red squares) and X-ray (blue squares) surface brightness radial profiles within the halo region, normalised at their maximum, are shown. The radio surface brightness is shown at 400 MHz, estimated from the same annular region used in Figure.~\ref{A2163_halo_SB}. The error bars are purely statistical, and the ridge and the residual of the tailed galaxies have been masked for extracting the profile. The blue solid line shows the zero line.}
    \label{r&X-prof}
\end{figure}

\section{Thermal and non-thermal correlation} \label{th-nonth}

Radio halos have been observed to be strongly morphologically correlated with the thermal gas in the ICM \citep[e.g.,][]{2014IJMPD..2330007B}. However, some radio halos do not show any correlation \citep[e.g.,][]{cova19,hoang19}, as in some cases there may be AGN lobes or radio relics projected along line-of-sight. Here, we present the radio and X-ray surface brightness profiles to check the correlation of their radial distribution from the cluster centre to the outskirts. The extended morphology of the radio halo is remarkably similar to that of the X-ray emission in A2163. 

We divided the diffuse halo emission into concentric circular annuli, centred on the cluster centre. The mean surface brightness and standard deviations were estimated within concentric rings. The radio surface brightness is decaying slowly compared to the X-ray surface brightness (in Fig.~\ref{r&X-prof}) (\textless 1100 kpc); however, at the outskirts, their  slope of decay approaches each other after 1200 kpc. This implies that at the centre, the distribution of CRe and/or the magnetic field is rather flat compared to the thermal ICM, and they approach a similar distribution in the outskirts. Therefore, after a certain distance, their radial evolutions resemble each other. In the next section, we will inspect this behaviour in more detail.

\begin{figure*}
    \includegraphics[width=\columnwidth, height= 6.9cm]{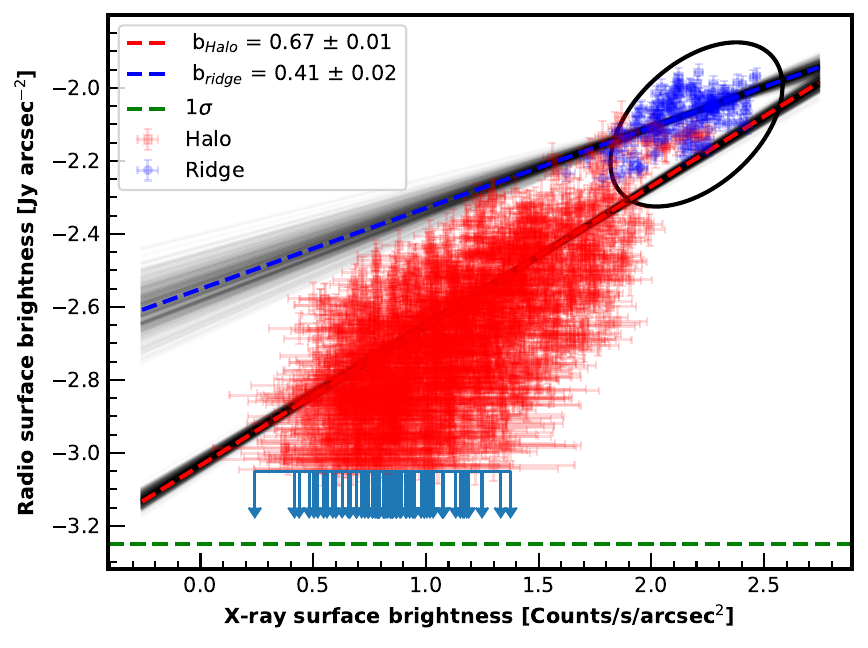}
    \includegraphics[width=\columnwidth, height=6.9cm]{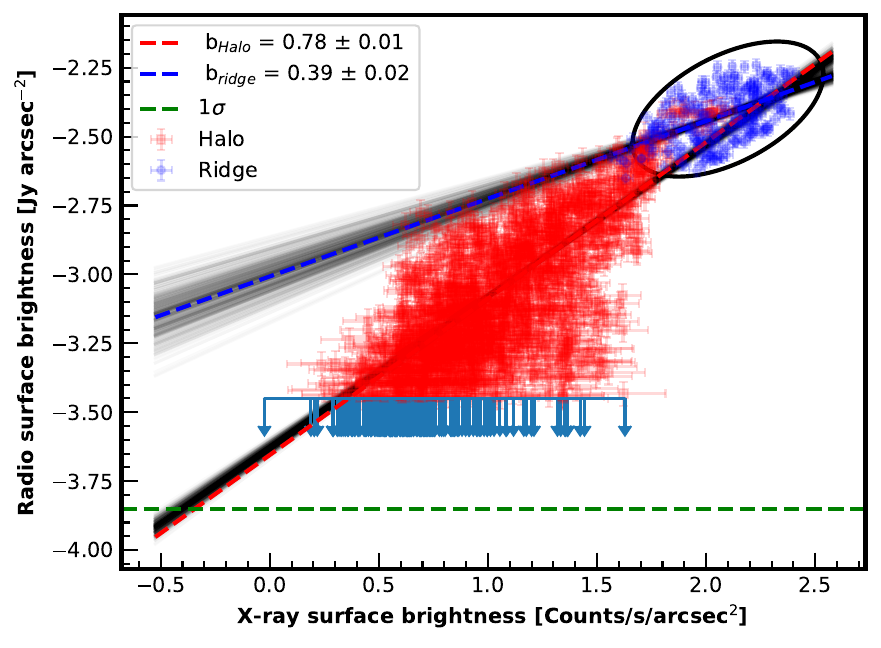}
    
    \caption{Radio Vs X-ray surface brightness plots for the halo in A2163 at 400 (left panel) and 650 MHz (right panel) are shown. The two different lines indicate the separate analysis for the halo and ridge. The black elliptical region indicates the data from the ridge. The red and blue square boxes show the detection of the radio halo above 3$\sigma_{\rm rms}$. The 2$\sigma_{\rm rms}$ are considered upper limits and indicated by cyan arrows. The green dashed line shows the 1$\sigma_{\rm rms}$ line at both frequencies.} 
    \label{ptp_plot}
\end{figure*}

\subsection{Radio and X-ray correlation study} \label{ptp}

Point-to-point (Ptp) analyses of radio (I$_{\rm R}$) and X-ray surface brightness (I$_{\rm X}$) have been carried out for many radio halos across the frequencies ranging from 144 MHz to 3 GHz. For giant radio halos, the correlation slopes are found to be predominantly sub-linear or, at most, linear \citep[e.g.,][]{2001A&A...369..441G, 2005A&A...440..867G,2019A&A...622A..20H,2019A&A...628A..83C,2020A&A...636A...3X,2021A&A...654A..41R,2021A&A...650A..44B,2021A&A...656A.154H,2022ApJ...933..218B,riseley22, 2024arXiv240218654B, riseley24,santra24b, kale25}. The relationship is generally described in the form of a power law:

\begin{equation}
    \log(\rm I_{\rm R}) = \rm a + \rm b\log(\rm I_{\rm X})
\end{equation}

where b is the slope of the correlation. The value of b determines whether the thermal components (gas density, temperature) of the ICM decay faster (b \textless 1) than the non-thermal components (relativistic electron populations, magnetic field) or vice versa (b \textgreater 1). The correlation provides significant physical insight into the particle acceleration mechanism responsible for the origin of the radio halo \citep{2001A&A...369..441G}. We are interested in obtaining the strength of the correlation (if any), which indicates how strongly the gas and non-thermal components are connected; the spatial distribution of the correlation slope from cluster centre to outskirts, i.e., whether one correlation is uniform throughout, or there are coherent areas of different correlations, which could indicate different physical processes at work and/or different environmental conditions; and the spectral dependence of the correlation slope, i.e., whether there is a change in correlation slope b with frequency. Finally, we are also interested in quantifying the slope b, as this provides key insights into the particle acceleration mechanism at work in the radio halo region.

The combination of sensitive high-resolution (10$''$) uGMRT data with XMM-\textit{Newton} X-ray observations enables a detailed examination of how the thermal and non-thermal components of the ICM are related. For this analysis, we use the publicly available software \texttt{PT-REX\footnote{\url{https://github.com/AIgnesti/PT-REX}}} (\texttt{Point-to-point TRend EXtractor}) \citep{2020A&A...640A..37I, 2022NewA...9201732I}, which samples radio and X-ray diffuse emission using a grid of beam-sized (10$''$) regions covering the entire extent of the diffuse source. Regions containing X-ray point sources or residual emission from tailed radio galaxies were masked. Only areas where both radio and X-ray surface brightness exceeded the 3$\sigma_{\rm rms}$ threshold were considered, while emission between 2--3$\sigma_{\rm rms}$ was treated as an upper limit during fitting. The determination of best-fitting parameters from the data set was carried out using the \texttt{MCMC}-based implementation \texttt{LinMix\footnote{For more information on LinMix see \url{https://linmix.readthedocs.io/en/latest/src/linmix.html}}} \citep{2007ApJ...665.1489K}. This method performs a Bayesian linear regression that incorporates uncertainties in both independent and dependent variables, accounts for intrinsic scatter ($\sigma_{\rm int}$), and handles upper limits on the dependent variable. The strength of the derived correlations was quantified using the \texttt{Pearson} (r$_{\rm p}$) and \texttt{Spearman} (r$_{\rm s}$) correlation coefficients. Radio brightness is reported in units of Jy~arcsec$^{-2}$, and X-ray brightness is given in Counts/s/arcsec$^{2}$.

We performed the correlation study separately for the ridge, radio halo, and, lastly, combining both, to understand whether the ridge and the halo have different physical properties or not. There is a positive correlation (Figure.~\ref{ptp_plot}) between the radio and X-ray surface brightness of the ridge, with a correlation coefficient of 0.60 and a correlation slope of $\sim$ 0.40, at both uGMRT frequencies (results are reported in Table.~\ref{ptp_results_tab}). Note that we have not reported any analysis for the ridge at 1.4 GHz, as the ridge is not distinguishable properly in the VLA low-resolution image. The southern part of the ridge lies very close to the cold front; therefore, we divided the ridge into two parts and checked the correlation for each part. We do not find the correlation slope to be stronger/weaker close to the cold front part. Such a sub-linear slope indicates that the non-thermal component of the ICM decays more slowly compared to the thermal gas, at least in the ridge regime (within 400 kpc from the cluster centre).

We have summarised the fitting results between the X-ray and radio surface brightness for the radio halo (only) in Table.~\ref{ptp_results_tab}. Figure.~\ref{ptp_plot} indicates that radio halo emissions span separate space in this plane, compared to the ridge, and a strong positive correlation has been found between radio halo emission and the ICM thermal emission, with a correlation coefficient of $\sim$ 0.80. We find a sub-linear slope of b$_{400 \rm MHz}$ = 0.67 $\pm$ 0.01, b$_{650 \rm MHz}$ = 0.78 $\pm$ 0.01, with an intrinsic scatter of 0.15 at both frequencies. This scatter likely arises as a result of both measurement uncertainties and intrinsic dispersion due to the physical properties of the halo, for example, inhomogeneous turbulence and fluctuations of the magnetic field and density of relativistic electrons. The correlation slope varies over the range of frequencies in our study. We have also checked the correlation over a grid size ranging from 10$''$ - 45$''$, however, we find that the correlation slope matches within 2$\sigma$ in every choice of the grid size. The radio halo appears to be more extended at lower frequencies; therefore, the areas chosen for the I$_{\rm R}$--I$_{\rm X}$ correlation analysis in Figure.~\ref{ptp_plot} vary with frequency. We also repeated the fitting procedure by restricting the analysis to only those regions where emission exceeds the $3\sigma$ threshold across all frequencies. This approach resulted in slope values of $b_{400\,\mathrm{MHz}} = 0.69 \pm 0.06$, $b_{650\,\mathrm{MHz}} = 0.80 \pm 0.08$. While these slopes matched well with those quoted in Table.~\ref{ptp_results_tab}, the general trend remains consistent: the slope becomes increasingly sublinear at lower frequencies.

When including the ridge, we find that for the total data points, the radio and X-ray surface brightness are correlated at 400 MHz and 650 MHz with a correlation coefficient of 0.80 and 0.82, respectively, having a correlation slope of b$_{400 \rm MHz}$ = 0.69 $\pm$ 0.01, b$_{650 \rm MHz}$ = 0.76 $\pm$ 0.03, and b$_{1.4 \rm GHz}$ = 0.84 $\pm$ 0.02. Therefore, the inclusion of the central ridge region does not significantly change the slope of the correlation (from sublinear to linear/super-linear), and we find that the correlation slope, as well as the correlation strength, varies slightly from 400 MHz to 1.4 GHz. Our slope at uGMRT frequencies is flatter and becomes steeper with an increase in frequency, and our estimate at 1.4 GHz is steeper, compared to previously reported by \citet{2001A&A...373..106F}, with b$_{1.4 \rm GHz}$ = 0.64 $\pm$ 0.05. We note that \citet{2001A&A...373..106F} used a least-squares fitting, which is less sophisticated for handling the intrinsic scatter, the error in both the x and y variables, and the upper limit of the dependent variable. Using the least-squares fit, we obtained a slope of 0.74 $\pm$ 0.06, consistent (within 1$\sigma$) with the results of \citet{2001A&A...373..106F}. Therefore, the inclusion of upper limits has affected the correlation slope, but the overall correlation remains sub-linear.

\begin{table*}
\centering
\caption{Best-fit slopes and Spearman (r$_{s}$) and Pearson (r$_{p}$) correlation coefficients of the radio and X-ray surface brightness data are summarised.}
 \begin{tabular}{@{}cccccccccc@{}}
 \hline\hline
 \multicolumn{0}{c}{}&\multicolumn{0}{c}{Frequency(MHz)}& \multicolumn{4}{c|}{3$\sigma$}& \multicolumn{4}{c|}{2$\sigma$} \\

\cline{3-10}
 & &slope (b) &$\sigma_{\rm int}$ & r$_{s}$ & r$_{p}$  & slope (b) &$\sigma_{\rm int}$ & r$_{s}$ & r$_{p}$ \\
\hline

 Halo + ridge & 400 & 0.67 $\pm$ 0.05 & 0.09 & 0.85 & 0.86 & 0.69 $\pm$ 0.01 & 0.10 & 0.84 & 0.87  \\

  &650 & 0.71 $\pm$ 0.03 & 0.10 & 0.86 & 0.86 & 0.76 $\pm$ 0.03 & 0.09 & 0.85 & 0.86 \\

  &1400 & 0.81 $\pm$ 0.02 & 0.08 & 0.81 &  0.85 & 0.84 $\pm$ 0.02 & 0.09 & 0.82 & 086  \\
  
\hline

Ridge & 400 & 0.41 $\pm$ 0.02 & 0.06 & 0.58 & 0.61 & - & -& - & - \\

& 650 & 0.39 $\pm$ 0.08 & 0.01 & 0.51 & 0.58 & - & - & - & -  \\

\hline

Halo & 400 & 0.65 $\pm$ 0.01 & 0.12 & 0.69 & 0.73 & 0.67 $\pm$ 0.01 & 0.12 & 0.68 & 0.75 \\

& 650 & 0.75 $\pm$ 0.02 & 0.13 & 0.71 & 0.76 & 0.78 $\pm$ 0.01 & 0.13 &  0.73 & 0.79  \\

\hline
\end{tabular}
\label{ptp_results_tab}
\end{table*}

\subsubsection{The radial variation of the strength}

A few studies have tried to investigate if the correlation slope of the relation varies with the radius from the cluster centre to the outskirts \citep[e.g.,][]{2022ApJ...933..218B, 2023A&A...678A.133B, 2024arXiv240218654B, trehaeven25}. The slope value indicates how the radio and X-ray components of the ICM are related. Therefore, a constant slope throughout the whole cluster extent would indicate that the radial variations of the thermal and non-thermal components are uniform. However, if we find a deviation in the slope values, it means that their relation is changing, and the value of the slope provides information on which component is increasing/decreasing with respect to the other. Here we also investigate the radial variation of the relation, both by deriving the correlation slope from the models given by \citet{2022ApJ...933..218B}, and comparing it with our data. To minimise potential biases, here we choose to study these possible slope changes as the ratio of the difference between the logarithm of the radio (X-ray) brightness in two consecutive annuli \citep[similar procedure to;][]{2023A&A...678A.133B}. We have chosen the width of each annulus to be equal to the beam width (only one case) of the image, and we have masked the ridge and residual emission of the tailed galaxies during the calculation.

\begin{figure}
	\includegraphics[width=\columnwidth]   
{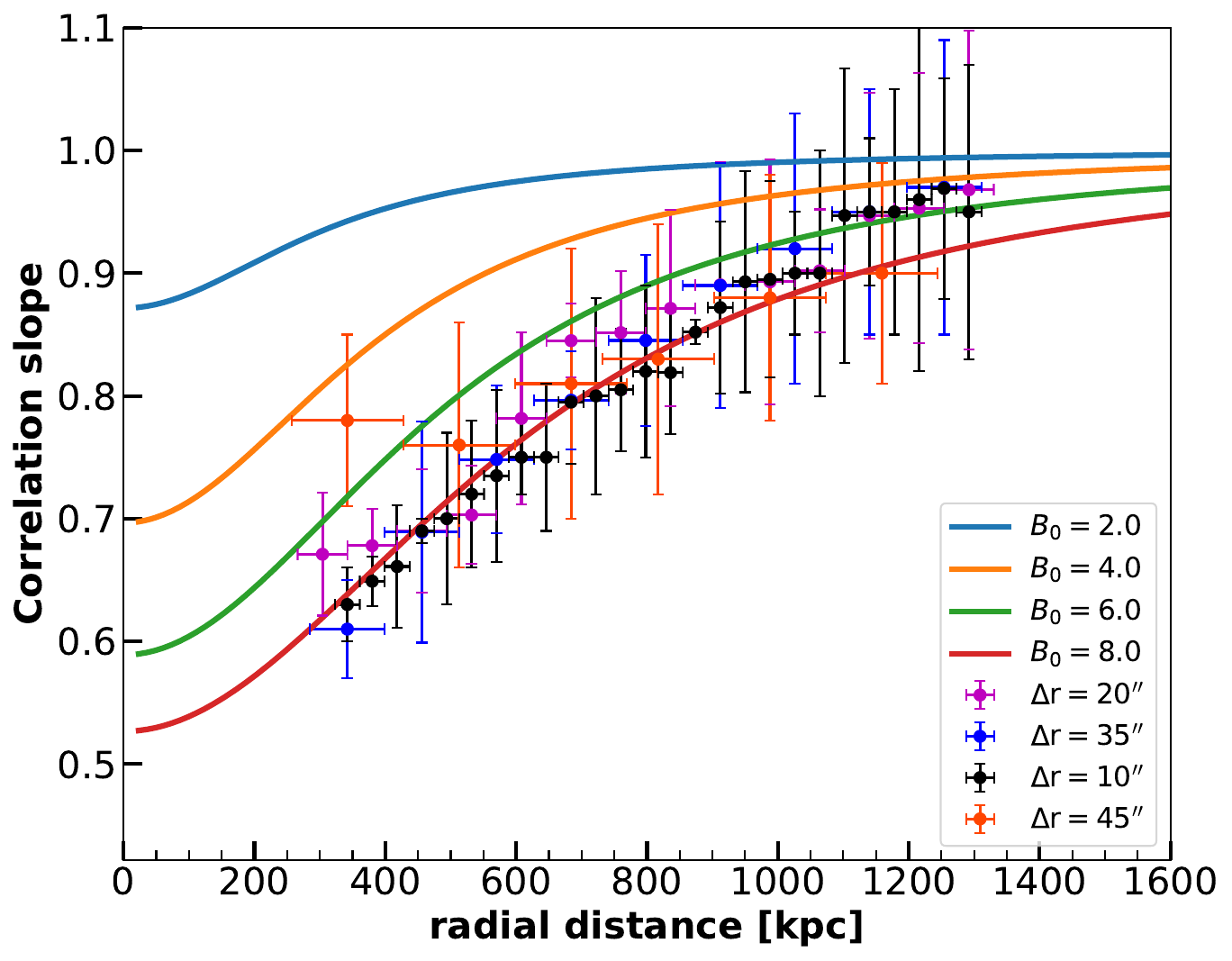}
        \caption{ Radio vs. X-ray correlation slope as a function of the radial distance is shown. The widths of the concentric sampling sectors depend on the beam of the radio image and the required SNR and are reported in the legend for four different resolutions. The solid lines are the model from the turbulent re-acceleration, adopted from \citet{2022ApJ...933..218B}, for different central magnetic fields.}       
    \label{corr_slope_model}    
\end{figure}

As shown in Figure.~\ref{corr_slope_model}, it is evident that the complex trends in the presented halo show that a uniform b is generally not representative of the whole cluster extent. The correlation slope shows a trend of being flatter close to the centre to becoming steeper along the outskirts is seen for all different beam sizes, suggesting that the radio emission steepens with the radius. In particular, b reaches $\sim$ linear values, indicating that the non-thermal component is decreasing very similarly to the thermal one. Therefore, the overall picture of A2163 is similar to what has recently been observed in Coma \citep{2022ApJ...933..218B}, other samples of merging clusters in the CHEX-MATE clusters \citep{2024arXiv240218654B}. Their slope profiles indicate a steeper trend in the outer regions than in the inner ones. We now consider the simple model (detailed description \citealt{2022ApJ...933..218B}) described in \citet{2024arXiv240218654B} for radio emissivity and X-ray emissivity, to compare with our observational results. This model (see also section.~\ref{sec:6.2}) assumes that the magnetic field strength follows B $\propto$ n$_{e}^{0.5}$ and cosmic-ray electrons are accelerated with a constant efficiency throughout the cluster volume, and the model computes the observable by integrating emissivities along the line of sight. While these assumptions provide a reasonable first-order match to observed trends, they do not account varying acceleration efficiency or radial evolution in the ratio of CR to thermal energy densities. We have selected the values of the central magnetic field similar to Coma; however, B$_{0}$ $\sim$ 6 $\mu$G shows close agreement for the radial variation of the correlation slope at the outskirts, and A2163 has a significantly higher ICM temperature, which implies a high plasma beta. Therefore, quantifying this radial variation requires further investigation. 

\subsection{Correlation between the spectral index and X-ray surface brightness}

A positive correlation between radio and X-ray emission may indicate the presence of correlation/anti-correlation between the other dependent quantities: spectral index and temperature. We also checked for any possible connection between the spectral index and the X-ray surface brightness in A2163, for the radio halo and the ridge. We extracted the surface brightness using the same \textit{XMM-Newton} image as in Section.~\ref{ptp}. We performed a linear regression between the spectral index and X-ray surface brightness to study any possible correlation/anti-correlation. A linear relation between the spectral index and X-ray surface brightness is assumed in the form of:

\begin{equation}
       \alpha = \rm a + \rm b\log (\rm I_{\rm X})
\end{equation}

\begin{figure*}
        \includegraphics[width=.60\columnwidth]{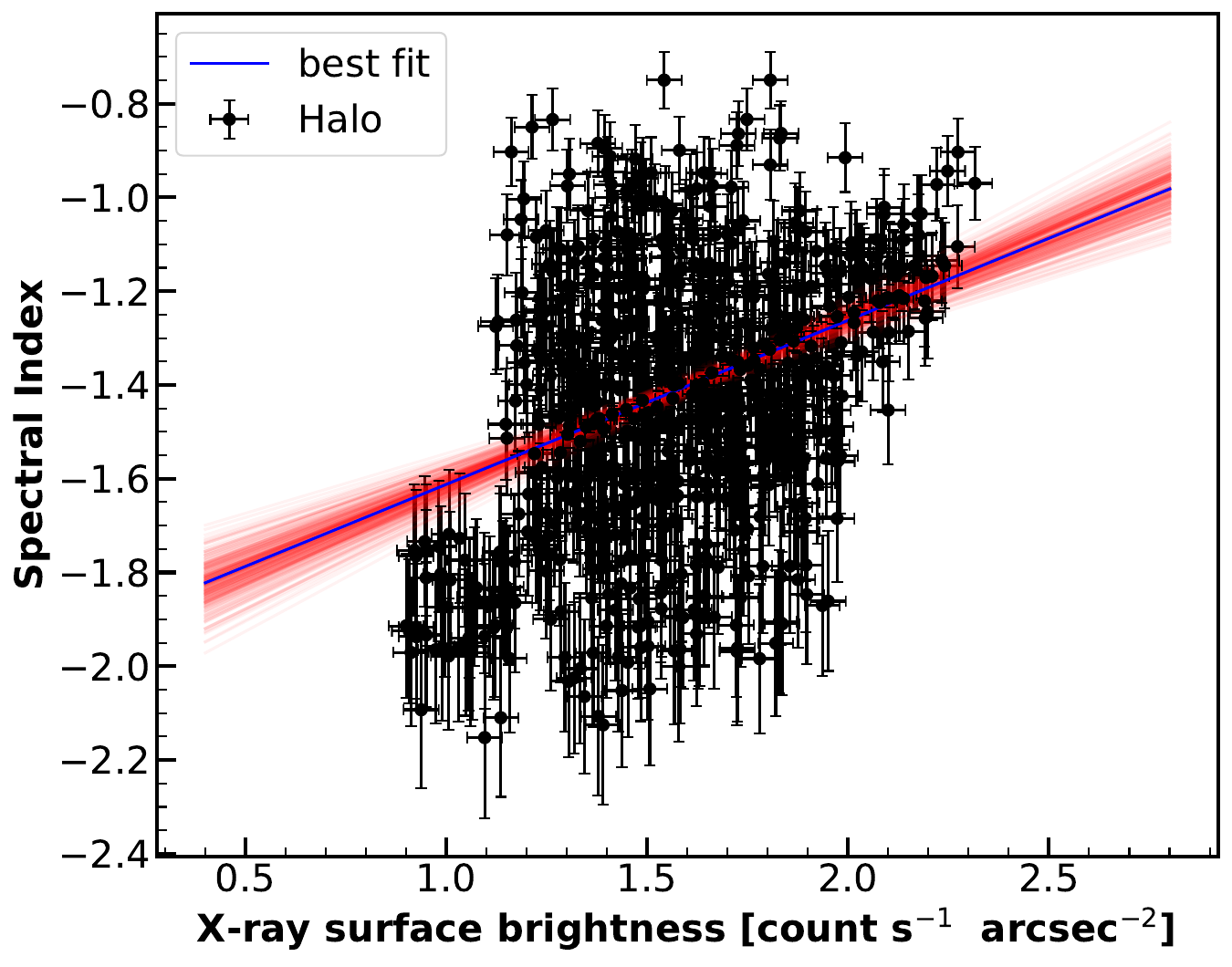}
        \includegraphics[width=.60\columnwidth]{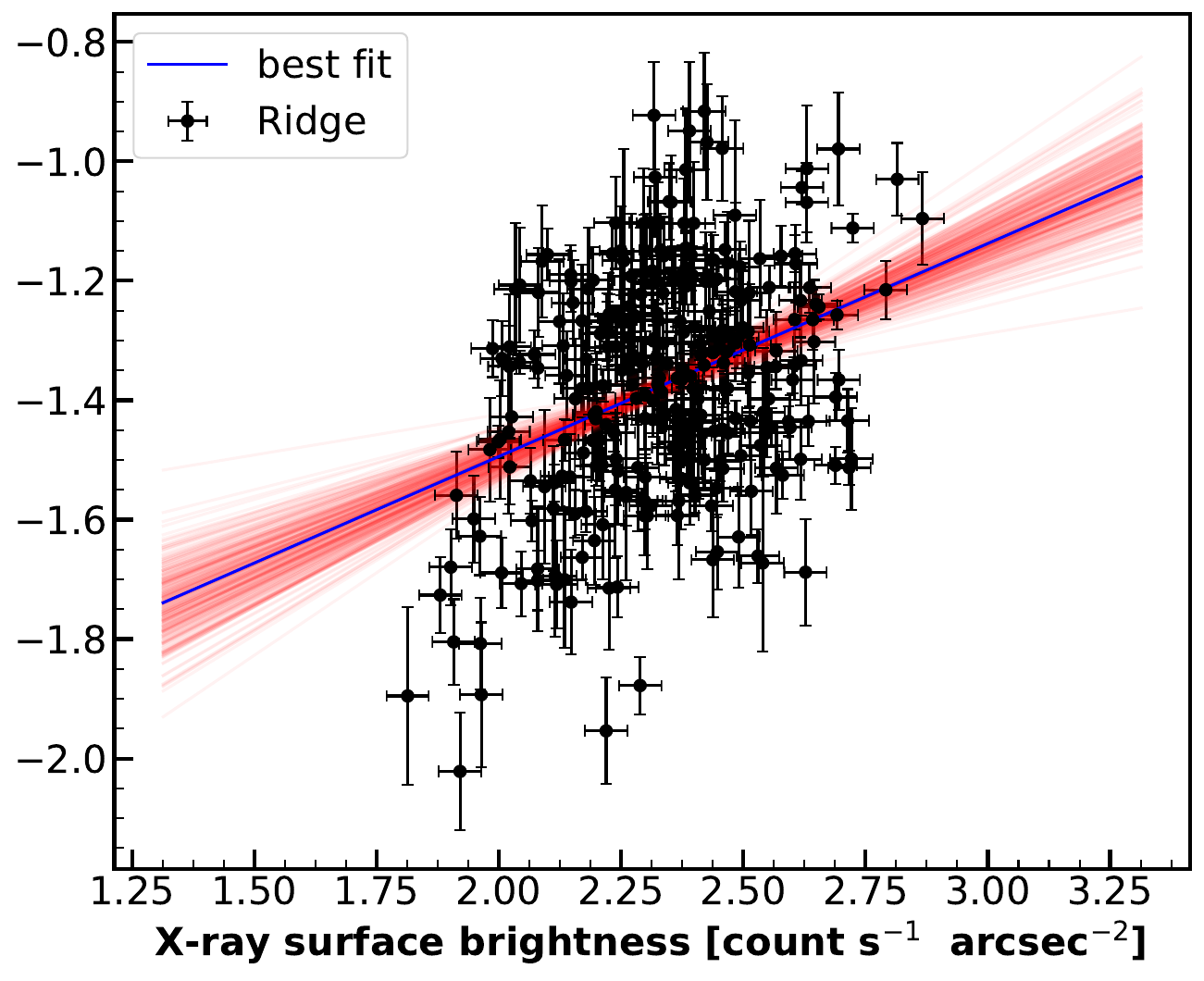}
        \includegraphics[width=.60\columnwidth]{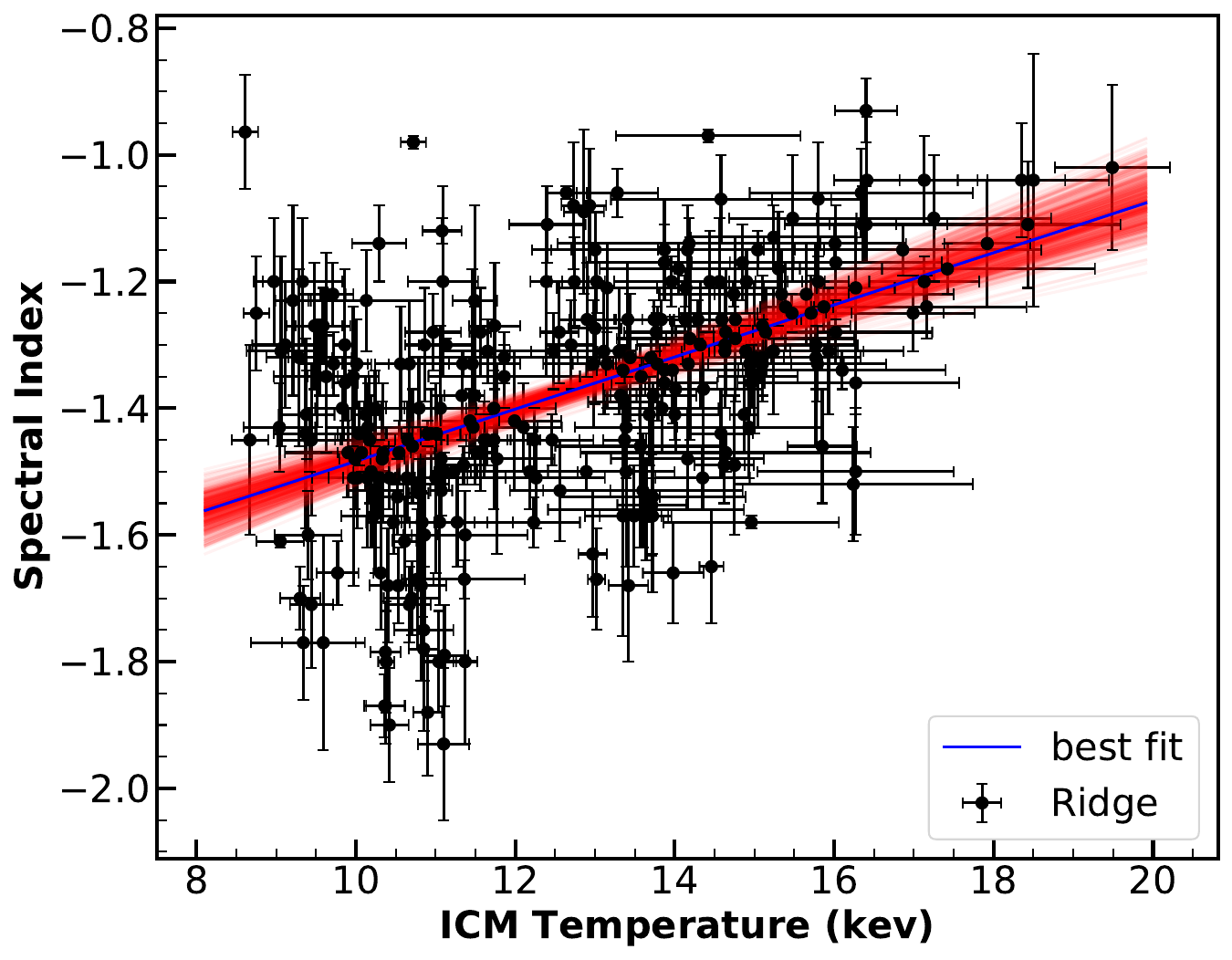}
        \caption{\textit{Left:} The X-ray surface brightness and the spectral index correlation for the A2163 radio halo are shown at 400 MHz. The red lines represent the posterior of the MCMC chain, and the blue line indicates the best fit for the regression.     
        \textit{Middle:} The same is shown here, but for the ridge separately. The correlation slopes are quoted in Table.~\ref{alpha-ix-t-corr-tab}.    
        \textit{Right:} $\alpha$$-$T relation for the radio ridge (with a correlation slope of 0.28$\pm$0.04) is shown. The temperature for the ridge is extracted from the temperature map, shown in Figure.~\ref{xray_SB_map}. }
        \label{alpha-ix-t-corr}
\end{figure*}

The large scatter of the data and the low r$_{s}$ and r$_{p}$ values indicate no correlation or anti-correlation. However, they appear to show a mild anti-correlation trend (Figure.~\ref{alpha-ix-t-corr}) both for the radio halo and the radio ridge. The significance of the correlation between these two quantities has been obtained via \texttt{Spearman} and \texttt{Pearson} correlation coefficients for the given data set (Table.~\ref{alpha-ix-t-corr-tab}). The anti-correlation trend indicates that the brighter X-ray-emitting regions correspond to flatter spectral indices. Most of the X-ray emission is produced at the centre of the cluster, in and surrounding the ridge region, where the flat-spectrum regions are sparsely distributed. 

To the best of our knowledge, the correlation between the spectral index and X-ray surface brightness has been studied for a small number of clusters: A2744 \citep[][]{2021A&A...654A..41R}, A2255 \citep{2020ApJ...897...93B} and CIG0217+70 \citep{2019A&A...622A..20H}, A2256 \citep{2022arXiv220903288R}, A2142\citep{riseley24}. In A2255, the two quantities exhibit a mildly positive correlation, whereas in A2744, both positive and negative correlations are observed, pointing to the presence of multi-component radio halos and a complex merger system. In contrast, CIG0217+70 does not show a significant correlation between the spectral index and radio surface brightness. Consequently, definitive evidence for either a correlation or an anti-correlation remains elusive. The lack of any significant correlation may indicate the different evolutionary stages of the radio halo; however, at the outskirts, where the X-ray emission is faint, the spectral index may become steep due to the imaging and deconvolution errors, possibly leading to an artificial anti-correlation.  

\begin{table}
  \centering
  \caption{Linmix best fitting slopes and Spearman and Pearson coefficient for the data plotted in $\alpha$ vs surface brightness plot }
  \begin{tabular}{@{}ccc@{}}
    \hline\hline
      &  Radio halo & Radio ridge  \\
    \hline
     Correlation slope &0.46 $\pm$ 0.10   & 0.37$\pm$ 0.09  \\
     Spearman Coff. & $-$0.37  & $-$0.43 \\
     Pearson Coff. & $-$0.35  & $-$0.39\\
   \hline
  \end{tabular}
  \label{alpha-ix-t-corr-tab}
\end{table}

\subsection{Correlation between the spectral index and Temperature}

A study by \cite{2006AN....327..565O} initially examined the relationship between the resolved spectral index and ICM temperature values for A2744, noting that regions with flatter-spectrum emission correspond to higher temperature areas. However, this finding was later challenged by \cite{2017ApJ...845...81P} using more detailed radio and X-ray observations, which found no strong correlation between these parameters. In contrast, \cite{2020ApJ...897...93B} identified a mild anti-correlation for the radio halo in A2255, while \cite{2024ApJ...962...40S} observed a weakly positive correlation for the radio halo in A521. Recently, \citet{2024arXiv240618983B} also presented the correlation between the spectral index and the thermodynamic quantities (pressure, entropy, temperature), reporting no correlation between them. For our analysis, we employed \texttt{LinMix} to explore potential correlations or anti-correlations specifically for the ridge. The Spearman correlation coefficient of 0.43 suggests a very weak positive correlation between the spectral index and the ICM temperature. These findings are in line with the notion that a fraction of the gravitational energy released during cluster mergers contributes to heating the thermal plasma and accelerating relativistic particles in the ICM. It is important to note that the thermal and non-thermal components of the ICM may evolve on different timescales. While shock heating of the thermal gas locally occurs on short timescales of the order of the shock crossing times, the acceleration of cosmic rays through turbulent processes can take significantly longer ($\sim$10$^{7} - 10^{8}$ yr). Moreover, these processes are embedded within the broader dynamical evolution of the cluster, which happens over $\sim$10$^{9}$ yr.

We have observed a hint of positive correlation in the ridge of A2163, as shown in Figure.~\ref{alpha-ix-t-corr}. High-temperature regions are expected to correspond with areas where turbulence has heated the gas, suggesting that the dynamics of the gas and the turbulent energy transferred to the particles should align with regions of flatter spectral index. Given that the ridge is located near the merger's epicentre, it serves as an ideal setting to test whether flatter spectral regions at the merging site are associated with higher-temperature zones. However, there can also be efficient damping of turbulent modes in the thermal plasma, reducing the energy available for particle acceleration \citep[e.g.,][]{2007MNRAS.378..245B}. This introduces a competition between increased turbulence generation and its dissipation, making the correlation a non-trivial one. \cite{2010ApJ...718..939K} also examined the validity of any correlation or anti-correlation between the spectral index and ICM temperature, and explained the lack of correlation with differing cooling times for thermal gas and non-thermal plasma. It is important to note that, while the study of this correlation across a significant number of radio halos is still limited, the microphysics of the ICM, such as how the collisional mean free path depends on temperature, also plays a crucial role in any observed correlation or anti-correlation. Here, a positive correlation between the spectral index and temperature is obtained with a statistically significant number of data points to exclude any systematic biases; however, more sophisticated numerical simulations and detailed plasma physics modelling are required to explore the existence of such a correlation theoretically.

\section{Discussion} \label{discussion}

\subsection{Source of the seed electrons} \label{seed-electron-source}

\citet{2023A&A...669A..50V} conducted a comprehensive simulation of cosmic rays ejected by radio galaxies, accounting for re-acceleration and adiabatic losses. Their findings indicate that relativistic electrons from radio galaxies efficiently disperse throughout the intracluster medium, driven by cluster-wide turbulence and the momentum from powerful jets, even with just a single jet injection event in the radio galaxy’s lifespan. Shortly after the injection from the jet, the interactions of the CRs with the magnetic field are significantly reduced (beyond 500 kpc). At the same time, cosmic-ray electrons introduced into the intracluster medium by jets gradually contribute to a reservoir of supra-thermal particles, which can later be re-accelerated by turbulence. \citet{2020ApJ...897..115S} estimated that the activity of about 30 radio galaxies over a cluster's lifetime would likely be sufficient to supply the intracluster medium with the necessary seed relativistic electrons and possibly magnetic fields for A2163. Nonetheless, the actual number of observed radio galaxies or diffuse lobes in the peripheral or central regions is considerably lower.

In our uGMRT images, the western diffuse lobes (D1) show a steep spectrum tail connected to the comparatively flatter spectrum radio halo region (Figure~\ref{spec_map}). The tailed galaxy T1 has a steep spectrum tail that connects with the radio halo emission, indicating a contribution of seed electrons to the radio halo volume. Using equation 3 from \citet{2024Galax..12...19V} and assuming no secondary electron production from hadronic processes, we estimate about 7 $\times$ 10$^{66}$ synchrotron electrons, considering $\gamma_{\rm min} = 1000$. For an A2163-type cluster (mass $\sim$ 10$^{15}$ M$_{\odot}$) with a radio halo spectral index of $-$1.3 and an extrapolated radio power of $\sim$ 5 $\times$ 10$^{25}$ W~Hz$^{-1}$ at 150 MHz, a single episode of activity from all radio galaxies would eject approximately 10$^{67}$ electrons. Thus, the activity of radio galaxies (T1, T2, T3) can account for the electron numbers needed to power diffuse radio sources. However, to fully understand the ICM dynamics and validate this as a quantitative explanation, further investigation is needed. \citet{2024Galax..12...19V} also analysed the area-filling factor of detectable radio emission versus frequency and radius from the cluster centre. Despite a large electron filling factor in the ICM, the detectable emission filling factor is smaller due to the fraction of high-energy electrons and the observing frequency. They found that the observable emission area-filling factor declines rapidly, especially at higher frequencies. About 1.2 Gyr after injection, electrons from all galaxies spread over large radii. However, cosmic rays dilute as they expand, and about 0.25 Gyr post-injection, less than 30\% of the projected area at 0.5 R$_{200}$ is covered by cosmic rays along the line of sight. The radio halo in A2163 extends to 1.6 Mpc from the centre, requiring CR electrons to travel significant distances. Thus, tailed galaxies and diffuse lobes in the cluster periphery could fill the entire cluster volume within the fading timescale. Additionally, cosmic rays from galactic winds or structure formation shocks, as seen by \citet{2018ApJ...856..112F}, could contribute.

\subsection{Thermal and non-thermal correlation}\label{sec:6.2}

Despite the extended morphology of the radio halo in A2163, the radio and X-ray surface brightnesses exhibit a positive correlation across all frequencies of our study. Excluding the ridge region, the correlation slope remains sublinear, approximately 0.8, suggesting that the radial decrease in non-thermal plasma density is slower compared to that of the thermal gas. Our analysis indicates an evolution in the correlation slope between 400 and 650 MHz. Variations in the correlation slope over frequencies are not common in radio halos, with notable exceptions such as CIG0217+70 \citep{2019A&A...622A..20H} and MACSJ0717+3745 \citep{2021A&A...646A.135R}. The estimated e-folding radius from the radio surface brightness profile (Table.~\ref{radio_SB_fit_param}) reveals a slight frequency dependence, indicating that the radio brightness at different frequencies declines at different rates. Consequently, this results in a frequency-dependent change in the correlation slope, and radial spectral steepening is anticipated, which is consistent with our spectral analysis presented in Section.~\ref{resv-spec-study}.

Under some assumptions on the magnetic field evolution, we can investigate whether the observed change in the correlation slope can be explained. Assuming that the magnetic field profile scales with thermal gas as B(r) $\propto n_{e}^{0.5} (r)$, the synchrotron emissivity scales as: 

\begin{equation}\label{eq:sync_emsvt}
    \epsilon_{\rm R} \propto \eta_{\rm acc} F \frac{\rm B^{2}}{\rm B^{2} + \rm B_{\rm IC}^{2}}
\end{equation}

where $\eta_{\rm acc}$ is the acceleration efficiency and $F$ is the turbulent energy flux in a unit volume, with: 

\begin{equation}\label{eq:turb_en_flux}
    F = \frac{1}{2} \rho \frac{\sigma_{v}^{3}}{L}
\end{equation}

Here $\sigma_{v}$ is the velocity dispersion on scale L, and $\rho$ is gas density. If we assume a constant temperature, Mach number, and  acceleration efficiency over the cluster volume, equation.~\ref{eq:sync_emsvt} can be expressed as:

\begin{equation}\label{therm-nontherm-eq}
    \epsilon_{\rm R} (r) \propto \rm X(r) \frac{\rm M_{\rm t}^{\alpha} (r)}{L(r)} \epsilon_{\rm X} (r) ^{\frac{1}{2}} \frac{1}{1+\left(\frac{\rm B_{\rm IC}}{\rm B(0)}\right)^{2} \left(\frac{\epsilon_{\rm X} (0)}{\epsilon_{\rm X} (r)}\right)^{\frac{1}{2}}}
\end{equation}

where $\epsilon_{\rm R}, \epsilon_{\rm X}$ are the radio and X-ray emissivities, B$_{\rm IC}$ is the equivalent CMB magnetic field, and X(r) = $\frac{\rm U_{\rm CRe}}{\rm U_{\rm th}}$ is the ratio of the energy densities of the cosmic rays to the thermal gas. M$_{\rm t}$ is the turbulent Mach number and $\alpha = 4$ for models based on transit-time-damping \citep{2007MNRAS.378..245B} and $\alpha = 3$ for models based on acceleration by incompressible turbulence \citep{2016MNRAS.458.2584B}. Therefore, different re-acceleration models would lead to sublinear slopes depending on the ICM conditions. Assuming X(r) and M$_{\rm t}$ are constant with radius, equation ~\ref{therm-nontherm-eq} would lead to a sub-linear ($\epsilon_{\rm R} \propto \epsilon_{\rm X}^{0.5}$) and linear correlations for B(0)$^{2}$/B$_{\rm IC}^{2}$ $>>$ 1 and  B(0)$^{2}$/B$_{\rm IC}^{2}$ $<<$ 1, respectively.

A steepening of the I$_{\rm R}$ - I$_{\rm X}$ correlation outside the core is expected as a consequence of the different relative weights of inverse Compton and synchrotron losses in a magnetic field declining with radius. We find that the correlation slope to be changed from $\sim$ 0.6 to $\sim$ 0.95, in line with this model prediction. However, none of the theoretical models which assume the constant turbulent Mach number over the cluster volume can reproduce the observed data (Figure.~\ref{corr_slope_model}), indicating a more complex situation with respect to that described by equation.~\ref{therm-nontherm-eq}. One problem is the departure from spherical symmetry; however, our findings may also suggest that the simplistic assumptions on the magnetic field profile and the variation of the X(r) need to be scrutinised. If we keep a magnetic field scaling B $\propto$ n$_{e}^{1/2}$, the likely possible scenario to explain the observed trend is to assume that the ratio of the CRe to thermal gas energy densities, X(r), decreases with radial distance from the cluster centre. Equation.~\ref{therm-nontherm-eq} suggests that a decrease of M$_{t}$ with radius would further increase the expected steepening of the radio to X-ray scaling, which aligns with our obtained results. Therefore, either the monotonic decrease of X(r) or M$_{t}$ can contribute to the change in the correlation slope. To date, the reported giant radio halos (highly dynamically active clusters) have shown steepening of the correlation slope in the outskirts \citep[e.g.,][]{2022ApJ...933..218B}; however, the ultra-steep spectrum radio halos show an opposite trend, flattening in the outskirts \citep[e.g.,][]{2022arXiv220903288R, 2024ApJ...962...40S}, very similar to radio minihalos \citep[e.g.,][]{2024arXiv240218654B}.

Departures from the assumed magnetic field profile will lead to increased complexity and variation in the decay slope of thermal to non-thermal emission. Theoretical \citep[e.g.,][]{1999A&A...348..351D, 2018SSRv..214..122D} and observational \citep[e.g.,][]{2004A&A...424..429M} evidence indicates that magnetic fields within galaxy clusters must decrease with increasing radial distance from the cluster centre, scaling as a function of the electron density. We assume that the magnetic field in galaxy clusters hosting radio halos follows the radial profile:

\begin{equation} \label{B-rad-dist}
    B(r) = B_{0} \left(\frac{n_{e}}{n_{0}}\right)^{\eta}
\end{equation}

where $B_{0}$ and $n_{0}$ are the central magnetic field and the central thermal electron density, respectively, $n_{e}(r)$ is the electron density at a given radius r from the cluster center, and $\eta$ is the scaling index of the magnetic field, expected to be in the range 0.3–1 \citep[references therein;]{2009A&A...499..679M}. Even assuming a simple scaling in the form equation.~\ref{B-rad-dist} for $\eta$ \textgreater 0.5, the steepening with radius, at distances where B \textless B$_{\rm IC}$, is expected to be more pronounced than that predicted by equation.~\ref{therm-nontherm-eq}, possibly in line with our findings. Furthermore, in this case, the Alfvén velocity is also not constant with radius, implying a change in acceleration efficiency in the case of some particular models \citep[e.g.,][]{2016MNRAS.458.2584B}.

The value of the scaling index can also be constrained from the rotation measure (RM) analysis \citep[e.g.,][]{2010A&A...522A.105G, 2022MNRAS.517.5232R,osinga25}, requiring polarisation observations, which will be a part of separate analysis. In conclusion, although we find a steepening of the ratio between radio and X-ray brightness with distance in line with model expectations in the case of B declining with radius, the data suggest more complex situations with respect to the case of constant acceleration efficiency and Alfvén speed.

\subsection{Turbulent re-acceleration and large-scale emission}

In a non-homogeneous turbulent re-acceleration scenario, the absence of a distinct spectral curvature in the integrated spectrum can be reconciled with significant hints of spectral fluctuations and radial spectral steepening in the emitting volume \citep[e.g.,][]{2021A&A...654A..41R}. Assuming homogeneous conditions, the frequency at which steepening occurs:

\begin{equation}
    \nu_{s} \propto \tau_{\rm acc}^{-2} \frac{\rm B}{(\rm B^{2} + \rm B_{\rm \rm IC}^{2})^{2}},
\end{equation}

where B is the magnetic field and $\tau_{\rm acc}$ is the re-acceleration time \citep{2007MNRAS.378..245B}. 
If the acceleration timescale $\tau_{\rm acc}$ is assumed to be constant (i.e., independent of B), the critical magnetic field at which the lifetime of cosmic-ray electrons is maximized for a given observing frequency is given by $B_{\rm cr} = B_{\rm IC} / \sqrt{3} \approx 2.73\ \mu$G. At this field strength, synchrotron and inverse Compton (IC) losses are balanced. In a scenario where the magnetic field decreases with radius, one expects the radio spectral index to be flatter at radii where $B(r) \approx B_{\rm cr}$ and to steepen beyond the radius $r_*$ where $B(r_*) = B_{\rm cr}$. Assuming the radial profile of the magnetic field follows Equation~\ref{B-rad-dist} (and B$_{\rm r = 0} = 6\mu$G), we estimate $r_*$ to be approximately 430 kpc, which is smaller than the observed steepening radius in Figure~\ref{rad-prof-halo}. This discrepancy suggests that the acceleration timescale $\tau_{\rm acc}$ may vary across the halo, and is similar to the conclusions in the previous section. Homogeneous turbulent re-acceleration models predict the spectral steepening to occur at frequencies ($\nu_{s}$) larger than a few times the critical frequency ($\nu_{c}$) of the high energy electrons \citep{2006AN....327..557C}, $\nu_{s}$ $\sim$ $\xi$$\nu_{c}$, where $\xi$ was estimated to be 6-8 by \citet{2012A&A...548A.100C}. However, if the magnetic field intensity and acceleration time scale vary in the emitting volume (inhomogeneous case) and along the line of sight, the spectrum may be stretched in frequency, and the curvature becomes less evident \citep{2021A&A...646A.135R, 2013MNRAS.429.3564D}. The similar inhomogeneous behaviour of the magnetic field and acceleration time may lead to the observed spectral fluctuations in the radio halo regions. \citet{2022ApJ...934..182N} reported evidence of the change in the re-acceleration coefficient D$_{pp}$ by constraining the radial surface brightness profile of the coma cluster, under the TTD regime. In the same work, the radial declination of $\tau_{\rm acc}$ up to $\sim$1 Mpc can also reproduce the observed spectral behaviour of Coma, indicating that the assumption of a homogeneous emitting medium may not be true in all cases.

\citet{2022SciA....8.7623B} reported that, in the outskirts (1--2 Mpc) of Abell 2255, the electrons are not directly accelerated from the thermal pool, but are likely re-accelerated from a pre-existing population of mildly relativistic particles. Simulations by \citet{2024ApJ...961...15N} have also shown that incompressible turbulence at large radii can provide sufficient energy for re-acceleration of these particles. Assuming that a fraction $\eta_{B}$ of the turbulent kinetic energy is converted into magnetic field amplification, the magnetic field strength can be estimated as:
\begin{equation}
    \frac{B^{2}}{8\pi} \sim \eta_{B} F_{\rm turb} t_{\rm eddy}
\end{equation}
where $t_{\rm eddy} = L / \sigma_{v}$ is the eddy turnover time, L is the turbulent injection scale, with $\eta_{B} \sim 0.03$--$0.05$ based on MHD simulations \citep{2016ApJ...817..127B, 2018MNRAS.474.1672V}. The particle acceleration efficiency can be expressed as \citep{2022SciA....8.7623B}:
\begin{equation}
    \eta_{\rm acc} \sim \frac{f\,L(\nu_{400})\,\nu_{400}}{F\,V} \left[1 + \frac{B_{\rm IC}^{2}}{8 \pi F (L/\sigma_{v}) \eta_{B}}\right]
\end{equation}
We adopted $f = 5$, $L = 500$ kpc, $\sigma_{v} = 550$ km\,s$^{-1}$, and a volume of $V = 6.15$ Mpc$^3$ (assuming the halo volume uniformly filled), based on typical merger-scale turbulence in clusters with $M_{500} \sim 10^{15}\ M_\odot$. A turbulent energy flux of $F \sim 2 \times 10^{44}$ erg~s$^{-1}$~Mpc$^{-3}$ is estimated using the equation.~\ref{eq:turb_en_flux}, with the density and $\beta$ parameter taken from \citet{1997ApJ...482..648S}. Under these conditions, we estimate an acceleration efficiency of $ \sim 0.0012$ (for $\eta_{B} = 0.03$), suggesting that about 0.12\% of the available turbulent energy is used for particle re-acceleration, one order of magnitude lower than the estimates for the external regions of A2255.

\citet{2024ApJ...961...15N} showed that large-scale emission extending to cluster outskirts can be explained if $\eta_{B} \sim 0.05$ of the turbulent flux drives dynamo action, for turbulent Mach numbers of 0.4 $-$ 0.7. In comparison, \citet{2017ApJ...843L..29E} estimated a turbulent Mach number of 0.32 $\pm$ 0.01 in A2163 via gas density fluctuations, although the radial variation in correlation slope suggests that Mach numbers may vary with radius. For M$_{s}$ $\sim$ 0.3 $-$ 0.4, $\eta_{B}$ can range from 0.05 $-$ 0.02, supporting the re-acceleration process.

\subsection{A megahalo in A2163?}

It is worth examining whether the confirmed non-detection of a megahalo in A2163 is due to observational limitations such as calibration issues, insufficient \textit{uv}-coverage at short spacings, limited noise sensitivity, imperfect subtraction of discrete sources, or the chosen sector geometry for the analysis. The megahalos reported by \citet{2022Natur.609..911C} have been observed in massive clusters (M$_{500}$ $\sim$6–11 $\times$ 10$^{14}$ M${\odot}$) at intermediate to low redshifts ($z \sim 0.17$–0.28). Given that A2163 is both very massive and relatively nearby, we investigate the potential presence of a megahalo in this system. According to \citet{2022Natur.609..911C} (see their Fig. 3), A2163 lies well within the region of the mass–redshift parameter space where megahalos are predicted to exist, even after accounting for redshift dimming effects.

\begin{figure}
	\includegraphics[width=\columnwidth]{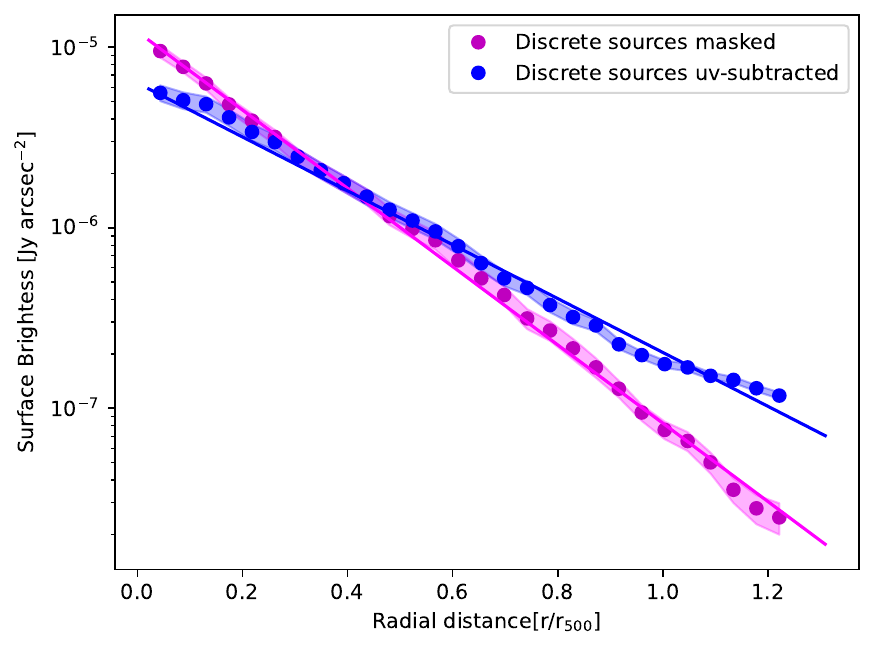}
        \caption{The radial profile of the radio surface brightness for the halo emission is shown at 400 MHz from two images, one with manual masking of discrete and unrelated sources (magenta), and the other with their subtraction (blue) in the \textit{uv}-domain. The solid lines show the exponential model (equation.~\ref{halo_model_eq}) fitted to the radial profile for the total halo emission.}        
    \label{A2163_halo_SB_comp}    
\end{figure}

The radial surface brightness profile does not exhibit the shallower outer trend, which characterises the megahalo profiles. Instead, the observed emission is well described by a single-component exponential fit. As noted in simulations by \citet{2017ExA....44..165D}, faint and extended radio emission can be partially lost if not properly modelled during calibration. This is further complicated by the fact that faint, large-scale emission in nearby clusters, on angular scales even larger than those in A2163, has proven difficult to recover using uGMRT. As discussed in Section~\ref{int_spec}, at the redshift of A2163, we are in principle sensitive to emission on scales up to $\sim$7 Mpc. Nonetheless, our uGMRT images (from both band 3 and band 4) show negative residuals, particularly in the western outskirts, suggesting a lack of visibility sampling in the \textit{uv}-plane, which likely limits our ability to detect very extended emission. However, this instrumental limitation does not undermine the main conclusions of our analysis, as the azimuthally-averaged radial profiles extend out to 1.2r$_{500}$. \citet{rajpurohit25} recently highlighted that systematic effects, such as selecting the sector centre and masking or subtracting discrete sources, can impact the detection or absence of a second component in radio halo profiles. We compared surface brightness profiles derived from source subtraction in both the \textit{uv} and image domains (by manually masking them). In Figure.~\ref{A2163_halo_SB_comp}, we have shown the comparison for the two processes, and the surface brightness profile corresponding to \textit{uv} subtraction shows a mild trend of flattening at larger radii. The bright halo emission dominates in the central regions, making the subtraction method less consequential. However, at larger radii, where the diffuse emission weakens, residual contributions from unmasked discrete sources become more prominent relative to the halo emission, and can mimic a second component. The sector geometry and centre selection on the surface brightness profile are two other factors that can influence the halo profile, resulting in flatter components \citep[e.g.,][]{rajpurohit25}.

\begin{figure*}
    \centering
    \includegraphics[width=17cm, height = 8cm]{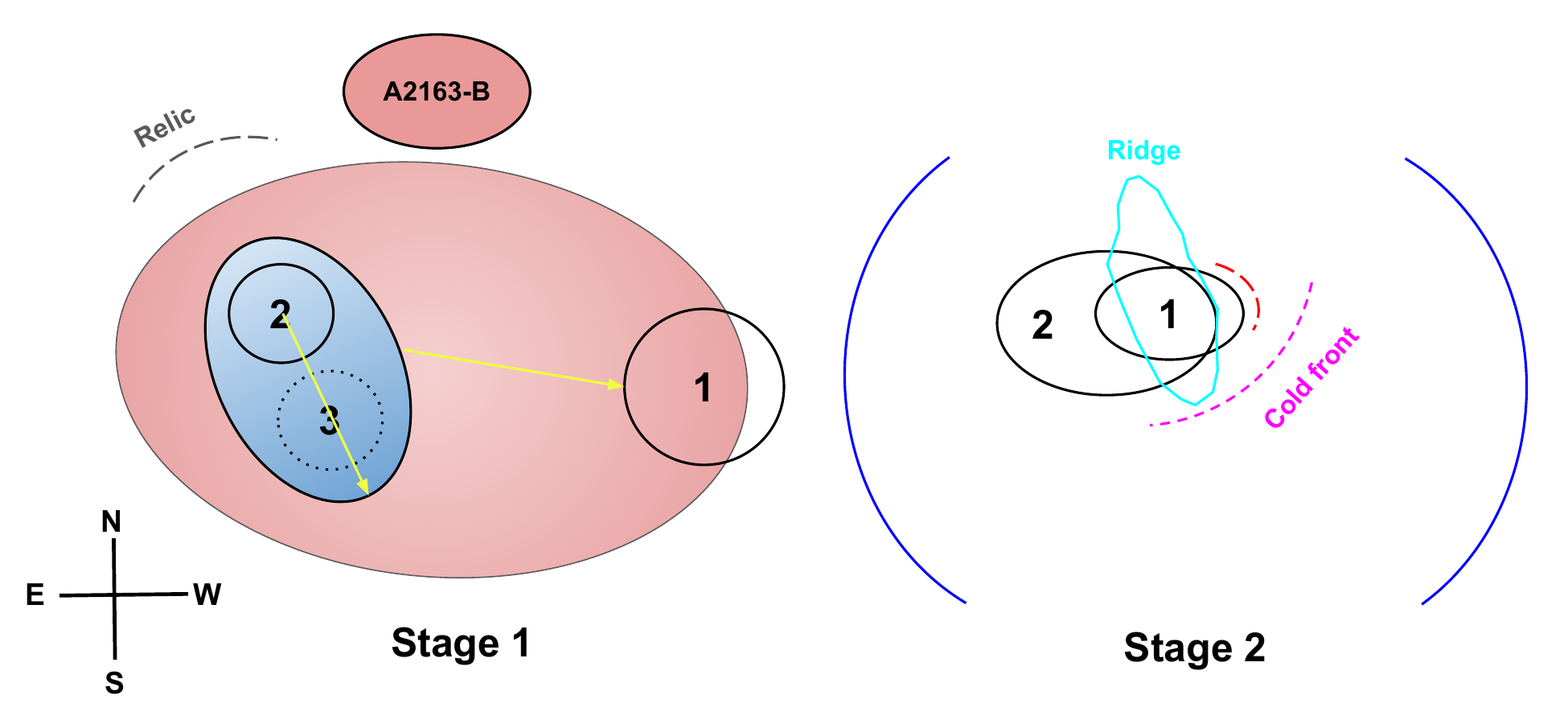}
    \caption{The schematic diagram for the merging scenario of cluster A2163 is shown. The northern subgroup is labelled as A2163-B, and in Stage 1, a merger between subgroups 2 and 3 is proposed; as a result, the relic (black dashed arc) is formed at the NE outskirts. The Dashed black circles correspond to a galaxy substructure not identified by the optical analysis. Circle 1 indicates the western subgroup. Stage 2 shows the major merger between subclusters 2 and 1 along the E-W direction. The red dashed line indicates the stripped cool gas detected by the \citet{2011A&A...527A..21B}, and the magenta dashed arc indicates the cold front close to the centre. The cyan region denotes the position of the ridge, and two blue arcs indicate the two outgoing shock waves created due to the merger.}
    \label{a2163_merger}
\end{figure*}

\subsection{The merger history}

A2163 exhibits considerable dynamical activity in its central region, as detailed in the X-ray analysis (Section.~\ref{xray-analysis}). \citet{2019ApJ...882...69G} identified three distinct structures within A2163: two are situated in the southern subcluster, while one is located in the northern, A2163-B. Their velocity dispersion estimates suggest that the southeastern brightest cluster galaxy (BCG) is part of the dominant subcluster, whereas the southwestern BCG belongs to a subordinate subcluster. Recent weak lensing analysis supports this, revealing that the peak mass distribution of the southern cluster is centred on the BCG, extending westward with a subordinate peak near the western BCG \citep{2024arXiv240702557F}. The estimated masses for the eastern and western subclusters are (9.2 $\pm$ 1.8) $\times$ 10$^{14}$ M$_{\odot}$ and (0.8 $\pm$ 0.6) $\times$ 10$^{14}$ M$_{\odot}$, respectively, consistent with the findings of \citet{2012A&A...540A..61S}. \citet{2011A&A...527A..21B} demonstrated that the northern subgroup is distinct from the main subcluster, as no evidence of interaction between them was found using the two-body analysis formalism.

Figure.~\ref{a2163_merger} presents a schematic diagram of this merger. \citet{2019ApJ...882...69G} previously suggested that the relic at the NE outskirts could be the result of a merger along the NE-SW direction (between subclusters 2 and 3). Similarly, \citet{2008A&A...481..593M} noted an extension in the galaxy distribution along this axis, with a bright maximum in the NE and a secondary maximum in the SW, which aligns with observations from \citet{2024arXiv240702557F}. It is most likely that a collision has already occurred within this subcluster, as also proposed by \citet{1995A&A...293..337E}. The high average temperature of the subcluster 2 supports this hypothesis, indicating that subcluster 2 is in a post-merger stage, with the merger occurring along the NE-SW axis (yellow arrow). Subsequently, a binary merger between the eastern (combination of 2 \& 3) and western subclusters (1) along the E-W direction generated the shock waves (blue arcs) reported by \citet{2018A&A...619A..68T}. The presence of a stripped cool core (red dotted line) near the cold front implies that A2163 may have recently accreted a subcluster along its East-West elongation \citep{2011A&A...527A..21B}. This stripping indicates that we are observing dense gas being stripped by less dense surrounding gas. The eastern subcluster, being denser than the western, results in hotter, shocked gas at the outer edges of the two colliding subclusters, as depicted in Figure.~\ref{a2163_merger}. In this scenario, the outer subcluster gas is halted by the collision shock, while the dense cores continue moving through the shocked gas. However, shock fronts (blue solid lines) in the central region have reached the cluster outskirts without penetrating the dense cores, which are moving through the shocked gas. These cores have likely passed their point of closest approach and are now separating. Thus, A2163 is considered a post-merger cluster where the dense cores have already moved past each other. Depending on the core velocities relative to the surrounding (previously shocked) gas, they might create additional bow shocks at some distance ahead. The observed 'ridge' (cyan colour) near the stripped cool core could be a shock created at a distance along the line of sight. 

\citet{2022ApJ...934..182N} showed that the fraction of merger kinetic energy converted into turbulence ($\eta_t$) depends on the mass ratio ($\xi_m$) of the merging subclusters. For a typical cluster mass of $M_{500} = 10^{15}\ M_\odot$ (comparable to A2163), their simulations suggest that mergers with $\xi_m = 0.2$ yield $\eta_t \sim 0.15$--$0.40$ depending on the acceleration timescale $t_{\rm acc}$. They also showed that in the TTD model, the diffusion coefficient scales as $D_{pp} \propto \xi_m^{-2/3}(1+\xi_m)^{-1} M^{1/3} $, implying that lower-mass-ratio mergers require higher $\eta_t$ to achieve similar re-acceleration. A2163, with $\xi_m \sim 0.087$, falls within this intermediate regime. However, these estimates rely on idealised assumptions: a binary, head-on merger (zero impact parameter), fully collisionless turbulence–particle coupling \citep{2011MNRAS.412..817B}, and uniform energy injection in a fixed volume. A2163, by contrast, exhibits morphological complexity and spectral steepening, indicating spatially variable turbulence and anisotropic dissipation. Thus, while the formalism offers a useful first-order framework, detailed modelling is required to robustly constrain $\eta_t$ in such a system.

\section{Summary} \label{sum}

We present the first detailed multi-frequency (300 $-$ 1400 MHz) analysis of the galaxy cluster A2163. These sensitive observations enable us to derive detailed spectral characteristics of the radio halo and the large-scale emission. Previous studies of this radio halo at frequencies below 1 GHz were limited by the poor resolution and insufficient \textit{uv}-coverage of the interferometer, which prevented the detection of the full extent of the halo. Our deep observations, combined with available X-ray observations (XMM-\textit{Newton}), provide profound physical insights into the thermal and non-thermal connections in the ICM, both in the central region and the outskirts. The overall findings are summarised as follows: 

\begin{enumerate}

\item We present the first deep and high to low resolution (5$''$ - 45 $''$) radio images of the galaxy cluster A2163. The halo emission is more extended than previously reported, $\sim$3.3 Mpc (up to r$_{500}$), at 400 MHz, and 3.0 Mpc at 650 MHz, respectively, along the east-west direction. We have also recovered the `ridge' emission with an LLS of 1 Mpc in length and $\sim$4 times brighter than the surrounding radio halo. The relic emission is observed to be extended up to $\sim$750 kpc (at 400 MHz) in the NE outskirts of the cluster. The tailed galaxies and three diffuse lobes have also been recovered well.  

\item The high significance detection of radio emission extending up to $\sim$ 1.6 Mpc has allowed us to investigate the potential presence of a second component in the radio halo. However, the surface brightness analysis up to a radius of r$_{500}$ reveals the non-existence of any second component, and the profile is well modelled with a single-component exponential fit. Additionally, a frequency-dependent change in the e-folding radius has been estimated, with a larger e-folding radius at low frequencies, highlighting variations in the decline of radio surface brightness across different frequencies. The non-detection or detection of any second component is highly sensitive to the masking/\textit{uv} domain subtraction of the discrete sources and other systematics.  

\item The high SNR of the halo emission at high resolution provides an excellent opportunity to identify surface brightness edges in the radio halo, akin to the discontinuities observed in the X-ray surface brightness profile. We have detected a discontinuity in the SW direction, approximately 3$'$ from the cluster centre, coinciding with the location of the X-ray detected shock. Additionally, the radio halo emission appears to be confined at the SW outskirts, marked by a sharp edge visible in the low-resolution image (at 6.5$'$ away from the cluster centre); however, the presence of -4$\sigma_{\rm rms}$ of negative regions can artificially originate this edge.

\item We measure a spectrum of the halo between uGMRT frequencies $\sim 1.36 \pm 0.08$ that is in line with a power-law $\sim 1.19 \pm 0.09$ obtained by combining measurements from the literature (153 -- 1400) MHz. The relic emission also follows a single power law, with a spectral index of $-$1.02 $\pm$ 0.03. The 1.4 GHz radio relic power aligns well with the P$_{1.4 \rm GHz}$ vs. LLS correlation for radio relics.

\item The spatially-resolved spectral index map suggests fluctuation over the extent of the radio halo, and the fluctuations average out with increasing beam size of the image. The ridge shows a uniform spectral index distribution, with steeper patches in the southern region. We have detected the radial spectral steepening from the cluster centre to the outskirts. Such behaviour indicates the inhomogeneous conditions (variation in B, acceleration time) in the emitting volume. 

\item The X-ray surface brightness map from the XMM-Newton observations shows elongated morphology along the E-W direction. The northern subgroup A2163-B is observed to be connected with a faint bridge of gas. Radio halo emission is morphologically well correlated with the thermal gas, and the ridge is co-located at the brightest part of the X-ray emission. The temperature map shows two distributions, east and west subclusters, each having temperatures $\sim$15 keV. The ridge is situated very close to the cold front, near the centre, suggesting that the ridge could be the tracer of the merger shock along the line of sight. 

\item A tight sub-linear correlation between the radio and X-ray surface brightness has been identified through a point-to-point analysis across the full extent of the radio halo. The correlation slope is flatter ($\sim$0.4) for the ridge compared to the halo ($\sim$0.7). We observe that the correlation slope varies from the centre to the outskirts of the cluster, in line with the model expectations. The observed trend of anti-correlation between the spectral index and X-ray surface brightness across the radio halo aligns with the conversion of gravitational energy into the non-thermal component of the ICM, as well as the radial spectral steepening. However, we also find that simple models do not match the details of the profiles, suggesting a non-constant Alfvén speed with distance or a non-constant ratio of CR electrons to the thermal energy budget.

\item We find that the tailed galaxies and diffuse lobes at the cluster periphery could serve as the source of seed electrons that permeate the entire cluster volume during their single injection epoch. Although the area-filling factor of relativistic electrons at uGMRT frequencies is relatively low, these lobes are capable of sustaining a sufficient population of supra-thermal electrons over their fading timescale. Particle acceleration at the cluster outskirts is highly inefficient, and is estimated to be around 0.12\% of the turbulent energy.

\item The multi-wavelength observations reveal a significant merger along the E-W direction with a mass ratio of 9:1. At the first stage, a collision between two similar subgroups is suggested along the SW-NE direction. During the second stage of the merger, as the cores of each sub-cluster come into the closest approach and then separate, the stripping of the cool gas occurs. In this case, the simulations suggest that $\sim$15-40\% of the merger kinetic energy is channelled into the turbulent energy.

\end{enumerate}

In conclusion, the extended diffuse radio emission in A2163 is detected on the scale of r$_{500}$ of the cluster, offering a unique opportunity to probe the non-thermal plasma conditions of the ICM in remarkable detail. This emission probes the presence of non-negligible cosmic rays and magnetic fields across the cluster volume, suggesting dynamical activity and efficient transformation of energy into non-thermal components. These results demonstrate the broader relevance of these processes, indicating they are not one of a kind or limited to any particular merging cluster. Our observational data provide evidence that the non-thermal components, including magnetic fields and relativistic particles, represent at least several percent of the thermal energy. However, this non-thermal energy constitutes only a portion of the overall kinetic energy from turbulent processes. Looking ahead, more sensitive observations with the Square Kilometre Array (SKA) will aim to explore particle acceleration mechanisms in the peripheral regions of clusters, areas which remain largely unexplored.

\section*{Acknowledgements}
We thank the anonymous referee for their constructive comments
that have improved the clarity of the paper. R.S. acknowledges Marco Balboni for many useful discussions. R.S., R.K., and R.J. acknowledge the support of the Department of Atomic Energy, Government of India, under project no. 12-R\&D-TFR-5.02-0700. R.K. also acknowledges the support from the SERB Women Excellence Award WEA/2021/000008. S.G. acknowledges that the basic research in radio astronomy at the Naval Research Laboratory is supported by 6.1 Base funding. G.B. acknowledges partial support from INAF Theory Grant ``Theory and simulations of non-thermal phenomena in galaxy clusters and beyond''. We thank the staff of the GMRT that made these observations possible. The GMRT is run by the National Centre for Radio Astrophysics (NCRA) of the Tata Institute of Fundamental Research (TIFR). The National Radio Astronomy Observatory is a facility of the National Science Foundation operated under a cooperative agreement by Associated Universities, Inc. This research made use of the NASA/IPAC Extragalactic Database (NED), which is operated by the Jet Propulsion Laboratory, California Institute of Technology, under contract with the National Aeronautics and Space Administration. Based on observations obtained with XMM-Newton, an ESA science mission with instruments and contributions directly funded by ESA Member States and NASA.

%%%%%%%%%%%%%%%%%%%%%%%%%%%%%%%%%%%%%%%%%%%%%%%%%%
\section*{Data Availability}
The data that support the figures and plots within this paper and the other findings of this study are available from the corresponding author upon reasonable request.

%%%%%%%%%%%%%%%%%%%% REFERENCES %%%%%%%%%%%%%%%%%%

% The best way to enter references is to use BibTeX:

\bibliographystyle{mnras}
\bibliography{mnras} % if your bibtex file is called example.bib

% Alternatively you could enter them by hand, like this:
% This method is tedious and prone to error if you have lots of references
%\begin{thebibliography}{99}
%\bibitem[\protect\citeauthoryear{Author}{2012}]{Author2012}
%Author A.~N., 2013, Journal of Improbable Astronomy, 1, 1
%\bibitem[\protect\citeauthoryear{Others}{2013}]{Others2013}
%Others S., 2012, Journal of Interesting Stuff, 17, 198
%\end{thebibliography}

%%%%%%%%%%%%%%%%%%%%%%%%%%%%%%%%%%%%%%%%%%%%%%%%%%

%%%%%%%%%%%%%%%%% APPENDICES %%%%%%%%%%%%%%%%%%%%%

\appendix

\section{Comparison of the fluxscale} \label{fluxscale-com}

\begin{figure}
	\includegraphics[width=\columnwidth]{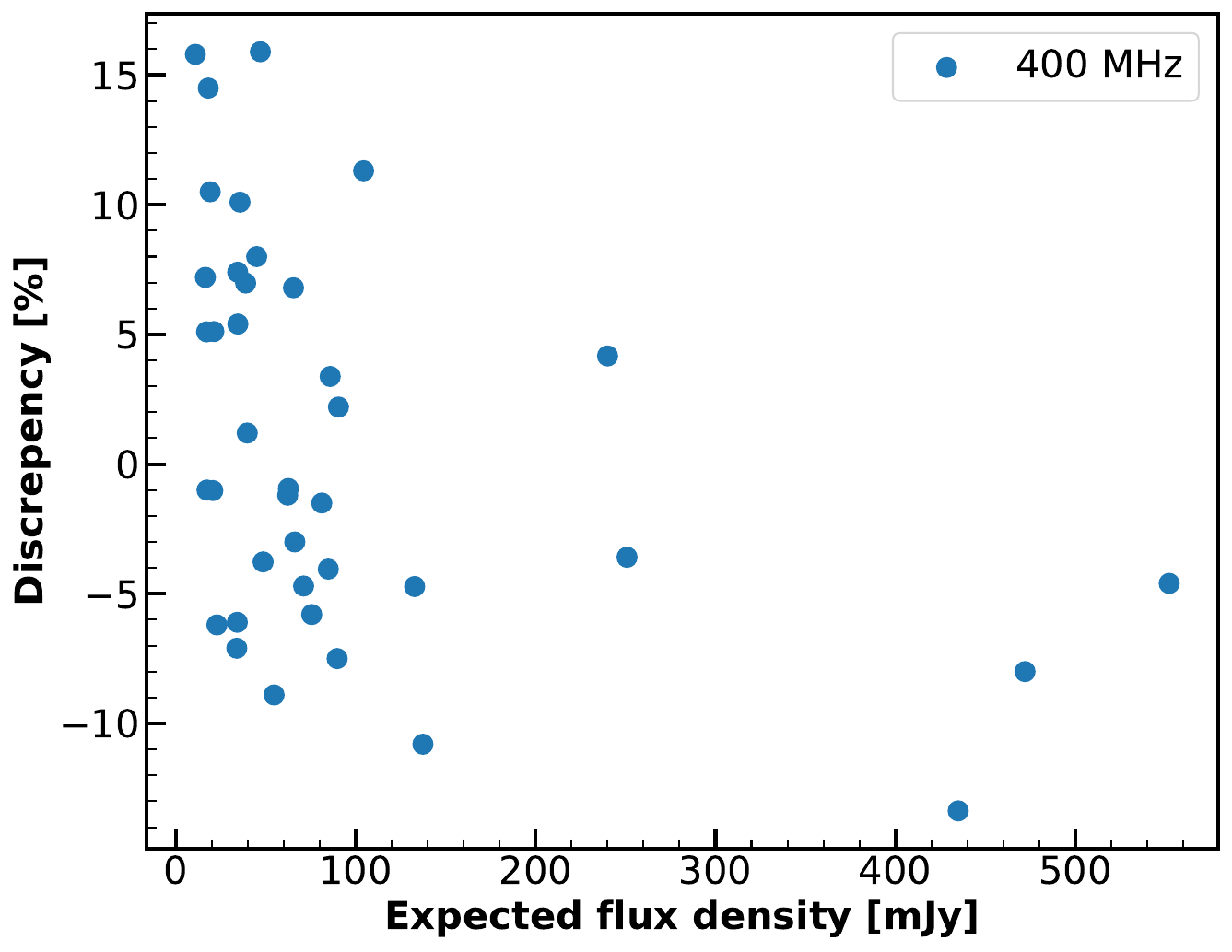}
    \caption{Comparison of flux densities for the compact sources detected in our images and TGSS-ADR1/NVSS at 400 MHz. The expected flux density at the observed frequency is calculated using the spectral index from the TGSS-ADR1 and NVSS flux density.}
    \label{fluxscale-plot}
\end{figure}

 Radio Frequency Interference (RFI) can distort calibration solutions and change the flux density scale by up to 25\% at very low frequencies, below 500 MHz. To verify our scale at 400 MHz, we used compact sources that had reliable measurements in both the TIFR GMRT Sky Survey (TGSS-ADR1; 150 MHz; \citealt{2017A&A...598A..78I}) and the NRAO VLA Sky Survey (NVSS; 1.4 GHz; \citealt{1998AJ....115.1693C}). For each source, an expected flux density at 400 MHz was obtained by interpolating a power-law spectrum between the TGSS and NVSS values. This comparison used compact sources, which were properly deconvolved in both catalogues, with positional coincidences within 15 arcseconds (one third of the NVSS resolution). Figure.~\ref{fluxscale-plot} displays the flux density comparison, with errors of 10\% for TGSS and 5\% for NVSS. The discrepancy between the expected flux density (from the spectral index measurement based on NVSS and TGSS frequencies) and the observed flux density (in our uGMRT images) is mostly within \textless 10\%.

\section{T-T plot between VLA and uGMRT} \label{A2163_tt}

\begin{figure}
\includegraphics[width=\columnwidth]{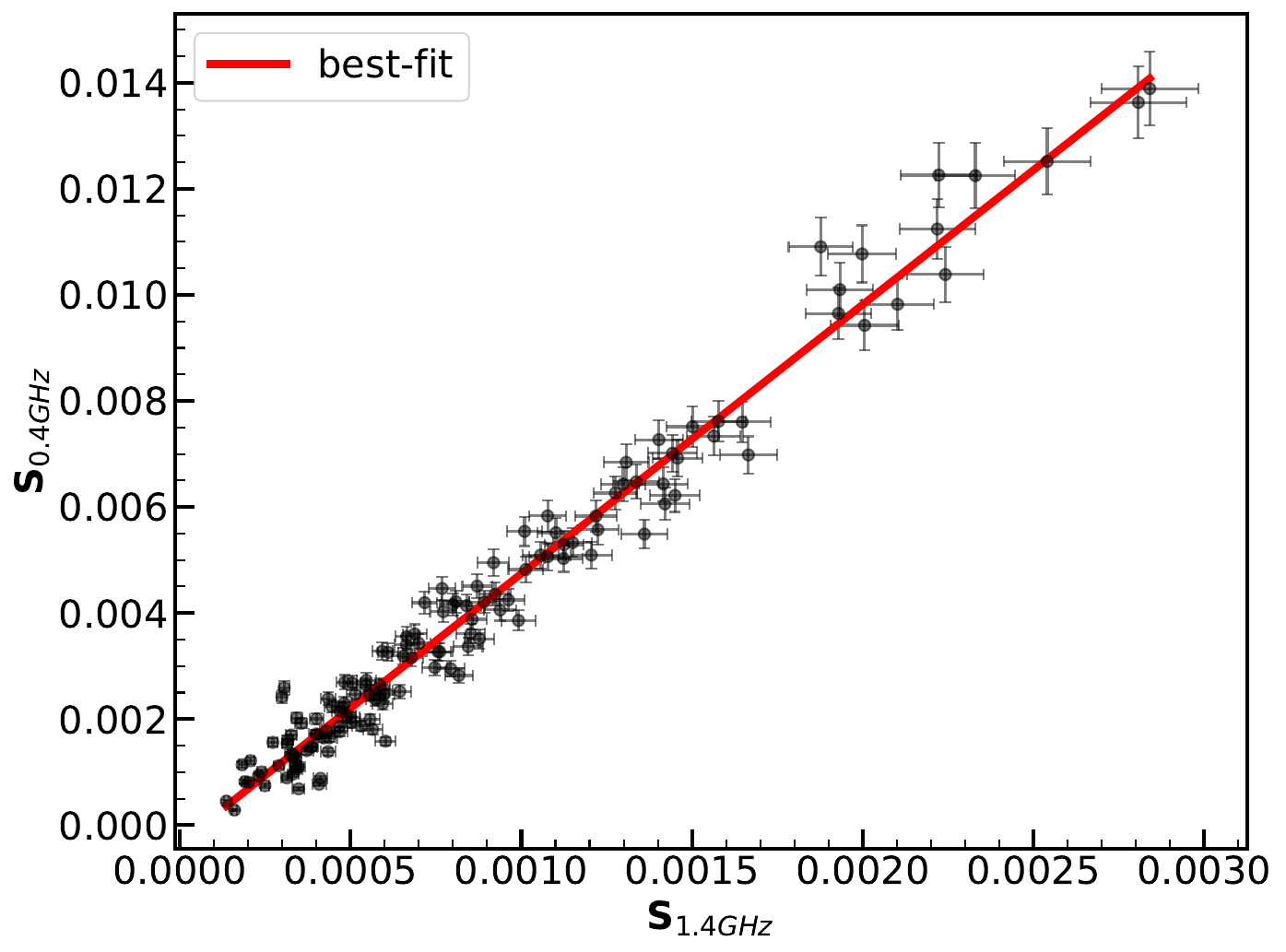}
    \caption{The T-T plot for the radio halo is generated from the VLA and uGMRT 400 MHz images at a common resolution of 35$''$. The red solid line indicates the best-fit linear regression to the data. The error bars on the data points are estimated using the equation.~\ref {eq-flux-err}. The reduced $\chi^{2}$ of the fit is 1.35.}
    \label{tt-plot}
\end{figure}

To investigate potential spatial variations in the spectral index across the halo and to test for any indications of missing flux, we used a T–T plot \citep{1962MNRAS.124..297T} constructed from the 400 and 1400 MHz images, following the approach of \citet{2013MNRAS.436.1286M}. T–T plots provide a robust method for comparing flux densities at two different frequencies obtained from separate instruments, as they help mitigate the effects of differing short-spacing coverage and foreground contamination. A mismatch in short spacing appears as a non-zero intercept on the y-axis. For this analysis, both the 400 MHz and 1400 MHz images were first convolved to the same angular resolution. The flux density was then measured on a pixel-by-pixel basis, with each pixel corresponding to the synthesised beam. A straight-line fit was applied to the data points in the plot (Figure~\ref{tt-plot}) using orthogonal-distance regression, yielding a reduced $\chi^{2} = 1.30$. The best-fit line has an intercept consistent with zero, implying no systematic flux offset between the two frequencies.

\section{Error map of the spectral index} \label{err-map}

\begin{figure*}
    \includegraphics[width=9.5cm, height = 8cm]{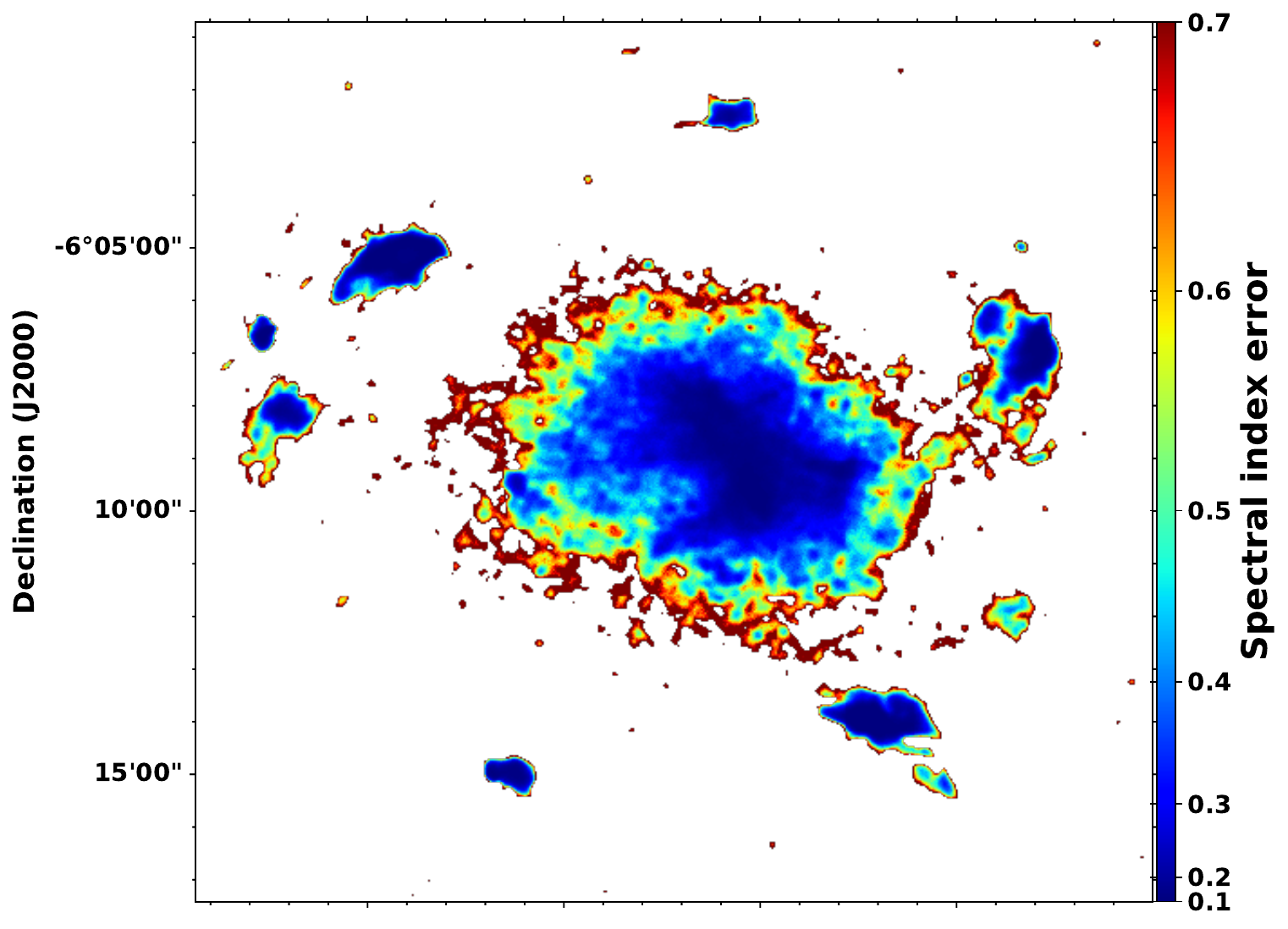}
    \includegraphics[width=8.2cm, height=8cm]{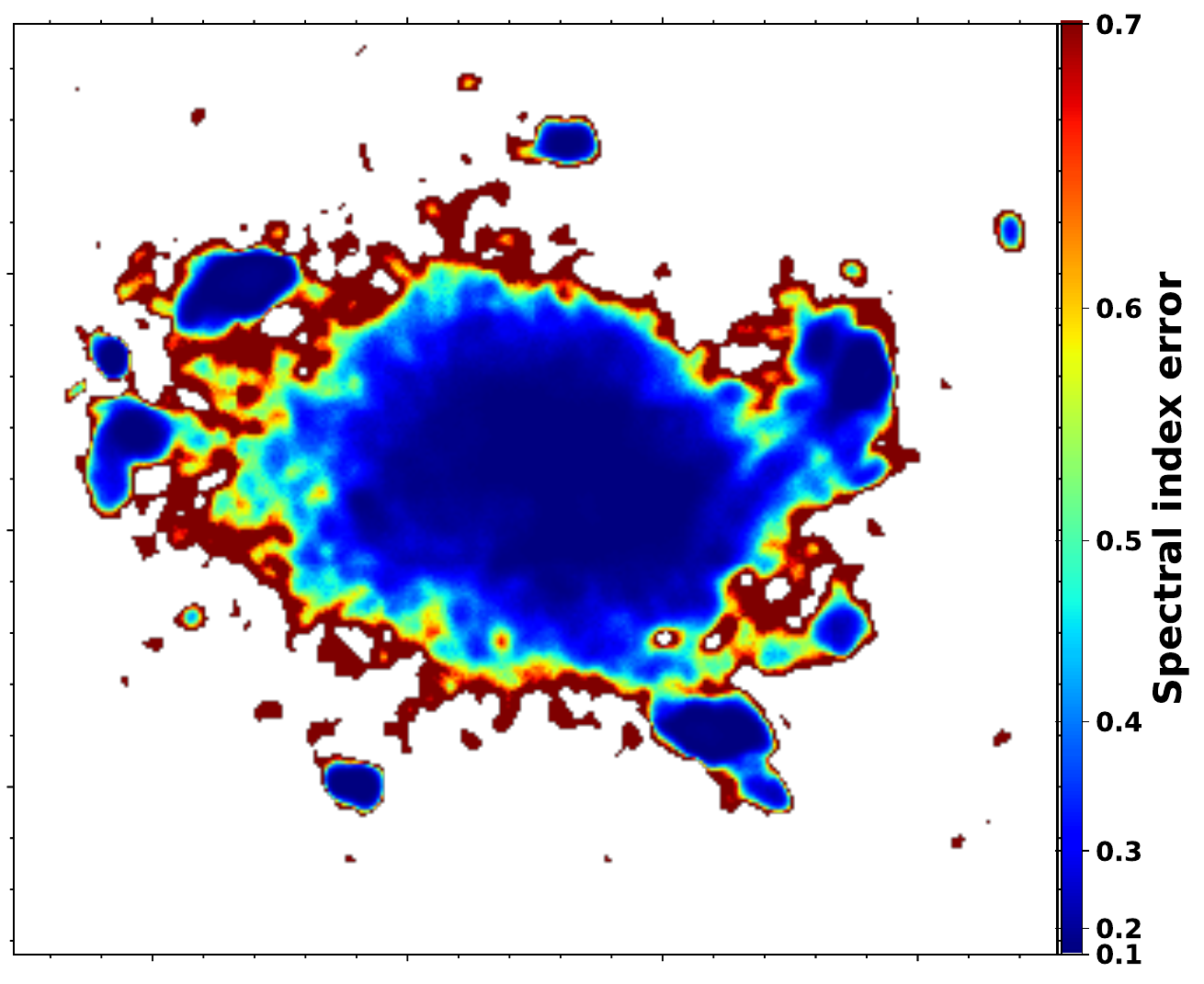}
    \includegraphics[width=9.5cm, height = 8cm]{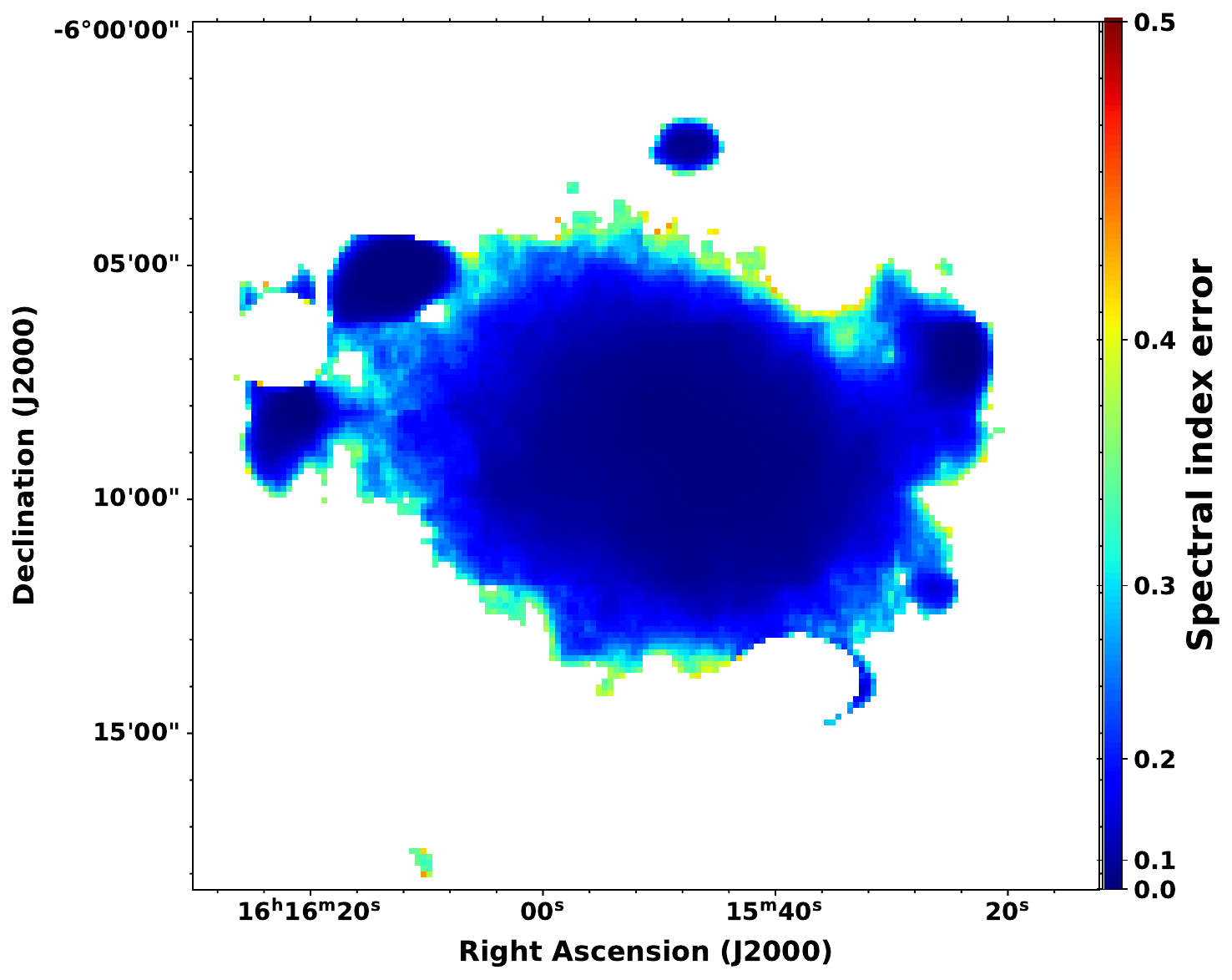}
    \includegraphics[width=8.2cm, height=8cm]{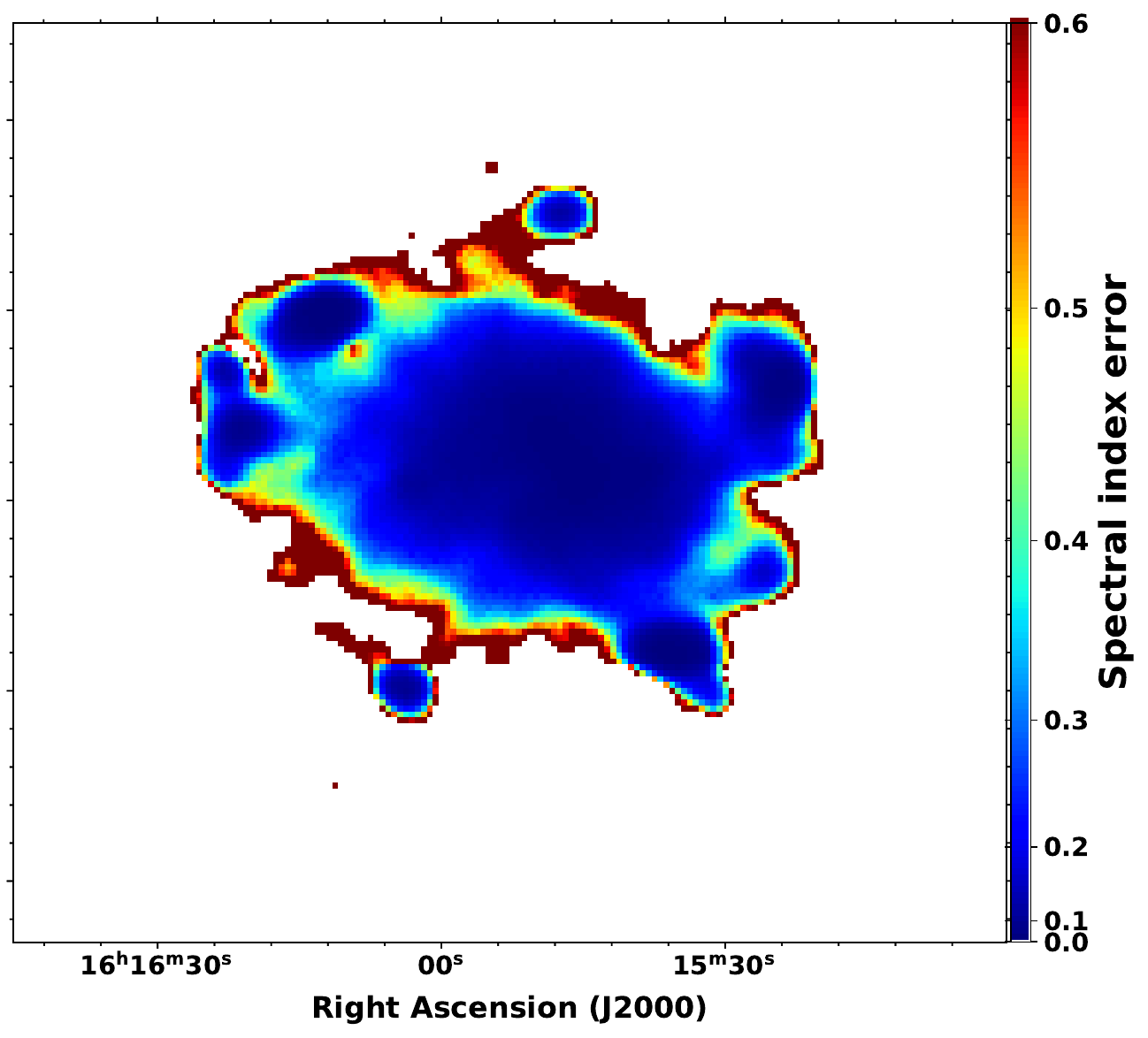}

    \caption{\textit{Upper left:} Spectral index error map between 400 and 650 MHz, at a resolution of 10$''$. In this resolution, the halo emission is not detected to a large extent. \textit{Upper right:} The same is shown but at 20$''$ resolution. The errors mostly dominate below 0.2 in the central region. \textit{Lower left:} The spectral index uncertainty map between the uGMRT 400 and VLA 1400 MHz is shown at 35$''$ resolution. The tailed galaxies T1 and T3 have been masked from both images.  \textit{Lower right:} spectral index error map between 400 and 650 MHz, at a resolution of 45$''$ }
    \label{spec_map_err}    
\end{figure*}

We show here the spectral index uncertainty maps, corresponding to Figure.~\ref{spec_map}.

%%%%%%%%%%%%%%%%%%%%%%%%%%%%%%%%%%%%%%%%%%%%%%%%%%

% Don't change these lines
\bsp	% typesetting comment
\label{lastpage}
\end{document}